\documentclass[usenatbib]{mn2e}
\usepackage{graphicx, txfonts, natbib}
\usepackage{epsf}
\usepackage{amssymb}
\usepackage{epstopdf}
\usepackage{longtable}
\newcommand{\degree}{\mbox{$^\circ$}}
\newcommand{\msun}{\mbox{M$_{\odot}$}}

\newcommand{\rsun}{\mbox{R$_{\odot}$}}

\DeclareMathAlphabet{\mathsc}{OT1}{cmr}{m}{sc}
\def\testbx{bx}%
\DeclareRobustCommand{\ion}[2]{%
\relax\ifmmode
\ifx\testbx\f@series
{\mathbf{#1\,\mathsc{#2}}}\else
{\mathrm{#1\,\mathsc{#2}}}\fi
\else\textup{#1\,{\mdseries\textsc{#2}}}%
\fi}
\newcommand{\ha} {\mbox{H$\alpha$}}
\newcommand{\hb} {\mbox{H$\beta$}}

\newcommand{\Feii} {\ion{Fe}{ii}}
\newcommand{\FeiiF}{[\ion{Fe}{ii}]}
\newcommand{\Caii} {[\ion{Ca}{ii}]}
\newcommand{\Ci} {\ion{C}{i}}

\newcommand{\Hei} {\ion{He}{i}}

\newcommand{\Oi} {[\ion{O}{i}]}
\newcommand{\Oinir} {\ion{O}{i}}
\newcommand{\Oii} {[\ion{O}{ii}]}

\newcommand{\Mgi} {\ion{Mg}{i}}
\newcommand{\Scii} {\ion{Sc}{ii}}

\begin{document}

\title[Optical and near infrared coverage of SN 2004et]
  {Optical and near infrared coverage of SN 2004et: physical parameters and comparison with other type IIP supernovae}

\author[K. Maguire et al.]
  {K.~Maguire,$^1$\thanks{E-mail: kmaguire11@qub.ac.uk}
  E.~Di Carlo,$^2$ S. J.~Smartt,$^1$  A.~Pastorello,$^1$ D. Yu.~Tsvetkov,$^3$ S.~Benetti,$^4$
    \newauthor
  S.~Spiro,$^{5,6}$    A. A.~Arkharov,$^{7,8}$
  G.~Beccari,$^{9}$ M. T.~Botticella,$^{1}$ E.~Cappellaro,$^4$
      S.~Cristallo,$^{2,10}$ 
   \newauthor
   M.~Dolci,$^2$
  N.~Elias-Rosa,$^{4,11}$ M.~Fiaschi,$^{12}$ D.~Gorshanov,$^7$ A.~Harutyunyan,$^{4,13}$ 
    \newauthor
V. M.~Larionov,$^{7,8}$  
   H.~Navasardyan,$^4$ A.~Pietrinferni,$^2$ G.~Raimondo,$^2$ G.~Di Rico,$^2$ 
  S.~Valenti,$^{1}$
     \newauthor  
    G.~Valentini,$^2$ 
     L.~Zampieri$^4$ \\
  $^1$Astrophysics Research Centre, School of Maths and Physics, Queen's University Belfast, Belfast BT7 1NN, UK\\
  $^2$INAF Osservatorio Astronomico Collurania di Teramo, Via Mentore Maggini, I-64100, Teramo, Italy\\
  $^3$Sternberg Astronomical Institute, University Ave. 13, 119992 Moscow, Russia\\
  $^4$INAF Osservatorio Astronomico di Padova, Vicolo dell' Osservatorio 5, I-35122 Padova, Italy\\
  $^5$Dipartimento di Fisica, Universit\'a di Roma Tor Vergata, Via della Ricerca scientifica 1, 00133 Roma, Italy\\
  $^6$INAF Osservatorio Astronomico di Roma, Via di Frascati 33, 00040 Monte Porzio Catone, Italy\\
   $^7$Pulkovo Central Astronomical Observatory, Pulkovskoe shosse 65, 196140, St. Petersberg, Russia \\ 
   $^8$Astronomical Institute of St Petersburg State University, Universitetskij Prospect 28, Petrodvorets, 198504 St. Petersburg, Russia\\ 
   $^9$ESA/ESTEC, Keplerlaab 1, 2200 AG Noordwijk, Netherlands\\
    $^{10}$Departamento de Fisica Teorica y del Cosmos, Universidad de Granada, Spain\\
 $^{11}$Spitzer Science Center, California Institute of Technology, 1200 E. California Blvd., Pasadena, California 91125, USA\\
   $^{12}$Dipartimento di Astronomia, Universit\'a di Padova, Vicolo dellÕ Osservatorio 2, I-35122 Padova, Italy \\
 $^{13}$Fundaci\'on Galileo Galilei - INAF, Apartado 565, E-38700 Santa Cruz de La Palma, Spain\\
    }
 
\maketitle

\begin{abstract}
We present new optical and near infrared (NIR) photometry and spectroscopy
of the type IIP supernova (SN), SN 2004et. In combination with already
published data, this provides one of the most complete studies of
optical and NIR data for any type IIP SN from just after explosion to +500 days.
The contribution of the NIR flux to the bolometric light curve is
estimated to increase from 15 per cent at explosion to around 50 per cent at the
end of the plateau and then declines to 40 per cent at 300 days. SN 2004et is one of the most luminous IIP SNe which has been well studied
and characterised, and with a luminosity of log L = 42.3 erg s$^{-1}$ and a $^{56}$Ni mass of 0.06 $\pm$ 0.04 \msun, it is 2 times brighter than SN 1999em. We provide parametrised bolometric corrections as a function of
time since explosion for SN 2004et and three other IIP SNe that have extensive optical and NIR data.
These can be used as templates for future events in optical and NIR surveys without full wavelength coverage.
We compare the physical parameters of SN 2004et with those of
other well studied IIP SNe and find that the kinetic energies span a range of 10$^{50}$--10$^{51}$ ergs.
We compare the ejected masses calculated from hydrodynamic models
with the progenitor masses and limits derived from prediscovery images. Some of the ejected mass estimates
are significantly higher than the progenitor mass estimates, with SN 2004et showing perhaps the
most serious mass discrepancy.
With the current models, it appears difficult
to reconcile 100 day plateau lengths and high expansion velocities with the low ejected masses of 5--6 \msun\
implied from 7--8 \msun\ progenitors.  The nebular phase is studied
using very late time HST photometry, along with optical and NIR spectroscopy. The light curve shows a clear flattening at 600
days in the optical and the NIR, which is likely due to
the ejecta impacting on circumstellar material. We further show that
the \Oi\ 6300, 6364 \AA\ line strengths in the nebular spectra
of four type IIP SNe imply ejected oxygen masses of 0.5--1.5 \msun.
\end{abstract}


\section{Introduction}
\label{intro}
Supernovae (SNe) are classified based on the elements present in their spectra, with type II SNe being distinguished from Type I SNe by the presence of hydrogen \citep[for review see][]{fil97}. The type of explosion is directly determined by the evolutionary status of the progenitor star, which is influenced by the initial mass, metallicity, rotation rate and presence of a binary companion \citep{sma09b}. The progenitor stars of some type II SNe have been shown to be red supergiants \citep{sma04, li05} and in some rare cases blue supergiants \citep{wal87, pas05}. Stellar evolutionary models predict that most single stars with masses in the range $\sim$ 8--30 \msun\ should end their lives as red supergiants and produce type II SNe \citep{eld04, heg03, hir04, lim03}. This is supported by the observational constraints on core-collapse supernova progenitors, although the lack of the detection of high mass progenitors is concerning \citep{sma09b}.

Type II SNe can be further subclassified based on the shape of their light curves. Those that display a linear decrease from peak magnitude are called type IIL SNe, while those that display an extended plateau of nearly constant luminosity are termed type IIP SNe. The plateau phase generally lasts $\sim$ 80--120 days before entering the exponential decay phase and is caused by the diffusion of thermal energy deposited by the shock wave and by the release of internal energy when hydrogen starts to recombine. During recombination, constant luminosity is achieved by the balance of the increase in radius caused by expansion and the inward movement (in Lagrangian coordinates) of the recombination front. It is observationally established that the majority of IIP SN progenitors are red supergiants that initially have masses above 8 \msun\ and up to at least 17 \msun. \cite{sma09} summarised the observed progenitor properties of the nearest IIP SNe and found a lack of high mass red supergiants in this sample. If these estimates of progenitor masses are accurate, it means long plateau phases are produced by ejected envelope masses as low as 6--7 \msun, which are in some cases much lower values than those calculated from light curve modelling \citep{ham03, nad03}. 

Type IIP SNe have been proposed as cosmological standard candles and their use as distance indicators has been demonstrated in \cite{ham02, nug06, poz09}. Therefore a detailed understanding of their pre-explosion parameters and explosion physics is of the utmost importance. Despite being the most common type of SN observed, there is a surprisingly small sample of IIP SNe to date with well monitored optical light curves and spectra from just after explosion through to the radioactive tail phase. In the past twenty years, there have been only four IIP SNe for which extended photometric and spectral monitoring observations have been published; SN 1990E \citep{sch93, ben94}, SN 1999em \citep{elm03, leo02, ham01}, SN 1999gi \citep{leo02b} and SN 2005cs \citep{pas06, tsv06, bro07, des08, pas09}. Coverage at near infrared wavelengths is even less common, with only one type IIP SN having been observed extensively both photometrically and spectroscopically at these wavelengths in the last ten years, SN 1999em \citep{ham01, elm03, kri08}.

SN 2004et was discovered in the nearby starburst galaxy, NGC 6946 by S. Moretti on 2004 September 27 \citep{zwi04}. At a distance of only 5.9 $\pm$ 0.4 Mpc \citep{kar00}, it was an ideal candidate for intensive follow-up observations.  A high resolution spectrum of SN 2004et was obtained with the Mt. Ekar 1.82-m telescope on 2004 September 28 that confirmed it was a type II event with spectra showing prominent H Balmer lines with P-Cygni profiles \citep{zwi04}. A total extinction of \textit{E(B-V)} = 0.41 $\pm$ 0.007 was also estimated by \cite{zwi04} from the equivalent width of the {\ion{Na}{ID} lines in the high resolution spectrum. SN 2004et (R.A.: 20$\rm^{h}$ 35$\rm^{m}$ $25\rm^{s}$.33, Decl.: +60$\degree$ 07' 17".7 (J2000)) was located in one of the spiral arms of the galaxy, which are known to be regions of high star formation. NGC 6946 is presently the galaxy with the highest number of SNe discovered to date, 9 SNe since the first in SN 1917A with the most recent being SN 2008S, which has been suggested by \cite{bot09} as a probable electron-capture event, although its nature has been debated \citep{smi09}.

\cite{sah06} reported optical photometric and spectroscopic observations of SN 2004et from 8--541 days post explosion. They noted that SN 2004et was at the brighter end of SNe IIP luminosities and estimated the mass of $^{56}$Ni synthesised during the explosion to be 0.06 $\pm$ 0.02 \msun. They also suggested that the steepening of the decline rates of the optical luminosity one year after explosion along with a blueshift in the emission lines at a similar epoch was an indication of dust formation. \cite{mis07} presented optical photometry of SN 2004et from $\sim$ 14--470 days post explosion and in agreement with \cite{sah06}, estimated an ejected $^{56}$Ni mass of 0.06 $\pm$ 0.03 \msun.

SN 2004et was detected at radio frequencies just 14 days post explosion \citep{sto04} and this early detection suggests the presence of appreciable circumstellar material (CSM) around the SN \citep{sah06}. The SN was extensively monitored by \cite{sto04} using the VLA at 22.4 and 8.4 GHz and by \cite{bes04} using MERLIN at 4.9 GHz. SN 2004et was also observed on 2005 January 02 using the GMRT at 1.4 GHz \citep{mis07} and an 8.4 GHz VLBI observation was obtained by \cite{mar07} on 2005 Feburary 20 showing a clear asymmetry in the emission structure that can be explained if the SN ejecta expanded in a clumpy CSM. The radio luminosity of SN 2004et was found to be among the highest for IIP SNe \citep{che06} with an estimated pre-supernova mass-loss rate of 1.5--3 $\times$ 10$^{-6}$ \msun\ yr$^{-1}$ (using a wind velocity of 10 km s$^{-1}$), which suggests a progenitor mass of $\sim$ 20 \msun.
X-ray data of SN 2004et were also obtained using the Chandra X-ray Observatory at 30, 45 and 72 days post explosion \citep{rho07, mis07}. The unabsorbed X-ray luminosity suggests a wind density parameter value of 2.5, which is a factor of 2 larger than the wind densities calculated for SN 1999em and SN 2004dj \citep{chu07}. This gives a pre-supernova mass-loss rate of 2--2.5 $\times$ 10$^{-6}$ \msun\ yr$^{-1}$, which is consistent with estimate from the radio observations.

\cite{kot09} presented mid infrared (MIR) observations of SN 2004et obtained with the Spitzer Space Telescope ranging from 64--1240 days post explosion, along with three very late-time optical spectra. They reported spectroscopic evidence for silicate dust formation in the ejecta of SN 2004et with a total mass for the dust of mass of a few times 10$^{-4}$ \msun, which would not make a major contribution to the total mass of cosmic dust. The most prominent spectral emission lines in the very late-time optical spectra are observed to display boxy profiles, which was suggested to be a signature of ejecta-CSM interaction.

\cite{li05} reported that the progenitor of SN 2004et had been identified in a pre-explosion optical image of NGC 6946 as a yellow supergiant. However, \cite{cro09} have shown that the source indicated as the progenitor by \cite{li05} is still visible in images taken after SN 2004et has faded and so cannot be the progenitor star. At the same time, \cite{cro09} also identified an alternative progenitor star in a pre-explosion \textit{i}' band image taken at the Isaac Newton Telescope with the Wide Field Camera. This star was not detected in pre-explosion \textit{V} or \textit{R} band images suggesting that the progenitor of SN 2004et is either intrinsically red or surrounded by dust. The luminosity of the progenitor was estimated using the \textit{I} band photometry and  \textit{R}--\textit{I} colour limits. Using the \textsc{STARS} stellar evolutionary model \citep{eld04}, the progenitor star was found consistent with a late K to late M-type supergiant with an initial mass of 8$_{-1}^{+5}$ \msun\ (see \cite{cro09} for more details).

In this paper, we present our own complete data set for SN 2004et making one of the most comprehensive sets of optical and NIR photometry and spectroscopy for a type IIP SNe, with data coverage from soon after explosion through to the radioactive tail phase. The information obtained from analysis of the light curve and spectra is compared to the progenitor information obtained from the pre-explosion images to explore if the resulting progenitor properties are consistent. Extensive NIR spectroscopy is rare for type IIP SNe and so we investigate what further results can be obtained from analysis of NIR spectral features. SN 2004et is also included in a wider sample of nearby type IIP SNe to determine how its properties fit in the overall picture for the core-collapse of massive, hydrogen-rich stars. 

\section{Observations and data reduction}
\label{photoptnir}
\subsection{Optical photometry}

Optical photometric observations of SN 2004et were carried out using a range of telescopes, beginning soon after its discovery (2004 September 27) until over three years post explosion (2008 January 18). Observations at very early times were taken with a 0.41-m telescope equipped with an Apogee AP47p CCD in \textit{BVRI} filters and the 1.82-m Copernico telescope equipped with AFOSC, located at Mt. Ekar-Asiago (Italy). Three epochs of data were  taken with the 3.58-m Telescopio Nazionale Galileo (TNG) using DOLORES, located at Roque de Los Muchachos Observatory, La Palma, Canary Islands, along with data from the 0.72-m TNT at the Collurania Observatory (Teramo, Italy).  Photometric observations were also obtained with the 0.7-m AZT2 telescope at the Sternberg Astronomical Institute, Moscow, the 0.6-m Z600 telescope at the Crimean Laboratory of SAI, Nauchny, Crimea, the 0.38-m KGB telescope at the Crimean Astrophysical Observatory, Nauchny, Crimea and the 1-m SAO Z1000 telescope at the Special Astrophysical Observatory of the Russian Academy of Sciences, Zelenchuk. Two late time observations were also taken with the 2.56-m Nordic Optical Telescope (NOT) at Roque de Los Muchachos Observatory, La Palma, Canary Islands using ALFOSC. 

All the images were trimmed, bias subtracted and flat-field corrected using the standard \textsc{iraf}\footnote{\textsc{iraf} is distributed by the National Optical Astronomy Observatories, which are operated by the Association of Universities for Research in Astronomy, Inc., under the cooperative agreement with the National Science Foundation.} tasks. A sequence of stars in the field of view of SN 2004et were calibrated with respect to the sequence star magnitudes given in \cite{mis07}. The sequence star magnitudes of \cite{mis07} were also confirmed independently using standard star observations obtained on some of the nights.  The errors on the reported magnitudes for SN 2004et were determined by adding in quadrature the errors on the instrumental magnitudes and the errors due to the magnitude calibration. Figure \ref{findingchart} shows a finding chart for SN 2004et, with the secondary standard stars marked.

Additional late time Hubble Space Telescope (HST) WFPC2 observations were obtained as part of the proprietary program GO-11229 (PI: Meixner) in July 2007 and January 2008, which have since been made public. These images were downloaded from the Space Telescope Science Institute (STScI) archive using the OTFR pipeline. Observations were obtained in two filters \textit{F606W} and \textit{F814W} that are equivalent to the Johnson \textit{V} and \textit{I} bands respectively.  PSF-fitting photometry was performed on the data using the HSTphot package \citep{dol00}. HSTphot corrects for chip-to-chip variations and aperture corrections. No colour corrections have been applied to convert to \textit{V} and \textit{I} filters because the colour transformations to the Johnson-Cousins magnitude system are not well constrained for late time SN spectra. The magnitude values quoted in Table \ref{opt_phot} are for the HST filters in the Vegamag system. One epoch of very late time \textit{BVRI} data was also obtained with the William Herschel Telescope (WHT) at Roque de Los Muchachos Observatory, La Palma, Canary Islands using the Auxiliary Port Imager (AUX).

\begin{figure}
\includegraphics[width=8.4cm]{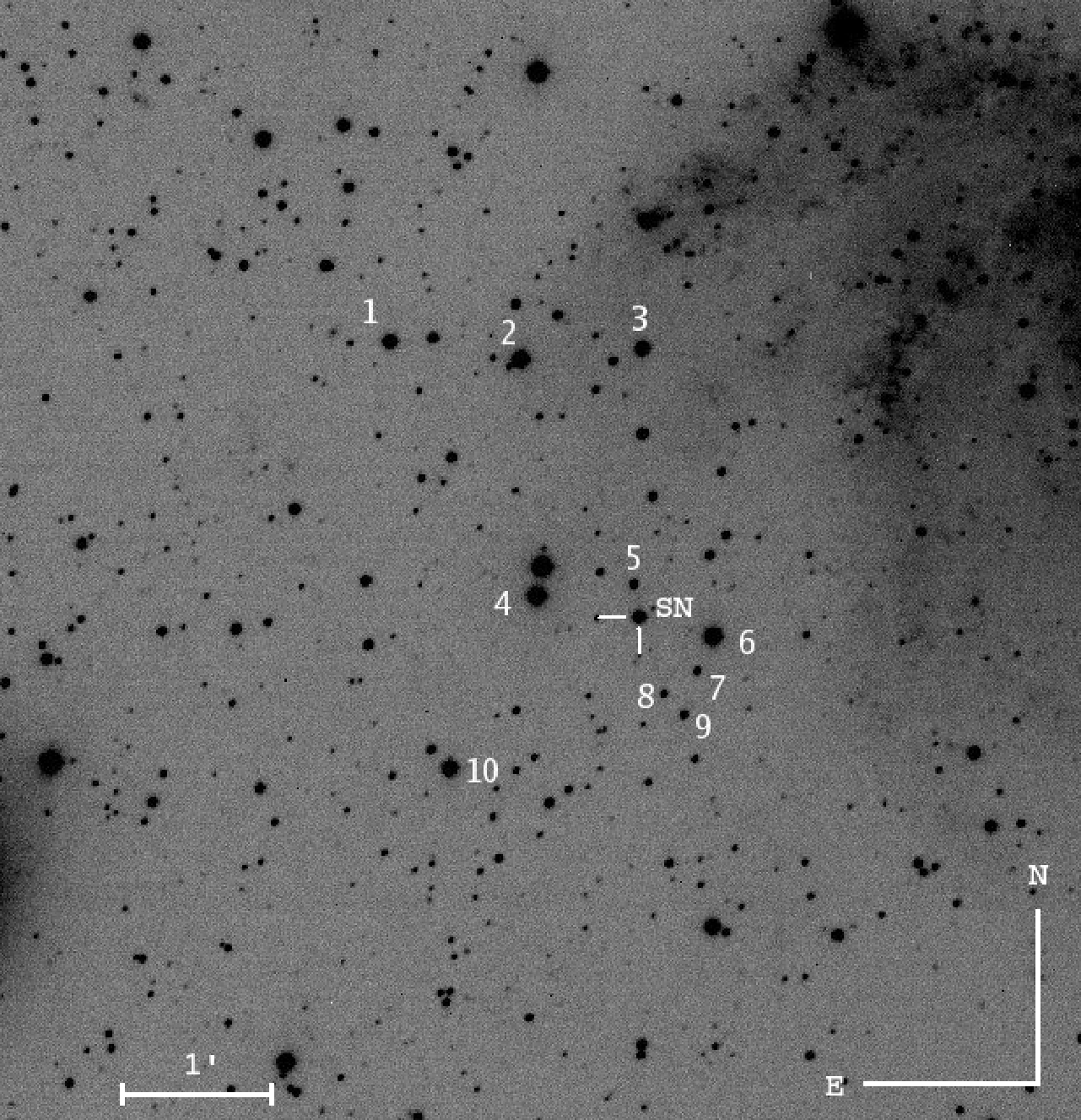}
\caption{The field of SN 2004et with the supernova and comparison stars marked. This image was taken with the 1.82m telescope at Mt. Ekar, Italy $\sim$ 3 weeks post explosion in the \textit{V} band.}
\label{findingchart}
\end{figure}

\begin{table*}
 \caption{Log of optical photometric observations of SN 2004et.}
 \label{opt_phot}
 \begin{tabular}{lcccccccc}
 \hline
 \hline
  Date  &  JD (2450000+)  & Phase* (days) &  	\textit{U}  	&   	\textit{B}  	&	    \textit{V} & \textit{R} & \textit{I}    &Source  \\
  \hline
29/09/2004  &53278.0& 7.5 &\ &12.94$\pm$0.01&  12.67$\pm$0.01&  12.35$\pm$0.02&  12.13$\pm$0.02 &TNT 0.72m \\
01/10/2004 & 53279.9   & 9.4       &    \ &	    12.90$\pm$0.04 &    12.60$\pm$0.02 &   12.28$\pm$0.02 &    12.01$\pm$0.02 &   0.41m+AP47p        \\   
04/10/2004  &53283.0& 12.5 &\ &12.98$\pm$0.01&  12.65$\pm$0.02&  12.27$\pm$0.01&  12.01$\pm$0.01 &TNT 0.72m \\
07/10/2004  &53286.0& 15.5&\ &13.02$\pm$0.01&  12.64$\pm$0.01&  12.26$\pm$0.02&  11.98$\pm$0.02  &TNT 0.72m\\
10/10/2004 & 53288.9   & 18.4       &    \ &	    13.00$\pm$0.02 &    12.58$\pm$0.02 &   12.23$\pm$0.01 &    11.98$\pm$0.02 &   0.41m+AP47p       \\   
15/10/2004 & 53294.2   &  23.7      &   13.04$\pm$0.04&   13.16$\pm$0.01 &    12.61$\pm$0.01 &   12.16$\pm$0.01 &    11.94$\pm$0.01 &     AZT2 0.7m      \\   
17/10/2004 & 53295.9   &  25.4     &   \ &	    13.27$\pm$0.02 &    12.63$\pm$0.01 &   12.22$\pm$0.01 &    11.92$\pm$0.01 &    0.41m+AP47p     \\   
19/10/2004 & 53297.1  & 26.5      &   \ &	                 \ &         \ &      12.24$\pm$0.01 &    11.94$\pm$0.01 & Ekar 1.82m+AFOSC  \\   
19/10/2004  &53298.0& 27.5&\ &13.41$\pm$0.02&  12.67$\pm$0.01&  12.22$\pm$0.01&  11.86$\pm$0.01  &TNT 0.72m\\
28/10/2004 & 53307.4  &   36.9     &   14.37$\pm$0.04&   13.79$\pm$0.01 &    12.79$\pm$0.01 &   12.26$\pm$0.01 &    11.95$\pm$0.01 &       AZT2 0.7m       \\   
30/10/2004 & 53309.4  &    38.9    &   14.51$\pm$0.03&   13.86$\pm$0.01 &    12.81$\pm$0.01 &   12.29$\pm$0.01 &    11.95$\pm$0.01 &     Z600 0.6m       \\      	  
02/11/2004 & 53312.4  &  41.9      &   14.70$\pm$0.09 &  13.93$\pm$0.01 &    12.83$\pm$0.01 &   12.27$\pm$0.01 &    11.95$\pm$0.01 &     AZT2 0.7m       \\   
05/11/2004 & 53315.4  &     54.9     &   14.85$\pm$0.10 &  14.04$\pm$0.01 &    12.87$\pm$0.01 &   12.30$\pm$0.01 &    11.97$\pm$0.01 &      Z600 0.6m     \\   
06/11/2004 & 53316.3  &    45.8    &   15.08$\pm$0.03&   14.10$\pm$0.01 &    12.95$\pm$0.01 &   12.40$\pm$0.01 &    12.01$\pm$0.01 &      Z600 0.6m    \\   
07/11/2004 & 53317.2  &     46.7   &   15.07$\pm$0.01&   14.08$\pm$0.01 &    12.88$\pm$0.01 &   12.32$\pm$0.01 &    11.95$\pm$0.01 &      Z600 0.6m     \\   
08/11/2004 & 53317.9   & 47.4       &   \ &	    14.12$\pm$0.01 &    12.89$\pm$0.01 &   12.36$\pm$0.01 &    11.93$\pm$0.01 &   0.41m+AP47p         \\   
08/11/2004 & 53318.5  &   48.0       &   15.09$\pm$0.17 &  14.08$\pm$0.01 &    12.89$\pm$0.01 &   12.38$\pm$0.01 &    11.97$\pm$0.01 &    Z600 0.6m    \\   
09/11/2004 & 53319.4  &   48.9   &   15.15$\pm$0.07 &  14.13$\pm$0.01 &    12.91$\pm$0.01 &   12.34$\pm$0.01 &    11.96$\pm$0.01 &    Z600 0.6m   \\   
10/11/2004  &53320.0& 49.5&\ & \ &  12.95$\pm$0.03&  12.37$\pm$0.03& \ &TNT 0.72m\\
10/11/2004 & 53320.4  &  49.9    &   15.25$\pm$0.09&   14.15$\pm$0.01 &    12.90$\pm$0.01 &   12.34$\pm$0.01 &    11.95$\pm$0.01 &    Z600 0.6m   \\   
14/11/2004 & 53323.9   &  52.6      &   15.38$\pm$0.02 &  14.25$\pm$0.01 &    12.91$\pm$0.02 &   12.36$\pm$0.01 &    11.91$\pm$0.01 & TNG+LRS       \\   
16/11/2004 & 53325.8   &  55.3     &    \ &	    14.30$\pm$0.01 &    12.99$\pm$0.01 &   12.39$\pm$0.01 &    11.92$\pm$0.01 &    Ekar 1.82m+AFOSC    \\   
16/11/2004  &53326.0& 55.5&\ &14.22$\pm$0.02&  12.94$\pm$0.03&  12.36$\pm$0.02&  11.92$\pm$0.03  &TNT 0.72m\\
18/11/2004  &53328.0&57.5 &\ &14.30$\pm$0.02&  12.99$\pm$0.02&  12.38$\pm$0.01&  11.93$\pm$0.01  &TNT 0.72m\\
19/11/2004 & 53329.0   &   58.5   &    \ &	    14.36$\pm$0.02 &    13.01$\pm$0.01 &   12.40$\pm$0.01 &    11.91$\pm$0.01 &    Ekar 1.82m+AFOSC      \\   
19/11/2004 & 53329.2  &  58.7   &   15.65$\pm$0.02  &14.32$\pm$0.01 &    12.97$\pm$0.01 &   12.36$\pm$0.01 &    11.96$\pm$0.01 &      Z600 0.6m      \\   
20/11/2004 & 53329.8   &  59.3      &     \ &	    14.39$\pm$0.01 &    12.98$\pm$0.01 &   12.38$\pm$0.01 &    11.93$\pm$0.01 &    0.41m+AP47p     \\   
21/11/2004 & 53331.2  &    60.7   &   15.60$\pm$0.07&  14.37$\pm$0.02 &    12.98$\pm$0.01 &   12.37$\pm$0.01 &    11.97$\pm$0.01 &       Z600 0.6m        \\   
22/11/2004  &53332.0&61.5 &\ &14.40$\pm$0.02&  13.02$\pm$0.03&  \ &  \ &TNT 0.72m\\
22/11/2004 & 53332.2  &  61.7  &    \ &	    14.45$\pm$0.04 &    12.95$\pm$0.03 &   12.31$\pm$0.02 &    11.92$\pm$0.03 &          KGB 0.38m      \\   
23/11/2004  &53333.0&62.5 &\ &14.38$\pm$0.02&  13.02$\pm$0.03&  12.42$\pm$0.02&  11.91$\pm$0.02  &TNT 0.72m\\
25/11/2004 & 53335.3  &    64.8  &   15.98$\pm$0.12 & 14.42$\pm$0.02 &    13.01$\pm$0.01 &   12.38$\pm$0.01 &    11.97$\pm$0.01 &      AZT2 0.7m     \\   
10/12/2004 & 53349.8   &  79.3     &   \ &	    14.73$\pm$0.01 &    13.14$\pm$0.01 &   12.45$\pm$0.01 &    11.93$\pm$0.01 &    Ekar 1.82m+AFOSC       \\   
13/12/2004 & 53352.7   &   82.2     &   \ &	    14.77$\pm$0.01 &    13.14$\pm$0.01 &   12.47$\pm$0.01 &     \ &          Ekar 1.82m+AFOSC    \\   
13/12/2004  &53353.0& 82.5 &\ &14.68$\pm$0.01&  13.16$\pm$0.01&  12.49$\pm$0.01&  11.97$\pm$0.02 &TNT 0.72m \\
14/12/2004  &53354.0& 83.5 &\ &14.68$\pm$0.01&  13.15$\pm$0.01&  12.52$\pm$0.02&  11.99$\pm$0.01 &TNT 0.72m \\
15/12/2004 & 53354.8   &  84. 3     &   \ &	    14.88$\pm$0.04 &    13.17$\pm$0.03 &   12.49$\pm$0.03 &    11.93$\pm$0.03 &  Ekar 1.82m+AFOSC       \\   
23/12/2004  &53363.0& 92.5 &\ &14.86$\pm$0.02&  13.24$\pm$0.02&  12.57$\pm$0.02&  11.99$\pm$0.02 &TNT 0.72m \\
28/12/2004  &53368.0& 97.5 &\ &14.93$\pm$0.03&  13.28$\pm$0.01&  12.61$\pm$0.01&  12.10$\pm$0.01 &TNT 0.72m\\
29/12/2004 & 53369.3  &  98.8  &   \ &\ &	    13.32$\pm$0.01 &   12.56$\pm$0.01 &    12.12$\pm$0.01  &    AZT2 0.7m       \\   
03/01/2005 & 53373.8   &  103.3      &   \ &	    15.14$\pm$0.04 &    13.41$\pm$0.01 &   12.69$\pm$0.01 &    12.16$\pm$0.01 &   Ekar 1.82m+AFOSC       \\   
05/01/2005  &53376.0& 105.5 &\ &15.18$\pm$0.04&  13.46$\pm$0.03&  12.83$\pm$0.04&  12.10$\pm$0.05&TNT 0.72m  \\
07/01/2005  &53378.0&107.5 &\ &15.23$\pm$0.02&  13.54$\pm$0.02&  12.80$\pm$0.02&  12.27$\pm$0.03 & TNT 0.72m\\
11/01/2005 & 53382.2  &   111.7 &   \ &	    15.47$\pm$0.01 &    13.70$\pm$0.01 &   12.82$\pm$0.01 &     \ &     AZT2 0.7m         \\   
14/01/2005 & 53384.8   &   114.3    &   \ &	    15.68$\pm$0.01 &    13.81$\pm$0.01 &   13.02$\pm$0.01 &    12.37$\pm$0.01 &   Ekar 1.82m+AFOSC        \\   
17/01/2005  &53388.0& 117.5 &\ &15.74$\pm$0.03&  13.95$\pm$0.03&  13.19$\pm$0.03&  12.56$\pm$0.03&TNT 0.72m  \\ 
18/01/2005 & 53389.3  &   118.8  &   \ &	    15.79$\pm$0.05 &    14.09$\pm$0.03 &   13.11$\pm$0.02 &    12.64$\pm$0.02 &      AZT2 0.7m        \\   
20/01/2005  &53391.5& 120.5&\ &16.06$\pm$0.04&  14.21$\pm$0.04&  13.42$\pm$0.04&  12.77$\pm$0.04 &TNT 0.72m \\
04/02/2005 & 53406.2  & 135.7    &   \ &	    17.34$\pm$0.05 &    15.64$\pm$0.03 &   14.47$\pm$0.03 &    13.86$\pm$0.02 &       AZT2 0.7m    \\   
10/02/2005  &53412.0&141.5 &\ &\ &                                                    15.66$\pm$0.06&\                               & \                              &TNT 0.72m \\
10/02/2005 & 53412.2  &   141.7  &   \ &	    17.36$\pm$0.09 &    15.75$\pm$0.06 &   14.56$\pm$0.02 &    13.97$\pm$0.02 &       AZT2 0.7m       \\   
03/03/2005 & 53432.5  & 162.0    &   \ &	    17.63$\pm$0.07 &    15.93$\pm$0.04 &   14.67$\pm$0.03 &    14.14$\pm$0.02 &      AZT2 0.7m       \\   
09/03/2005 & 53439.2   &  168.7    &   \ &	    17.74$\pm$0.02 &    15.97$\pm$0.01 &   14.89$\pm$0.01 &    14.06$\pm$0.01 &    Ekar 1.82m+AFOSC   \\   
17/03/2005 & 53446.6  & 176.1   &   \ &	    17.55$\pm$0.06 &    16.12$\pm$0.04 &   14.81$\pm$0.03 &    14.26$\pm$0.03 &      AZT2 0.7m      \\   
18/03/2005 & 53447.0   & 176.5       &   \ &	    17.75$\pm$0.01 &    16.04$\pm$0.01 &   14.96$\pm$0.01 &      \ &    Ekar 1.82m+AFOSC     \\   
27/03/2005 & 53456.5  &  186.0   &   \ &	    17.77$\pm$0.05 &    16.14$\pm$0.03 &   14.92$\pm$0.03 &    14.39$\pm$0.02 &      AZT2 0.7m      \\   
29/03/2005 & 53458.5  &  188.0   &   \ &	    17.77$\pm$0.08 &    16.21$\pm$0.05 &   14.92$\pm$0.03 &    14.41$\pm$0.02 &      AZT2 0.7m       \\   
08/04/2005 & 53468.5  &  198.0   &   \ &	    17.78$\pm$0.17 &    16.24$\pm$0.12 &   14.96$\pm$0.05 &    14.47$\pm$0.06 &       AZT2 0.7m      \\   
13/04/2005 & 53473.5  &   203.0  &   \ &	    17.87$\pm$0.05 &    16.35$\pm$0.03 &   15.06$\pm$0.03 &    14.53$\pm$0.03 &      AZT2 0.7m  \\   
18/05/2005 & 53509.5  &    239.0  &   \ &	    17.94$\pm$0.07 &    16.69$\pm$0.04 &   15.46$\pm$0.03 &    14.99$\pm$0.03 &       AZT2 0.7m       \\   
15/06/2005 & 53537.0   &  266.5      &   \ &	    18.35$\pm$0.02 &    16.97$\pm$0.02 &   15.76$\pm$0.01 &    14.99$\pm$0.03 &  Ekar 1.82m+AFOSC     \\   
02/07/2005 & 53554.0   &  283.5     &   \ &	    18.44$\pm$0.03 &    17.13$\pm$0.02 &   15.88$\pm$0.01 &    15.24$\pm$0.02 &      Ekar 1.82m+AFOSC        \\   
13/07/2005 & 53565.4  &   294.9       &   \ &	   18.48$\pm$0.07 &    17.39$\pm$0.04 &   16.11$\pm$0.03 &    15.74$\pm$0.03 &       AZT2 0.7m	   \\	
\end{tabular}
\end{table*}

\begin{table*}
 \contcaption{}
 \begin{tabular}{lcccccccc}
 Date  &  JD (2450000+)  & Phase* (days) &  	\textit{U}  	&   	\textit{B}  	&	    \textit{V} & \textit{R} & \textit{I}    &Source  \\
  \hline
14/07/2005 & 53566.5  &   296.0          &   \ &	   18.52$\pm$0.03 &    17.20$\pm$0.03 &   16.14$\pm$0.03 &    15.83$\pm$0.03 &     SAO Z1000 1m	  \\  
28/07/2005 & 53580.4  & 309.9    &   \ &	   18.62$\pm$0.08 &    17.33$\pm$0.03 &   16.27$\pm$0.03 &    15.99$\pm$0.12 &      AZT2 0.7m	     \\   
04/08/2005 & 53587.4  &  316.9       &   \ &	   18.70$\pm$0.06 &    17.53$\pm$0.04 &   16.34$\pm$0.03 &    16.08$\pm$0.03 &      AZT2 0.7m		   \\	
10/08/2005 & 53593.4  &   322.9    &   \ &	   18.75$\pm$0.06 &    17.47$\pm$0.05 &   16.40$\pm$0.03 &    16.19$\pm$0.03 &       AZT2 0.7m		   \\	
16/08/2005 & 53599.4  &  328.9 &   \ &	    18.86$\pm$0.07 &    17.59$\pm$0.10 &   16.49$\pm$0.03 &    16.23$\pm$0.04 &       AZT2 0.7m   	    \\   
24/08/2005 & 53607.4  &  336.9   &   \ &	    18.84$\pm$0.13 &    17.76$\pm$0.08 &   16.65$\pm$0.03 &    16.42$\pm$0.03 &       AZT2 0.7m  	    \\   
28/08/2005 & 53611.2   &  340.7      &   20.17$\pm$0.03 & 18.91$\pm$0.02 &    17.78$\pm$0.02 &   16.71$\pm$0.01 &    16.12$\pm$0.01 &   TNG+LRS      \\   
30/08/2005 & 53613.4  &   342.9&   \ &	    18.84$\pm$0.05 &    17.77$\pm$0.03 &   16.68$\pm$0.03 &    16.53$\pm$0.05 &    Z600 0.6m       \\   
31/08/2005 & 53614.5  &   344.0 &   \ &	    18.67$\pm$0.07 &    17.87$\pm$0.05 &    \ &          16.43$\pm$0.09 &     Z600 0.6m       \\   
07/09/2005 & 53621.3  & 350.8  &   \ &	    18.80$\pm$0.08 &    17.95$\pm$0.12 &   16.76$\pm$0.04 &    16.66$\pm$0.06 &    AZT2 0.7m      \\   
14/09/2005 & 53628.1   & 357.6       &   \ &	    19.08$\pm$0.03 &    17.97$\pm$0.02 &   17.03$\pm$0.02 &    16.39$\pm$0.02 &    NOT+ALFOSC      \\   
14/09/2005 & 53628.4  &  357.9  &   \ &	    18.95$\pm$0.06 &    17.84$\pm$0.06 &   16.81$\pm$0.04 &    16.61$\pm$0.05 &     AZT2 0.7m        \\   
03/10/2005 & 53647.4  &  376.9 &   \ &	     \ &       18.32$\pm$0.18 &   17.15$\pm$0.05 &      \ &            AZT2 0.7m    \\   
10/10/2005 & 53654.0   &  383.5    &   \ &	    19.25$\pm$0.02 &    18.24$\pm$0.01 &   17.39$\pm$0.02 &    16.71$\pm$0.09 &   Ekar 1.82m+AFOSC         \\   
25/10/2005 & 53669.0   & 398.5       &   \ &	    19.40$\pm$0.02 &    18.42$\pm$0.01 &   17.48$\pm$0.01 &    16.76$\pm$0.02 &      Ekar 1.82m+AFOSC       \\   
01/11/2005 & 53676.4  &  405.9  &   \ &	    19.44$\pm$0.06 &    18.54$\pm$0.04 &   17.54$\pm$0.03 &    17.32$\pm$0.13 &        Z600 0.6m      \\   
03/11/2005 & 53677.8   & 407.3       &   \ &	    19.47$\pm$0.03 &    18.56$\pm$0.01 &   17.56$\pm$0.01 &    16.86$\pm$0.04 &    Ekar 1.82m+AFOSC      \\   
07/11/2005 & 53682.3  & 411.8   &   \ &	    19.45$\pm$0.05 &    18.60$\pm$0.04 &   17.63$\pm$0.03 &    17.62$\pm$0.14 &        Z600 0.6m       \\   
09/11/2005 & 53684.3  &  413.8 &   \ &	    19.27$\pm$0.09 &    18.62$\pm$0.10 &   17.71$\pm$0.04 &     \ &            Z600 0.6m     \\   
21/05/2006 & 53876.2   &   605.7     &  \ &\ & 21.43$\pm$0.04 &    20.81$\pm$0.02 &                         \ &       NOT+ALFOSC      \\   
30/09/2006 & 54008.0   &  737.5      &   \ &	 \ &  22.45$\pm$0.08 &                      \ &      \ &     NOT+ALFOSC      \\ 
08/07/2007 & 54289.5  &  1019.0         & \   &   \ &23.01$\pm$0.01$^{a}$     &   \ & 22.85$\pm$0.02$^{a}$    &HST+WFPC2  \\
12/08/2007& 54324.5  &  1054.0       &\ &23.62$\pm$0.06&23.06$\pm$0.04&22.27$\pm$0.05&21.99$\pm$0.06& WHT+AUX\\
18/01/2008 & 54484.8 &   1214.3	& \ & \   &23.24$\pm$0.01$^{a}$       &   \ & 22.13$\pm$0.02$^{a}$    &HST+WFPC2 \\
 \hline
\end{tabular}
\begin{flushleft}
* relative to the epoch of date of explosion (JD = 2453270.5)  \\ 
$^{a}$HST (F606W $\simeq$ \textit{V} band, F814W $\simeq$ \textit{I} band) \\
 \end{flushleft}
\end{table*}

\subsection{Near infrared photometry}

SN 2004et was observed in the NIR \textit{JHK} bands with SWIRCAM mounted at the focal plane of the 1.08-m AZT-24 telescope, at Campo Imperatore Observatory (Italy) on 27 epochs between 2004 September 30 and 2006 September 21. The AZT-24 telescope is operated jointly by the Pulkovo observatory (St. Petersburg, Russia) and INAF Observatorio Astronomico di Roma/Collurania (Campo Imperatore, Italy). SWIRCAM is equipped with a Rockwell PICNIC array with 256 $\times$ 256 pixels with a pixel size of 1.04". One epoch of \textit{JHK} data was also obtained on 2005 July 24 at the 3.58-m Telescopio Nazionale Galileo (TNG), using NICS with a 1024 $\times$ 1024 pixel Rockwell array with a pixel size of 0.25". At each epoch, multiple dithered images of the SN were acquired along with multiple dithered sky images. The sky images were median combined to eliminate stars. The resulting sky frames were subtracted from the SN frames for each band. The sky-subtracted SN images were then aligned and co-added.  The data were processed using standard tasks in \textsc{iraf}. 

The photometric calibration was carried out using the magnitudes of nearby stars taken from the Two Micron All Sky Survey (2MASS) photometric catalogue. Point spread function (PSF) fitting magnitude measurements were performed using \textsc{iraf}'s \textsc{daophot} package. Differential photometry of the SN magnitude was then carried out by comparison with the calibration of the 2MASS standards. The 2MASS filters differ from those of SWIRCAM and NICS, particularly in the \textit{K$_{s}$} bands but the colour terms derived from the reference stars were small, which implies only a negligible difference in photometric systems. The resulting SN magnitudes and associated errors are given in Table \ref{nir_phot}. 

Late time NIR data had been obtained with HST using NICMOS in July 2007 and January 2008 as part of the proprietary program GO-11229 (PI: Meixner). Multiple exposures were taken in each of the filters,  \textit{F110W}, \textit{F160W} and \textit{F205W}, that had been offset by sub-pixel amounts, which could then be `drizzled' \citep{fru02} to improve the spatial sampling of the PSF. PSF-fitting photometry was performed on the NIR `drizzled' data using \textsc{iraf}'s \textsc{daophot} package. The filters used, \textit{F110W}, \textit{F160W} and \textit{F205W} are roughly equivalent to the \textit{J}, \textit{H} and \textit{K} filters respectively. However no conversion equations have been applied to convert to \textit{J}, \textit{H} and \textit{K} filters since the colour transformations are not well constrained for late time SN spectra (in fact no NIR spectra at this epoch exist for accurate S-correction). The values quoted in Table \ref{nir_phot} are for the HST filters in the Vegamag system.

\begin{table*}
\begin{minipage}{126mm}
 \caption{Near infrared photometric observations of SN 2004et.}
 \label{nir_phot}
 \begin{tabular}{@{}lccccc}
  \hline
  \hline
  Date  &  JD (2450000+)  & Phase* (days)   &   	\textit{J}   	&   	\textit{H}    	&	    \textit{K$_{s}$}      \\
    \hline
30/09/2004& 53278.8&  8.3&  11.89$\pm$0.04&     11.58$\pm$0.05&  11.33$\pm$0.05 \\
01/10/2004& 53279.8&  9.3&  11.94$\pm$0.04&     11.55$\pm$0.06&  11.31$\pm$0.05 \\
02/10/2004& 53280.9&  10.4&  11.93$\pm$0.05&   11.54$\pm$0.06	&	     \\
03/10/2004& 53281.9&  11.4&  11.85$\pm$0.04&   11.48$\pm$0.05&  11.32$\pm$0.05 \\
05/10/2004& 53283.9&  13.4&  11.79$\pm$0.04&   11.46$\pm$0.05&  11.27$\pm$0.05 \\
07/10/2004& 53285.8&  15.3&  11.76$\pm$0.04&   11.44$\pm$0.06&  11.19$\pm$0.06 \\
08/10/2004& 53286.8&  16.3&  11.74$\pm$0.04&   11.40$\pm$0.06&  11.16$\pm$0.06 \\
09/10/2004& 53287.7&  17.2&  11.81$\pm$0.04&   11.37$\pm$0.09&  11.03$\pm$0.08 \\
18/10/2004& 53296.9&  26.4&  11.64$\pm$0.04&   11.36$\pm$0.08&  10.92$\pm$0.08 \\
20/10/2004& 53298.8&  28.3&  11.59$\pm$0.05&   11.25$\pm$0.06&  10.97$\pm$0.07 \\
22/10/2004& 53300.8&  30.3&  11.60$\pm$0.04&   11.18$\pm$0.08&  10.96$\pm$0.06 \\
23/10/2004& 53301.8&  31.3&  11.54$\pm$0.05&   11.22$\pm$0.08&  11.03$\pm$0.06 \\
27/10/2004& 53305.8&  35.3&  11.55$\pm$0.03&   11.18$\pm$0.05&  10.96$\pm$0.05 \\
17/11/2004& 53326.8&  56.3&  11.42$\pm$0.05&   11.07$\pm$0.05&  10.93$\pm$0.05 \\
18/11/2004& 53327.9&  57.4&  11.44$\pm$0.04&   11.08$\pm$0.05&  10.88$\pm$0.05 \\
20/11/2004& 53329.7&  59.2&  11.44$\pm$0.04&   11.09$\pm$0.05&  10.88$\pm$0.05 \\
21/11/2004& 53330.7&  60.2&  11.43$\pm$0.04&   11.07$\pm$0.05&  10.84$\pm$0.05 \\
23/11/2004& 53332.8&  62.3&  11.48$\pm$0.04&   11.12$\pm$0.05&  10.89$\pm$0.05 \\
25/11/2004& 53334.8&  64.3&  11.45$\pm$0.04&   11.06$\pm$0.04&  10.87$\pm$0.05 \\
07/12/2004& 53346.7&  76.2&  11.47$\pm$0.05&   11.11$\pm$0.08&		      \\
13/12/2004& 53352.9&  82.4&  11.43$\pm$0.04&   11.08$\pm$0.05&  10.86$\pm$0.05 \\
15/12/2004& 53354.7&  84.2&  11.46$\pm$0.04&   11.08$\pm$0.05&   	          \\
04/01/2005& 53374.7&  104.2&  11.64$\pm$0.04&  11.30$\pm$0.05&  11.08$\pm$0.05 \\
10/01/2005& 53380.8&  110.7&  11.69$\pm$0.05&  11.36$\pm$0.04&  11.16$\pm$0.05 \\
03/02/2005& 53404.7&  134.2&  13.21$\pm$0.04&  12.83$\pm$0.06&		      \\
10/02/2005& 53411.7&  141.2&  13.37$\pm$0.04&  13.06$\pm$0.07&		      \\
24/07/2005& 53576.6&306.1& 15.36$\pm$0.14$^{a}$   &  15.69$\pm$0.12$^{a}$&15.50$\pm$0.16$^{a}$ \\
21/09/2006& 54000.0&  729.5&  $>$ 20.2			      \\
08/07/2007&54289.5&   1019.0& 22.08$\pm$0.02$^{b}$             &22.57$\pm$0.08$^{b}$ & $>$ 22.3$^{b}$     \\
19/01/2008&54485.6& 1215.1 &22.30$\pm$0.03$^{b}$             &22.70$\pm$0.10$^{b}$   & 21.18$\pm$0.06$^{b}$ \\
 \hline
 \end{tabular}
 \begin{flushleft}
*relative to the epoch of date of explosion (JD = 2453270.5)    \\
$^{a}$TNG+NICS \\
$^{b}$HST (F110W $\simeq$ \textit{J} band, F160W $\simeq$ \textit{H} band, F205W $\simeq$ \textit{K} band) \\
\end{flushleft}
 \end{minipage}
\end{table*}

\section{Photometric Evolution}

\subsection{Light curve}
\label{lightcurve101}

\begin{figure}
\includegraphics[width=8.4cm]{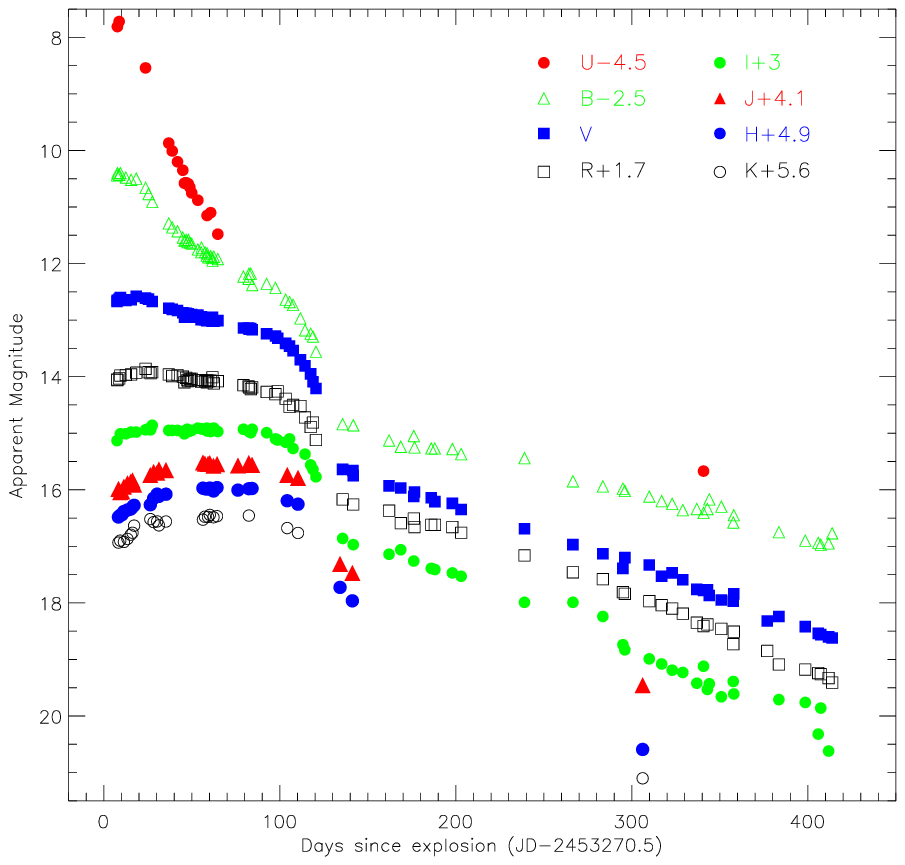}  
\caption{\textit{UBVRIJHK} light curves of SN 2004et, from soon after core-collapse to +414 days post explosion. The light curves are constructed using the data in Table \ref{opt_phot} and Table \ref{nir_phot} and have not been corrected for reddening.}
\label{lightcurve_all}
\end{figure}

Figure \ref{lightcurve_all} shows the optical and NIR light curve (\textit{UBVRIJHK}) of SN 2004et from just after explosion to +414 days post explosion. Very early time \textit{R} band magnitudes were reported by Klotz and collaborators in \cite{yam04}. Nothing was visible to a limiting magnitude of 19.4 $\pm$ 1.2 at the SN position on 2004 September 22.017 UT, when imaged by the robotic TAROT telescope. Pre-discovery detection of SN 2004et on September 22.983 UT exists, with the SN having a magnitude of 15.17 $\pm$ 0.16 \citep{yam04}. The explosion epoch of SN 2004et has therefore been well-constrained and the explosion epoch is taken as 2004 September 22.0 (JD 2453270.5), the same as in \cite{li05}. 

The characteristic plateau of type IIP SNe is visible up to 110 $\pm$ 15 days, before showing a sharp decline onto the radioactive tail, whose luminosity is powered by the decay of $^{56}$Co to $^{56}$Fe. The expected decline rate during the nebular phase is 0.98 mag per 100 days for complete $\gamma$-ray trapping \citep{pat94}. The decline rates per 100 days for the \textit{BVRI} bands were calculated using a least squares fit, during the early nebular phase ($\sim$ 136--300 days) and found to be $\gamma_{B}$ = 0.66 $\pm$ 0.02, $\gamma_{V}$ = 1.02 $\pm$ 0.01,  $\gamma_{R}$ = 0.92 $\pm$ 0.01 and $\gamma_{I}$ = 1.09 $\pm$ 0.01. These values are close to the expected decline rate and are consistent with the early nebular phase flux being dominated by radioactive decay.

After 300 days, steeper decline rates were seen in the \textit{BVR} bands and a comparison of the decline rates is shown in Figure \ref{latetimeslope}. The decline rates based on data from $\sim$ 136--296 days only, are shown as a solid line for each of the bands and are extrapolated out to 800 days. The decline rates using data from between $\sim$ 296--414 days are $\gamma_{B}$ = 0.85 $\pm$ 0.02, $\gamma_{V}$ = 1.17 $\pm$ 0.02,  $\gamma_{R}$ = 1.33 $\pm$ 0.01 and $\gamma_{I}$ = 0.93 $\pm$ 0.02 and are shown as the dashed lines in Figure \ref{latetimeslope}. Note that the late time \textit{I} band photometry shows a large scatter due to the intrinsic differences in the instrumental \textit{I} band filters and this could affect our estimates of the slope of the \textit{I} band light curve. The points after 600 days shown in Figure \ref{latetimeslope} were not included in the determination of the slopes of the \textit{V} and \textit{R} bands. The \textit{BVR} bands became steeper during this period while the slope of the \textit{I} band became slightly less steep. This steepening could imply either the leakage of $\gamma$ rays because the SN had become transparent to $\gamma$ rays at this stage or dust formation in the SN ejecta. To determine the cause of the steepening of the \textit{BVR} bands, further analysis of the photometry and spectroscopy was carried out and is detailed in Section \ref{dust}.

\begin{figure}
\includegraphics[width=8.4cm]{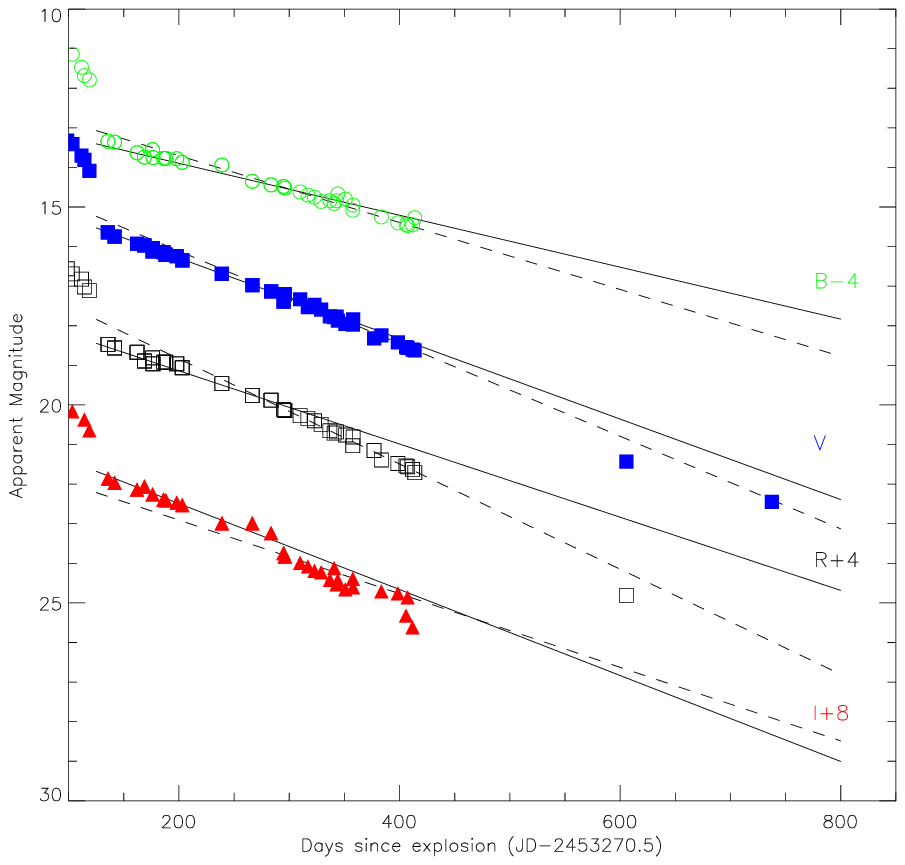}
\caption{\textit{BVR} light curves of SN 2004et during the tail phase. The solid lines indicate the decline rates calculated from $\sim$ 136--300 days post explosion, while the dashed lines indicate the decline rates calculated using the data from $\sim$ 295--415 d. A steepening of the \textit{BVR} light curves is seen while the slope of the \textit{I} band light curve becomes marginally less steep.}
\label{latetimeslope}
\end{figure}

 \subsection{Colour Curves}
The extinction to SN 2004et, \textit{E(B-V)} was estimated to be 0.41 $\pm$ 0.07 mag using {\ion{Na}{ID} equivalent widths \citep{zwi04}. Reddening corrected colour curves for SN 2004et using the optical and NIR data detailed above are shown in Figure \ref{colour_opt} and Figure \ref{colour_NIR}. The optical colour curves are compared to those of other IIP SNe: SN 1999em \citep{leo02}, SN 2003hn \citep{oli08,kri08}, SN 2005cs \citep{pas09} and SN 1990E \citep{sch93, ben94}.  The explosion epoch, distance and extinction for each SN is detailed in Table \ref{cc}. Few data are available in the NIR for type IIP SNe, particularly at late times so in this case for comparison we can only use SN 1999em, SN 1987A, SN 2005cs and SN 2002hh. SN 2002hh was a highly reddened SN and \cite{poz06} adopted a two component model for the extinction with total A$_{V}$ of 5 mag, which we use in Figure \ref{colour_NIR}.

As is expected, SN 2004et became monotonically redder after explosion for $\sim$ 100 days as it cooled. This can be seen in both the optical and NIR comparison colour curves. Later on when SN 2004et enters the nebular phase, the colour curves become bluer. The colour evolution of SN 2004et is most similar to `normal' type IIP SNe, SN 1999em and SN 1990E, while SN 2005cs showed a red peak at $\sim$ 100 days, characteristic of low-luminosity SNe \citep{pas04, pas09}. The \textit{V-H}, \textit{V-K} and \textit{J-K} colours for SN 2004et are more blue at $\sim$ 300 days than for SN 1987A and SN 2002hh. The \textit{V-J} colour evolution appears to be similar to that of SN 1987A and SN 2002hh, which suggests the difference lies in the $H$ and $K$ bands being more blue than for the other SNe. This could be due to weaker features of the Brackett series of H in the spectra of SN 2004et, but we do not possess spectra of a high enough signal-to-noise to quantitatively determine if this is the cause.

\begin{table}
 \caption{Table of properties of our sample of type IIP SNe.}
 \label{cc}
 \begin{tabular}{lcccc}
   \hline
   \hline
 SN &Explosion epoch& D (Mpc)& \textit{E(B-V)} &References \\
  \hline
 1987A&2446850.5&49.9$\pm$1.2 $\times$ 10$^{-3}$&0.19&1\\
1990E&2447932.0& 18.2$\pm$2.4&0.48& 2, 3\\
1999br&2451278&14.1$\pm$2.6&  0.02& 4 \\
1999em&  2451475.6& 11.7$\pm$1.0  & 0.06&5\\
1999gi & 2451518.3 & 10.0$\pm$0.8 &0.21&6\\
2002hh&2452577.5 & 5.9$\pm0.4$&see text &  7, 8\\
2003hn& 2452857.0&   17.8$\pm$1.0  &0.19&9, 10 \\
2004A & 2453011.5& 20.3$\pm$3.4 &0.06&11, 12\\
2004et& 2453270.5&5.9$\pm0.4$&0.41& 13 \\
2005cs&2453549.0& 7.1$\pm$1.2 & 0.05& 14\\
2006my&2453943&22.3$\pm$2.6 & 0.03 & 15\\
\hline \\
 \end{tabular}
  \medskip \\
       REFERENCES -- (1) \cite{ash87}; (2) \cite{sch93}; (3) \cite{ham03}; (4) \cite{pas04}; (5) \cite{leo02}; (6) \cite{leo03}; (7) \cite{poz06}; (8) \cite{tsv07}; (9) \cite{oli08}; (10) \cite{kri08}; (11) \cite{hen06}; (12) \cite{tsv08b}; (13) this work; (14) \cite{pas09} (15) \cite{sma09}
\end{table}

\begin{figure}
\includegraphics[width=8.4cm]{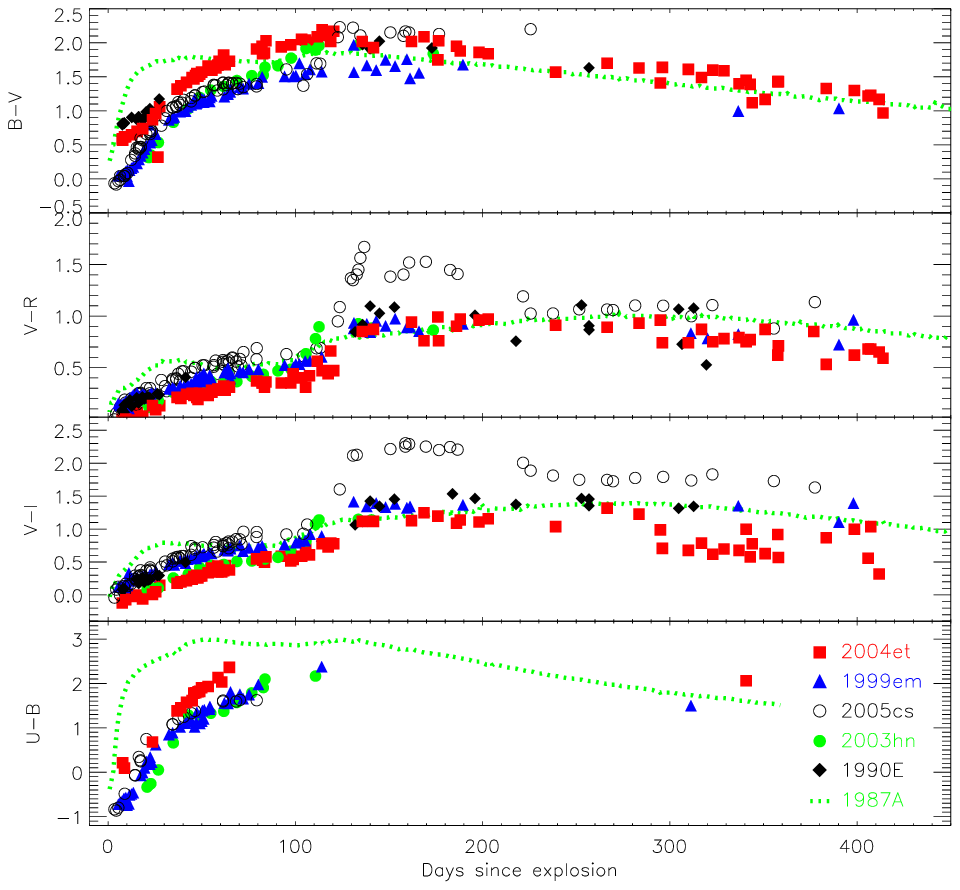}
\caption{Optical colour evolution of SN 2004et compared with those of other type IIP SNe; SN 1999em, SN 2003hn, SN 2005cs, SN 1990E and SN 1987A. }
\label{colour_opt}
\end{figure}

\begin{figure}
\includegraphics[width=8.4cm]{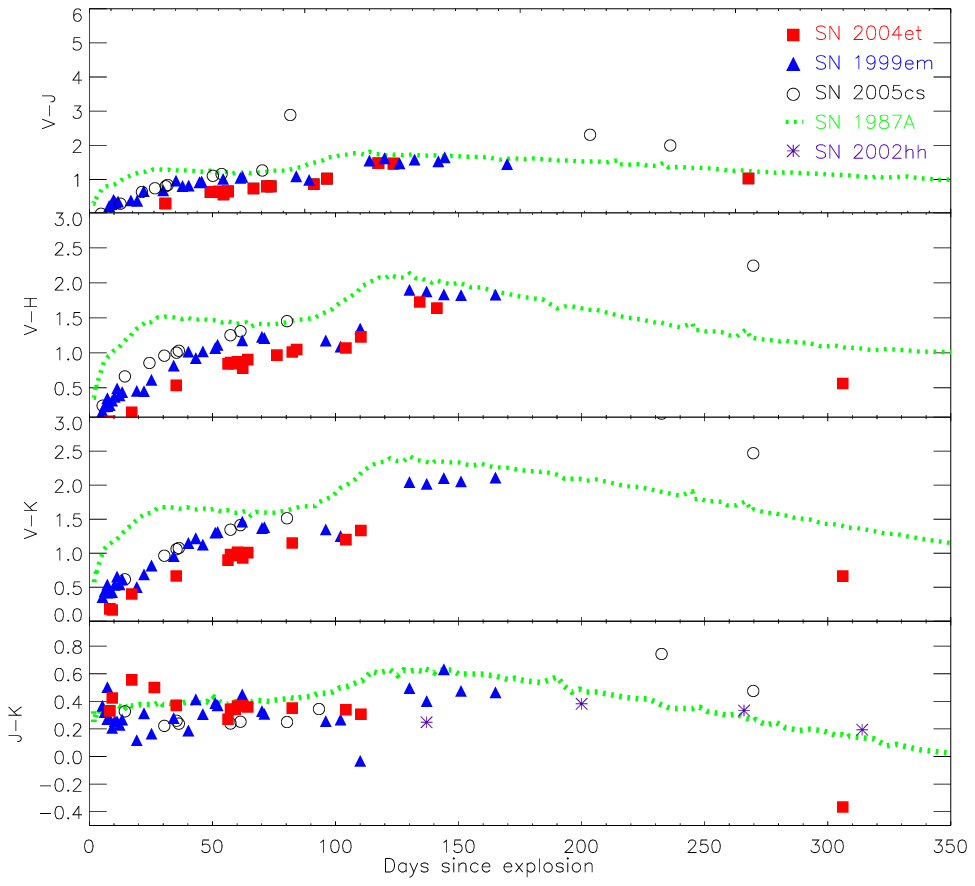}
\caption{Near infrared colour evolution of SN 2004et compared with those of SN 1999em, SN 2005cs and the peculiar SN 1987A. }
\label{colour_NIR}
\end{figure}

\subsection{Bolometric light curve and $^{56}$Ni Mass}
\label{bolsection}

\begin{figure}
\includegraphics[width=8.4cm]{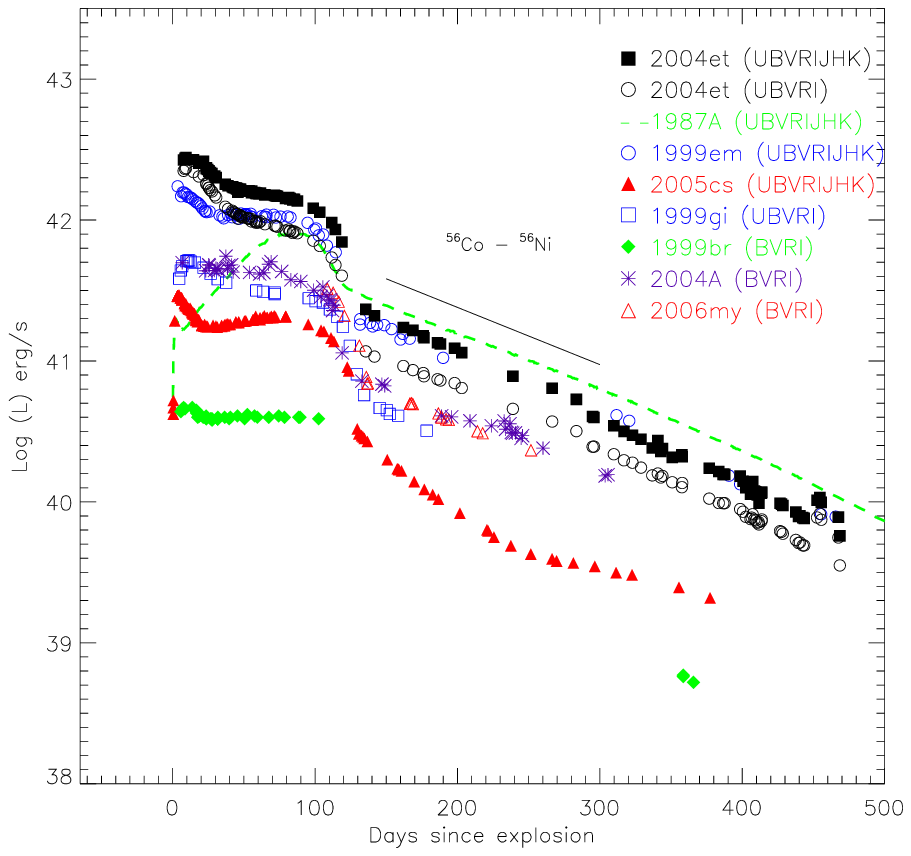} 
\caption{Comparison of bolometric light curve of SN 2004et with other type IIP SNe.}
\label{com_opt}
\end{figure}

\begin{figure}
\includegraphics[width=8.4cm]{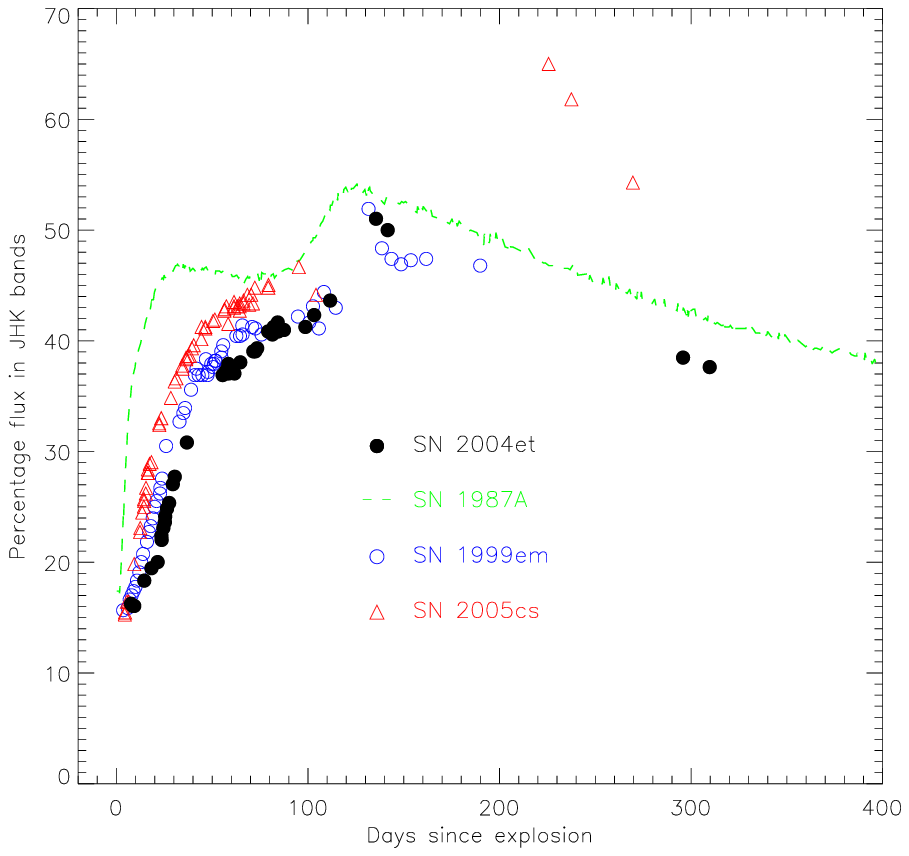}
\caption{Flux contribution from NIR bands as a percentage of the total flux from optical and NIR bands for a selection of type IIP SNe with NIR data.}
\label{percent_flux}
\end{figure}

The bolometric light curve of SN 2004et was computed by correcting the observed magnitudes for reddening and converting to flux values before integrating the combined flux from the \textit{UBVRIJHK} bands excluding the overlapping regions of the filters \citep[see also][]{val08}. This integrated flux could then be converted to a luminosity using a distance of 5.9 Mpc (see Section \ref{intro}) and the extinction detailed in the previous section. Our \textit{UBVRIJHK} data are supplemented by optical data from \cite{mis07}. The bolometric luminosity was only calculated for epochs where V band data were obtained. If data were not obtained in the other bands at the same epoch, the contribution was calculated by interpolating the data from adjacent nights. 

Figure \ref{com_opt} shows a comparison with other type IIP SNe, some of which also include contributions from NIR data. Two pseudo-bolometric light curves of SN 2004et are shown, one that includes the contribution from the NIR bands (filled squares) and one that does not (open circles). When the NIR is included the luminosities of the plateau and the tail phase are significantly increased. The bolometric light curve of SN 2004et is one of few type IIP SNe that includes the contribution of NIR light, which is seen to make a significant contribution (up to 50 per cent). The percentage flux contribution of the \textit{JHK} bands compared to the total flux from the optical to the NIR bands for SN 2004et and three other type IIP SNe, SN 1987A, SN 1999em and SN 2005cs is shown in Figure \ref{percent_flux}. The contribution of NIR flux to the total flux of SN 2004et is most similar to the `normal' type IIP, SN 1999em. It can be seen that particularly at early times the NIR flux contribution of SN 1987A is significantly greater than that of other type IIP SNe due to the faster cooling of the ejecta of this SN. When no NIR data is available for a type IIP SN, the NIR contribution of SN 1987A is sometimes used as a correction to the bolometric light curve of SNe due to its excellent coverage \citep[e.g.][]{mis07}. Figure \ref{percent_flux} demonstrates however that using a correction based on SN 1987A could lead to an overestimate of the bolometric luminosity, especially at very early times.

\cite{sah06} computed a \textit{UBVRI} bolometric light curve of SN 2004et and found it to have one of the highest luminosities among type IIP SNe. They did not apply any correction for the NIR flux. \cite{mis07} constructed a \textit{UVOIR} bolometric light curve using a correction for the lack of NIR data, which was obtained by a comparison with the bolometric light curve of SN 1987A. They used a smaller distance estimate to SN 2004et ( 5.5 $\pm$ 1.0 Mpc), which results in smaller luminosity values. Despite the smaller distance used, their bolometric light curve displayed higher luminosities at early times than the light curve of SN 2004et shown in Figure \ref{com_opt}. The peak luminosity of the \textit{UBVRIJHK} bolometric light curve of SN 2004et given in \cite{mis07} was log L $\sim$ 42.55 erg s$^{-1}$ compared to 42.44 erg s$^{-1}$ for our data in Figure \ref{com_opt}. This difference at early times is most likely due to their overestimate of the NIR luminosity from the comparison with SN 1987A.

\cite{ber09} calculated the bolometric correction (BC) to transform from \textit{V} band magnitudes of SN 1987A, SN 1999em and SN 2003hn to \textit{UBVRIJHK} bolometric luminosities and provided parametrised corrections as a function of SN colour for the average of these three SNe. Here we present the BC for four type IIP SNe with extensive optical and NIR coverage as a function of time. These SNe are SN 1987A (a peculiar type IIP), SN 2005cs (a low luminosity SN), and SN 1999em and SN 2004et, which both show a very similar bolometric light curve shape.  We have rearranged equation 1 of \cite{ham03} to produce this equation for the BC at any given time t,

\begin{equation}
\label{bol_corr}
\displaystyle{B(t) = -2.5\  log_{10} L(t) - V(t) + A_{total} (V) + 5\ log_{10} D - 8.14}
\end{equation}

where L(t) is the bolometric luminosity in ergs s$^{-1}$, V(t) is the V band magnitude, A$_{total}$ is the total V band extinction toward the SN, D is the distance to the SN in cm and the constant is to convert from Vega magnitudes to values of luminosity in cgs units. Using the bolometric light curve for each SN, which has been constructed using the method detailed above, the BC as a function of time can be parametrised by fitting a polynomial to the data. The polynomial coefficients for the \textit{V} and $R$ bands during the plateau phase are given in Table \ref{bc_para1} and Table \ref{bc_para2} respectively. 

The BC relative to the \textit{V} and $R$ bands during the photospheric phase are shown in Figure \ref{plot_bolcorr1}. For the $V$ band, the scatter of the BC for each SN is relatively large with SN 1987A showing a significantly different evolution during the plateau phase. For the $R$ band however, the evolution of the BC is much more homogeneous for the SNe, particularly after $\sim$ 50 days and we suggest that the $R$ band BC may be more suitable for computing the luminosities of a SN when there is ambiguity about the subtype of the type IIP SN being studied.

The coefficients a$_i$ of the polynomial for each SN and band can be used in Equation \ref{bol_corr3} for each epoch of data to give the BC as a function of time, 

\begin{equation}
\label{bol_corr3}
\displaystyle{BC(t) = \sum_{i=0}^na_i\ t^i}
\end{equation}

where n is the number of coefficients for the particular band and SN being studied. Equation \ref{bol_corr2} can then be used to give the bolometric luminosity of a SN with only single band photometric data, 

\begin{equation}
\label{bol_corr2}
\displaystyle{log  _{10}L(t) = - 0.4\ [V(t) -  A_{total} (V) + BC(t)\ - 5\ log_{10} D + 8.14]}
\end{equation}

where the parameters have the same meaning as detailed above. Equation \ref{bol_corr2} is independent of the choice of zero point for the Vega magnitude system and is dependent only on the values of the magnitude, distance, extinction and time since explosion. 

\begin{figure}
\includegraphics[width=8.4cm]{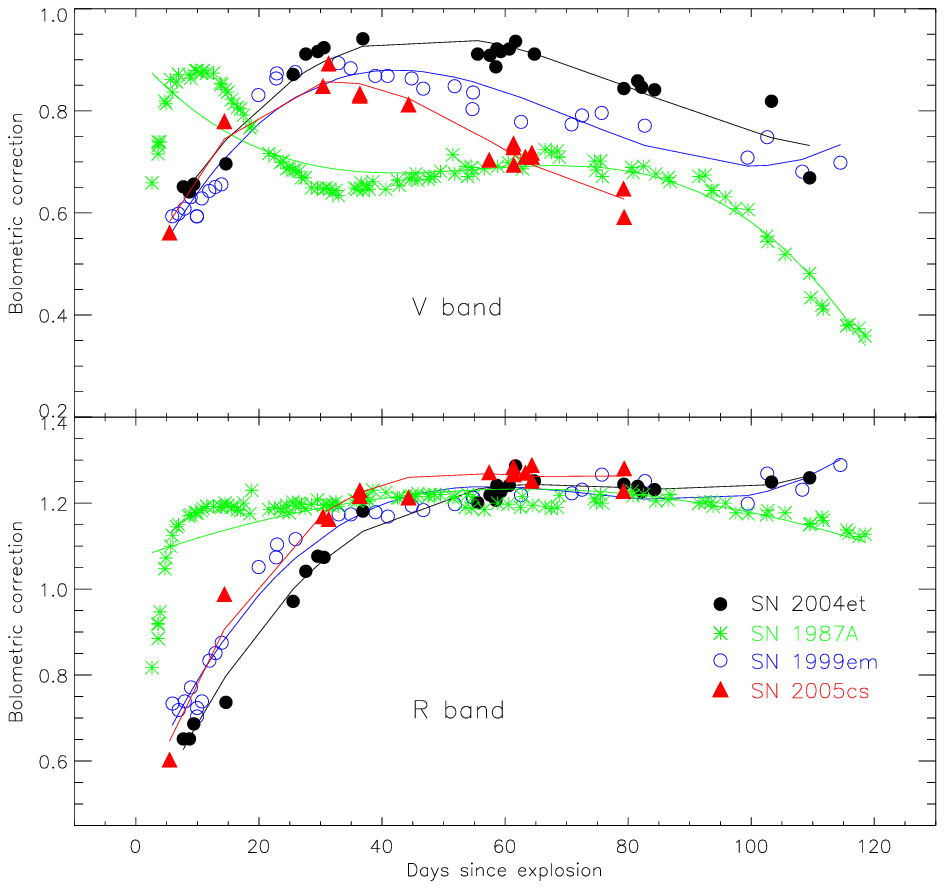} 
\caption{Bolometric correction from the $V$ band (upper panel) and $R$ band (lower panel) to the \textit{UBVRIJHK} bolometric light curve as a function of time for SN 1987A, SN 1999em, SN 2004et and SN 2005cs during the photospheric phase. The bolometric correction of each SN is fitted with a lower order polynomial (solid lines).}
\label{plot_bolcorr1}
\end{figure}

\begin{figure}
\includegraphics[width=8.4cm]{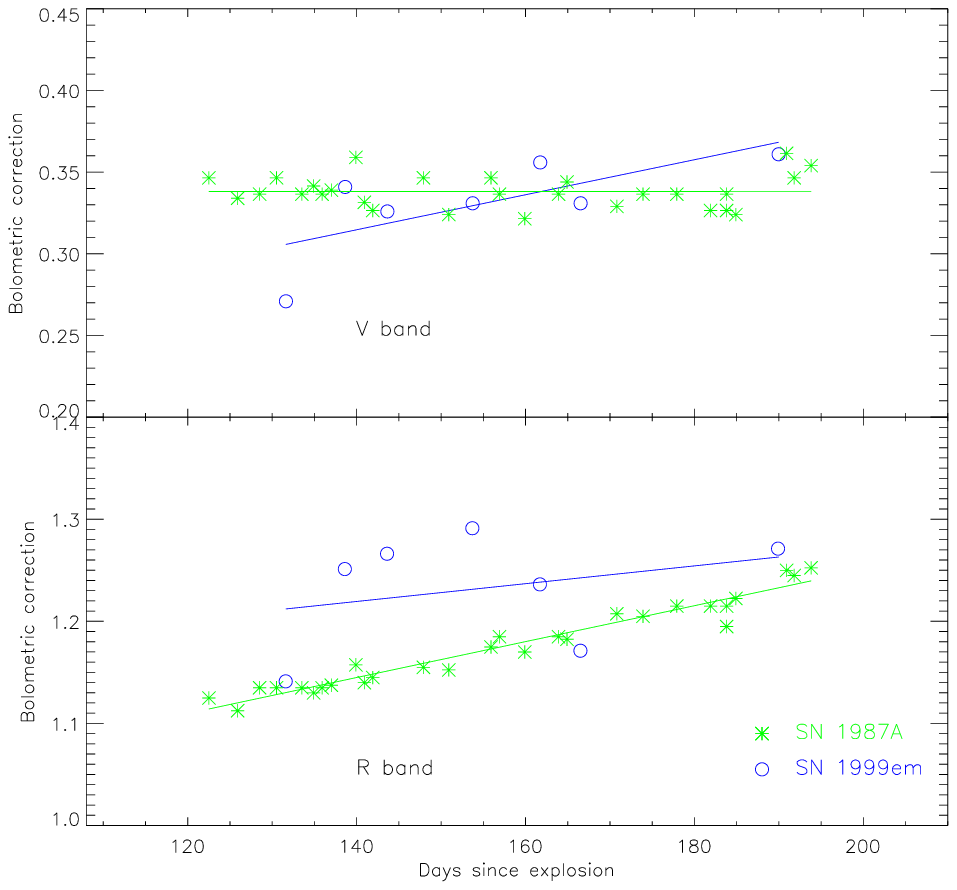} 
\caption{Bolometric correction from $V$ band (upper panel) and $R$ band (lower panel) to the \textit{UBVRIJHK} bolometric light curve as a function of time for SN 1987A and SN 1999em during the early nebular phase. The bolometric correction of each SNe is fitted with a linear fit (solid lines).}
\label{plot_bolcorr3}
\end{figure}

Figure \ref{plot_bolcorr3} shows the BC for the $V$ and $R$ bands to the \textit{UBVRIJHK} bolometric light curves of SN 1987A and SN 1999em during the early nebular phase ($\sim$ 120--200 days). SN 2004et and SN 2005cs did not have sufficient data at NIR wavelengths at these epochs to be included in the plot. The BC is relatively similar for both SNe, which is consistent with the results of \cite{ham01} and \cite{ber09}. By applying the parametrised BC given in Table \ref{bc_para3} to data of other SNe with sparser photometric coverage during the early nebular phase, an estimate of the \textit{UBVRIJHK} luminosity can be calculated and hence an estimate of the ejected $^{56}$Ni mass of a less well studied type IIP SN can be determined. We find an average BC during early nebular phase of 0.33 $\pm$ 0.06 mag using SN 1987A and SN 1999em, which is systematically larger than the value of 0.26 $\pm$ 0.06 mag obtained by \cite{ham01}. The probable cause of this discrepancy is the different methods used to compute the bolometric light curve; we integrated the combined flux over the \textit{UBVRIJHK} bands as detailed above, while \cite{ham01} fitted Planck functions at each epoch to the reddening-corrected $BVIJHK$ magnitudes.

Information on the bolometric light curve is useful for constraining the amount of radioactive material ejected in the explosion. 
We estimated the $^{56}$Ni mass of SN 2004et using three methods; a comparison with the luminosity of SN 1987A, the steepness parameter method of \cite{elm03} and the tail luminosity method of \cite{ham03}. 

The $^{56}$Ni mass can be estimated by a comparison of the bolometric luminosities of SN 2004et and SN 1987A during the early nebular phase, before the possible formation of dust ($\sim$ 120--250 days post explosion). The $^{56}$Ni mass of SN 1987A was 0.075 $\pm$ 0.005 \msun\ \citep{arn96} and based on a comparison of their \textit{UBVRIJHK} luminosities, we find SN 2004et to have a mass of  $^{56}$Ni of 0.057 $\pm$ 0.03 \msun. Using this method \cite{sah06} found a $^{56}$Ni mass of 0.048 $\pm$ 0.01 \msun. Given that for SN 2004et they did not include the NIR contribution, their value is a lower limit for the $^{56}$Ni mass.

Despite the lack of good data coverage at the transition from the plateau to the tail phase, we used the equations of \cite{elm03a} that correlate the steepness of the rate of the decline from plateau to tail phase of the bolometric light curve and the mass of $^{56}$Ni. The smaller the $^{56}$Ni mass produced in the explosion, the steeper the decline rate is. Using this method, we estimated the steepness parameter to have a value of 0.068, which gives a $^{56}$Ni mass of 0.057 $\pm$ 0.02 \msun.  \cite{sah06} found a $^{56}$Ni mass of 0.062 $\pm$ 0.02 \msun\ and \cite{mis07} a $^{56}$Ni mass of 0.056 $\pm$ 0.016 \msun\ using this steepness method. 

The $^{56}$Ni mass of SN 2004et can also be estimated using the assumption that the tail luminosity of SN 2004et is dominated by the radioactive decay of $^{56}$Ni to $^{56}$Co \citep{ham03} and is found to have a value of 0.05 $\pm$ 0.02 \msun\ using the BC of \cite{ham01} of 0.26 $\pm$ 0.06 mag. When our BC, calculated during the early nebular phase of 0.33 $\pm$ 0.06 mag is used, we obtain a $^{56}$Ni mass of 0.06 $\pm$ 0.02 \msun. The final assumed $^{56}$Ni mass for SN 2004et used in this paper is taken as the average of the three methods listed above, which gives a value of 0.056 $\pm$ 0.04 \msun.

 \section{Spectroscopy}
\subsection{Optical spectroscopy}

Optical spectroscopic observations of SN 2004et were carried out at the Mt. Ekar 1.82-m Copernico Telescope (Asiago, Italy) with AFOSC and at the TNG using DOLORES, as detailed in Table \ref{tab_opt}. The data were trimmed, bias corrected, and a normalised flat-field was applied before the spectra were extracted using tasks in \textsc{iraf}'s \textsc{ctioslit} package. Wavelength calibration was performed using arc lamp spectra. The instrumental response curves as a function of wavelength were determined using spectrophotometric standard stars. The spectra were then flux calibrated to the photometry taken at the closest epoch. The resolution of each spectra was estimated by measuring the full width half maximum (FWHM) of the night sky lines and the values are given in Table \ref{tab_opt}.

An optical spectrum of SN 2004et taken at 56 days post explosion is shown in Figure \ref{opt_lineid}. The key features of the plateau phase of the SN are marked. The identification of the lines was made by comparing the spectrum to the template plateau phase spectrum of SN 1999em \citep{leo02} and SN 1998A \citep{pas05}, for which detailed line identifications were carried out. The P-Cygni profile that is characteristic of a fast expanding ejecta is most clearly visible in \ha, but can also be seen in other elements marked in the spectrum.

Figure \ref{opt_specevo1} shows the optical spectroscopic evolution of SN 2004et during the photospheric phase, from +24 to +115 days. The spectra at +24 d and +31 d from \cite{sah06} were retrieved from the SUSPECT\footnote{http://bruford.nhn.ou.edu/$\sim$suspect/index1.html} archive to show the early time evolution of SN 2004et. As the photospheric phase progresses, the blue region of the spectra becomes dominated by metal lines, which add to the increasing absorption troughs of the H Balmer series. The absorption troughs of the H lines become deeper as the photospheric phase progresses before decreasing as the SN enters the nebular phase.

An optical nebular phase spectrum, taken at +401 days post explosion is shown in Figure \ref{opt_nebid}. The most prominent emission lines have been marked following the line identifications of \cite{leo02} for SN 1999em. The spectrum is seen to be dominated by forbidden emission lines such as \Oi\ 6300, 6364 \AA, \FeiiF\ 7155 \AA, \Caii\ 7291, 7324 \AA\ and a strong \ha\ emission line. The \ha\ and \Caii\ 7291, 7324 \AA\ doublet have comparable luminosities. The individual components of the \Caii\ doublet are not resolved. Figure \ref{opt_specevo2} shows the overall spectroscopic evolution of SN 2004et during the nebular phase. The spectrum of +464 days is taken from \cite{sah06}. At early nebular phases, it is seen that the most prominent feature is the \ha\ emission line. The blue part of the nebular spectra are dominated by \Feii\ and \FeiiF\ lines. \cite{utr09} investigated possible asymmetry in the structure of SN 2004et by fitting three Gaussian components to each of the lines of the \Oi\ 6300, 6364 \AA\ doublet in a +301 day spectrum. Their results showed a non-negligible contribution from asymmetric components and from this, they inferred a bipolar structure for SN 2004et, which has also been shown for SN 1987A, SN 1999em and SN 2004dj \citep{utr07}.

A comparison between an optical photospheric spectrum of SN 2004et and other IIP SNe, SN 1999gi, SN 1999em, SN 2004A and SN 2005cs, all taken at $\sim$ 40 days post explosion is shown in Figure \ref{com_opt_spec}. The spectrum of SN 2004et is taken from \protect \cite{sah06}, while for SN 2004A see Appendix B. The other spectra are from the references given in Table \ref{cc}. SN 2004et has the most blue-shifted absorption minima of all the SNe of this sample, which suggests a higher expansion velocity for the outer ejecta. However it will be seen in Section \ref{expansion}, that the photospheric expansion velocity of SN 2004et calculated from weak metal lines is similar to that of SN 1999em.

\begin{figure}
\includegraphics[width=8.4cm]{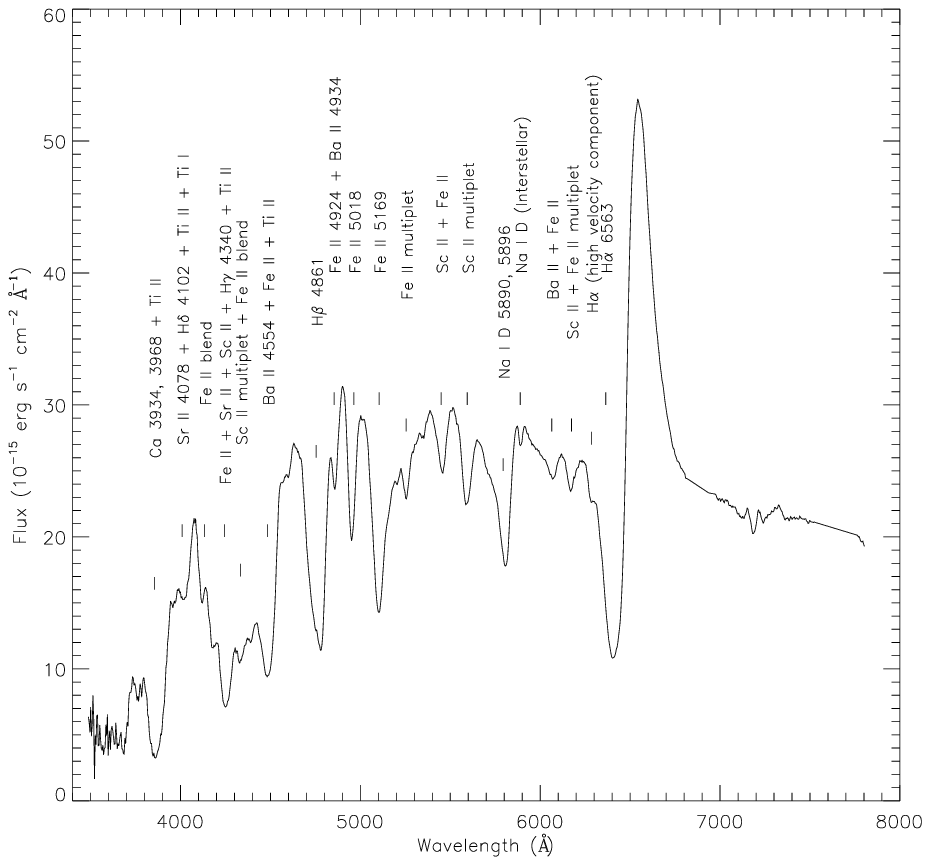}
\caption{Optical spectrum of SN 2004et at 56 days post explosion (JD 53326.0). The spectrum has been corrected for the recessional velocity of the host galaxy of 48 km s$^{-1}$. The prominent features are labelled.}
\label{opt_lineid}
\end{figure}

\begin{figure}
\includegraphics[width=8.4cm]{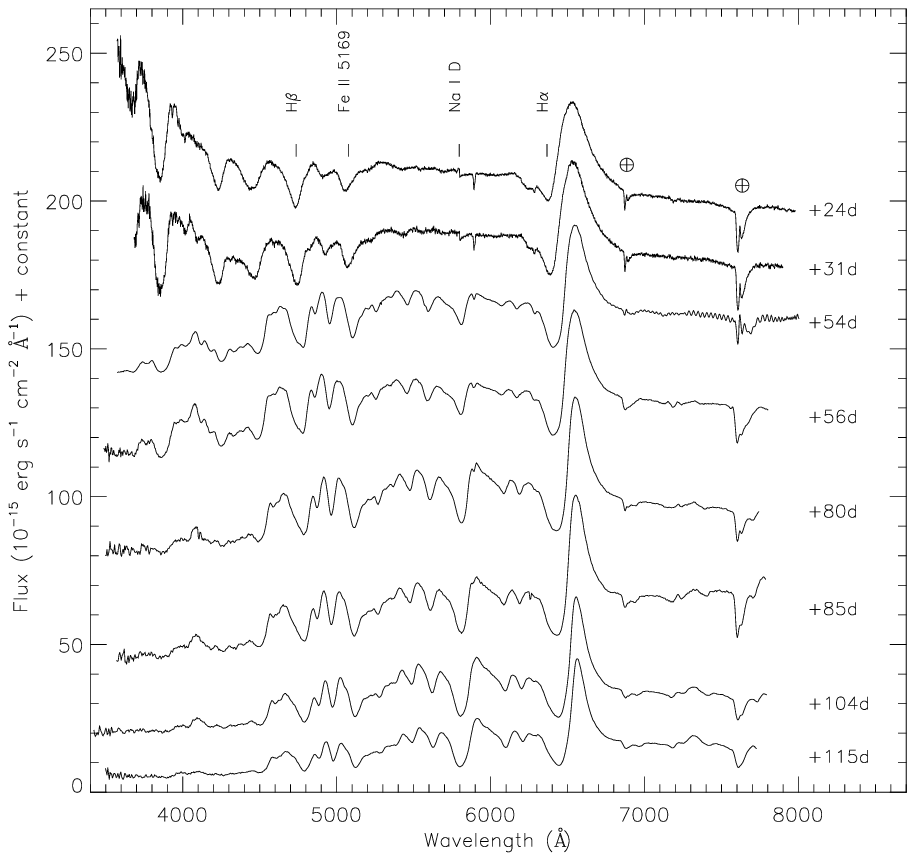}
\caption{Optical spectroscopic evolution of SN 2004et during the photospheric phase with days marked since the explosion epoch (JD 2453270.5 ). The spectra have been corrected for the recessional velocity of the host galaxy of 48 km s$^{-1}$. Some key emission and absorption features are labelled. The spectra of +24 d and +31 d are taken from \protect \cite{sah06}.}
\label{opt_specevo1}
\end{figure}

\begin{figure}
\includegraphics[width=8.4cm]{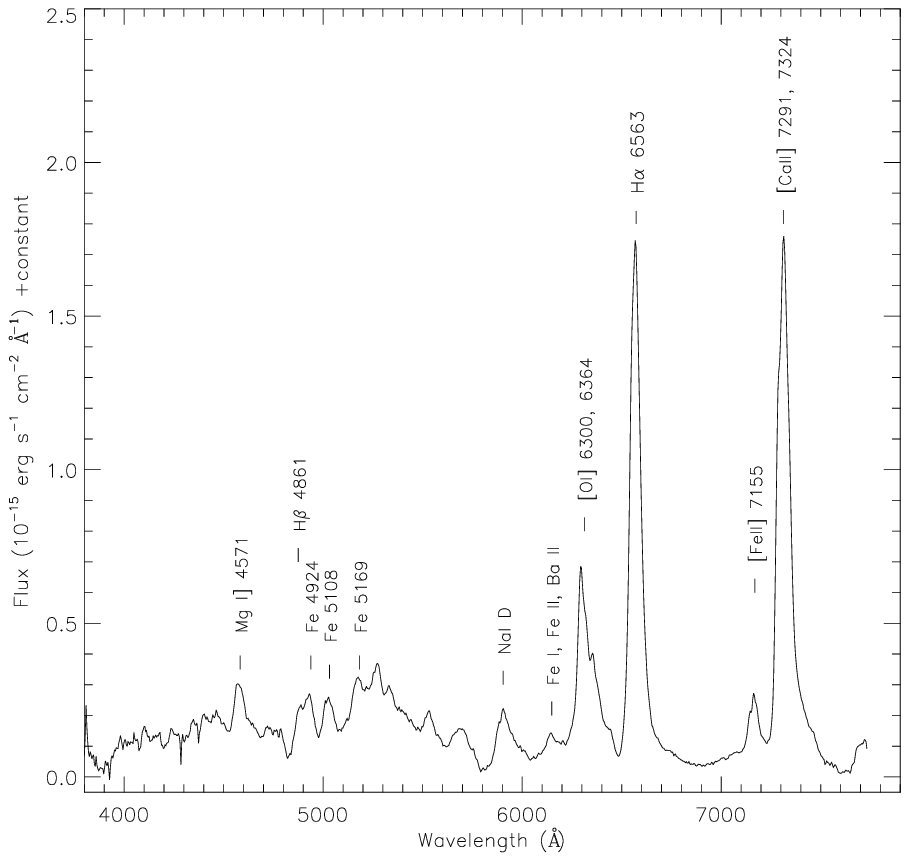}
\caption{Optical spectrum of SN 2004et at 401 days post explosion. The spectrum has been corrected for the recessional velocity of the host galaxy of 48 km s$^{-1}$. The prominent features are labelled.}
\label{opt_nebid}
\end{figure}

\begin{figure}
\includegraphics[width=8.4cm]{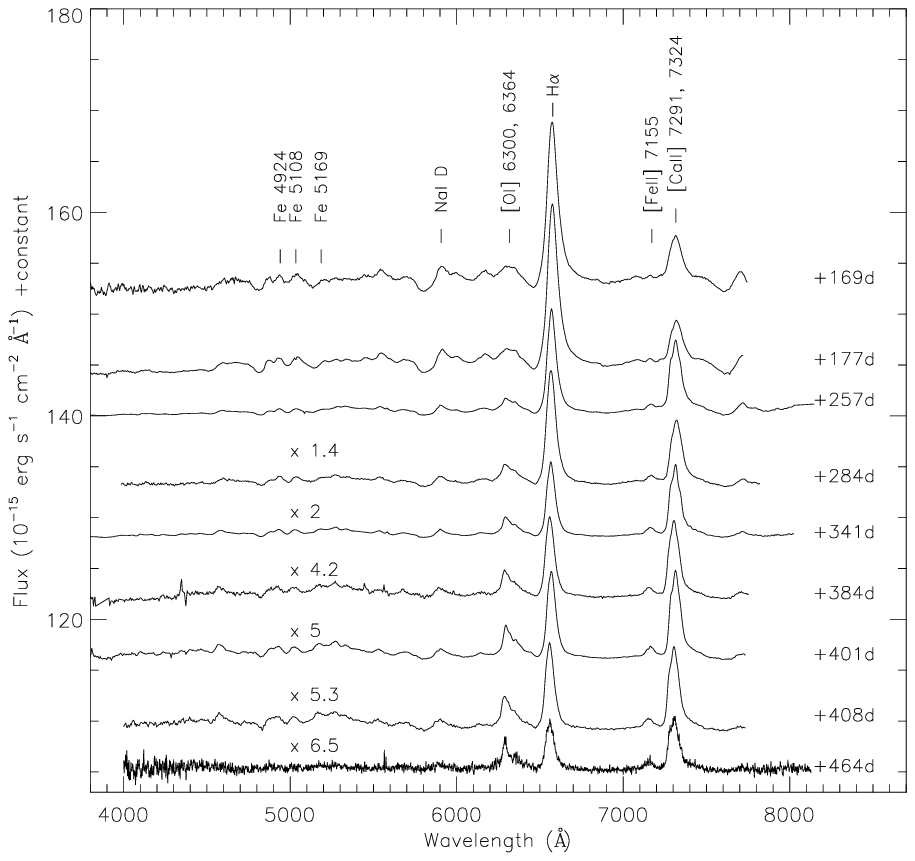}
\caption{Optical spectroscopic evolution of SN 2004et during the nebular phase with days since the explosion epoch marked next to spectra. Emission lines are seen to be the dominant features during this phase.}
\label{opt_specevo2}
\end{figure}

\begin{figure}
\includegraphics[width=8.4cm]{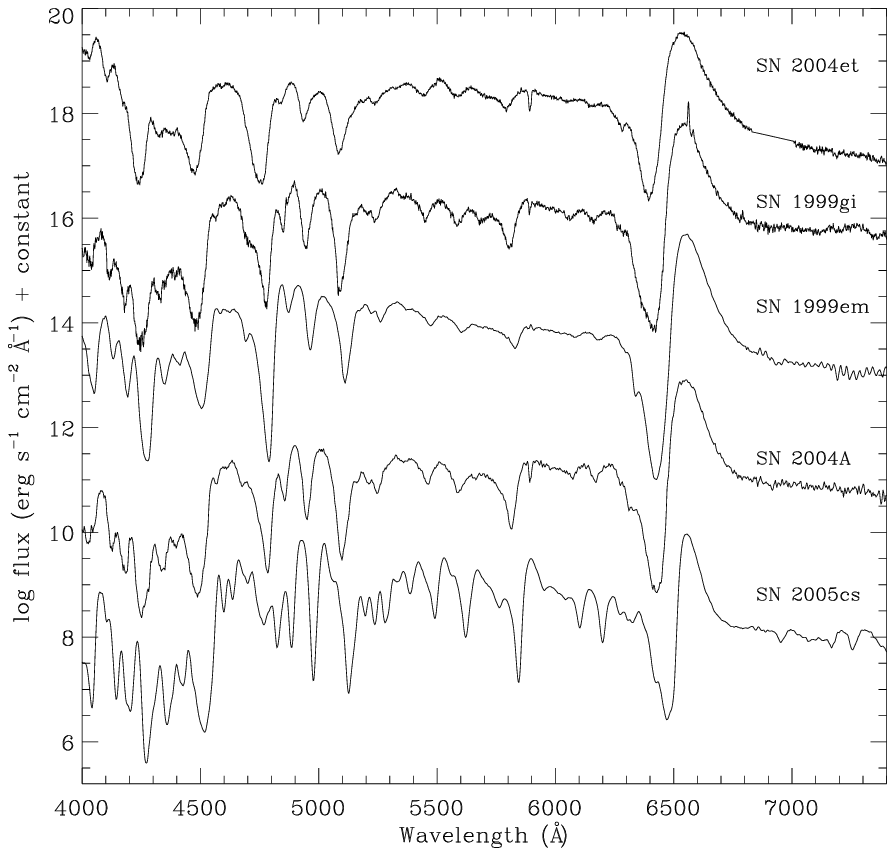}
\caption{Comparison of coeval optical photospheric spectra of type IIP SNe at $\sim$ 39 days post explosion. The spectrum of SN 2004et is taken from \protect \cite{sah06} and the spectrum of SN 2004A is from Appendix B. The other spectra are taken from the references given in Table \ref{cc}.}
\label{com_opt_spec}
\end{figure}

\begin{table*}
\begin{minipage}{126mm}
 \caption{Log of optical spectroscopic observations.}
 \label{tab_opt}
 \begin{tabular}{@{}lccccc}
  \hline
  \hline
  Date  &  JD (2450000+)  & Phase* (days)   &   Telescope +instrument	&   	Range (\AA)    	&	Resolution (\AA)   \\
    \hline
14/11/2004& 53324.5&  54&   TNG+LRS (LR-B)            &   3580--8000 & 16\\
16/11/2004& 53326.3&  55.8&  Mt. Ekar 1.82m+AFOSC&   3500--7750& 24\\
10/12/2004& 53350.2&  79.7&  Mt. Ekar 1.82m+AFOSC&    3800--7750&24\\
15/12/2004& 53355.3&  84.8&  Mt. Ekar 1.82m+AFOSC&    3600--7720&24\\
03/01/2005& 53374.2&  103.7&  Mt. Ekar 1.82m+AFOSC& 3500--7750&25\\
14/01/2005& 53385.3&  114.8&  Mt. Ekar 1.82m+AFOSC&   3500--7700&33\\
09/03/2005& 53439.7& 169.2& Mt. Ekar 1.82m+AFOSC&    3800--7700&24\\
18/03/2005& 53447.6& 177.1 &  Mt. Ekar 1.82m+AFOSC&   3800--7700&25\\
05/06/2005& 53527.7&  257.2&  TNG+LRS (LR-B)          &  3700--8070  &13\\
02/07/2005& 53554.6&  284.1&Mt. Ekar 1.82m+AFOSC&    4000--7820&36\\
28/08/2005& 53611.6&  341.1& TNG+LRS (LR-B)         &    3400--8020& 13\\
10/10/2005& 53654.5&  384.0&  Mt. Ekar 1.82m+AFOSC&   3500--7750&24\\
27/10/2005& 53671.4& 400.9&  Mt. Ekar 1.82m+AFOSC &    3800--7730&25\\
03/11/2005& 53678.3&407.8& Mt. Ekar 1.82m+AFOSC&    4000--7730&25\\
\hline
 \end{tabular}
 \medskip
* relative to the epoch of date of explosion (JD = 2453270.5)    
 \end{minipage}
\end{table*}

\subsection{Near infrared spectroscopy}

NIR spectroscopic observations were obtained at 22 epochs from +7--65 days post explosion at the 1.08-m AZT-24 telescope Campo Imperatore Observatory (Italy) with SWIRCAM, as detailed in Table \ref{tab_near}. A late time spectrum was obtained with the TNG+NICS on 25 July 2005 (+306 days post explosion).

The reduction of the NIR spectra was carried out using \textsc{iraf}'s \textsc{onedspec} package. Pairs of spectra (labelled spectrum A, spectrum B) were obtained immediately after each other with the SN at two different positions along the slit. Each pair of images were subtracted from each other (A-B) so that the sky background was removed from the 2-D spectrum of the SN. Three pairs of images were obtained for each epoch and for each grism. These subtracted pairs were then added together to produce a 2-D spectrum of higher signal-to-noise. From these combined images, the SN spectrum was then extracted and any residual background that remained in the wings of the aperture was also subtracted off.  

No arc lamp spectra were available so the sky lines in the original non-subtracted images were used to perform the wavelength calibration for the early time spectra. This was carried out by extracting along the same line as the SN but in the other image of the pair so that only the background along that line remained. These sky spectra were then used as `arcs' due to the well known features of atmospheric OH$^{-}$ ions.  A further check of the wavelength calibration was carried out using the telluric features present in the SN spectra. 

The late time spectrum from the TNG was obtained using the low resolution Amici prism with a wavelength range of 0.8--2.5 $\mu$m. The resolution of this spectrum is not given in Table \ref{tab_near} because the Amici prism has a nearly constant resolving power and so the dispersion varies by more than a factor of 3 over the spectral range. Due to this very low resolution the arc lamp lines are blended so a look-up table\footnote{http://www.tng.iac.es/instruments/nics/spectroscopy.html} detailing pixel coordinate against wavelength is used to perform the wavelength calibration. The pixel coordinates are converted to wavelengths using this table and then a further correction is performed to the fit using prominent emission lines of the SN. The SN spectrum was divided by a G-type telluric standard star to remove the strong NIR telluric features. The instrumental response of the system was found by comparing the standard star observation to that of a solar type spectrum. Then the SN spectrum was multiplied by this response function to obtain the flux calibrated spectrum.  Finally the flux calibration of the SN spectra was checked against the NIR photometry from the same epoch and adjusted if necessary. 

Figure \ref{IJ_bands_selec2} shows the spectral evolution of the NIR \textit{IJ} ($\sim$ 0.85--1.35 $\mu$m) band during the photospheric phase. The spectral features show the characteristic P-Cygni profiles of a fast moving ejecta, which are also visible in the optical spectra during the photospheric phase. Some lines of the Paschen series of H are seen, along with \Ci, \Hei\ and \Oinir\ lines. The wavelength region between $\sim$ 1.8 $\mu$m to just before 2 $\mu$m was lost due to strong atmospheric absorption in this region. The spectra are low resolution so not many spectral features can be determined unambiguously. The resolution of each spectra was calculated from the slit width and the dispersion and these values are quoted in Table \ref{tab_near}. 

The spectral evolution of the NIR \textit{HK} band ($\sim$1.45--2.2 $\mu$m) is shown in Figure \ref{HK_bands}. Five members of the Brackett series of H have features that are visible in the \textit{HK} band spectra. The \textit{HK} band does not show any major evolution between +7 days and +63 days. Like the \textit{IJ} band spectra, the resolution of the spectra is low so not many spectral features other than the H Brackett series, \Hei\ and possibly \Mgi\ can be identified. The feature identification of both the \textit{IJ} and \textit{HK} bands was carried out following \cite{fas01} and \cite{ger00}.

Figure \ref{IJ_comp} shows a comparison of two photospheric phase \textit{IJ} band spectra of SN 2004et with other type IIP SNe for which NIR spectra were available at roughly coeval epochs, SN 1997D \citep{ben01}, SN 1999em \citep{ham01} and SN 2005cs \citep{pas09}. To ensure a realistic comparison between spectra of different resolutions, the spectra have been smoothed to the resolution of the worst spectrum. The lowest resolution spectra are those of SN 2004et with a resolution of 38 \AA. The comparison spectra have been convolved with a Gaussian with a FWHM equal to this resolution and then the smoothed spectra have rebinned to a common dispersion relation. Note that the telluric features in the spectrum of SN 2005cs have not been well removed because the telluric standard spectrum was taken a few days after that of SN 2005cs \citep{pas09}.

At $\sim$ 20 days, the spectra of SN 2004et and SN 1999em show mainly the features of the Paschen series of H. For the later epoch, we note that SN 1997D and SN 2005cs had lower expansion velocities and so the separate components making up the feature at $\sim$ 1.1 $\mu$m are identifiable, unlike in the spectra of SN 2004et. The NIR spectra of this selection of type IIP SNe do show many similarities, with the H Paschen series being particularly prominent in all the spectra. Four combined optical and NIR spectra of SN 2004et are shown in Figure \ref{all_bands}. Some of the optical spectra used in this figure were taken from \cite{sah06} due to their larger wavelength coverage and matching epochs with our NIR data.

The late time NIR spectrum of SN 2004et is compared to that of SN 2005cs \citep{pas09} and SN 2002hh \citep{poz06} in Figure \ref{amici_comp}. The line identifications of \cite{ham01} and \cite{poz06} were adopted for the nebular spectrum. As in the photospheric phase spectra of SN 2004et the prominent features are those of the Paschen series of H, along with the other heavier elements marked in Figure \ref{amici_comp}. Like in the spectrum of SN 2002hh, the first overtone band of CO is also seen in SN 2004et at wavelengths longer than 2.3 $\mu$m. This is known as a indicator of dust formation and is discussed in more detail in Section \ref{dust}.

 \begin{figure}
 \includegraphics[width=8.4cm]{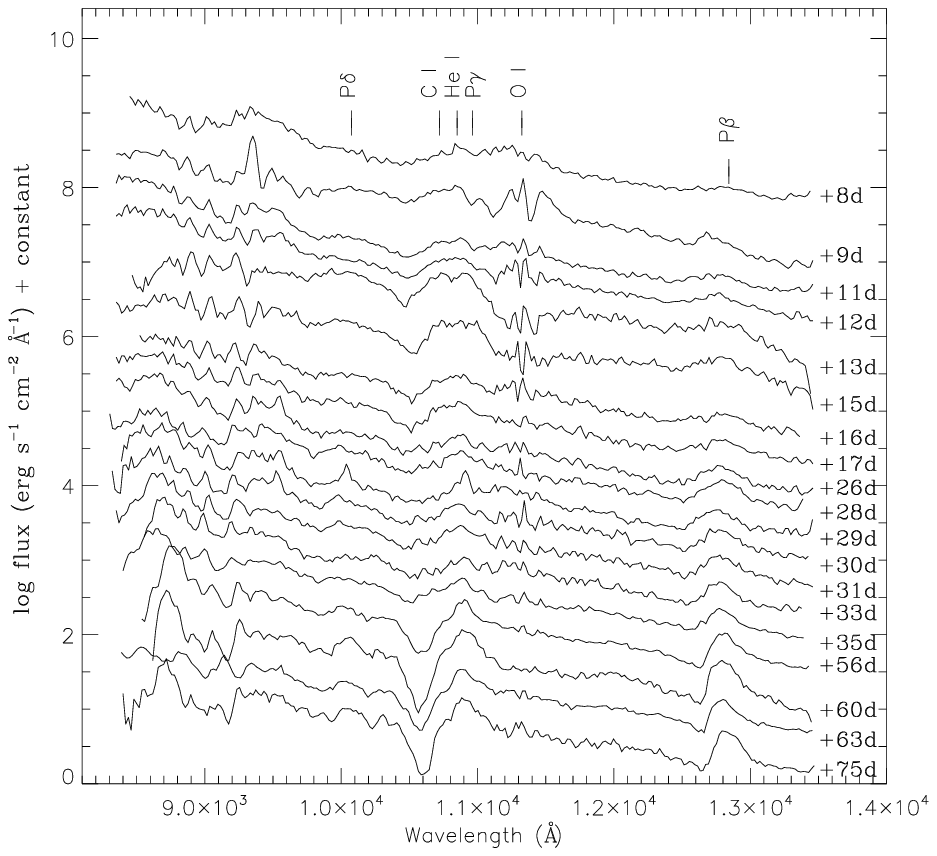}
 \caption{\textit{IJ} band spectral evolution of SN 2004et during the photospheric phase from +8 days to +75 days post explosion. The positions of the most prominent spectral features are marked.}
\label{IJ_bands_selec2}
\end{figure}

\begin{figure}
\includegraphics[width=8.4cm]{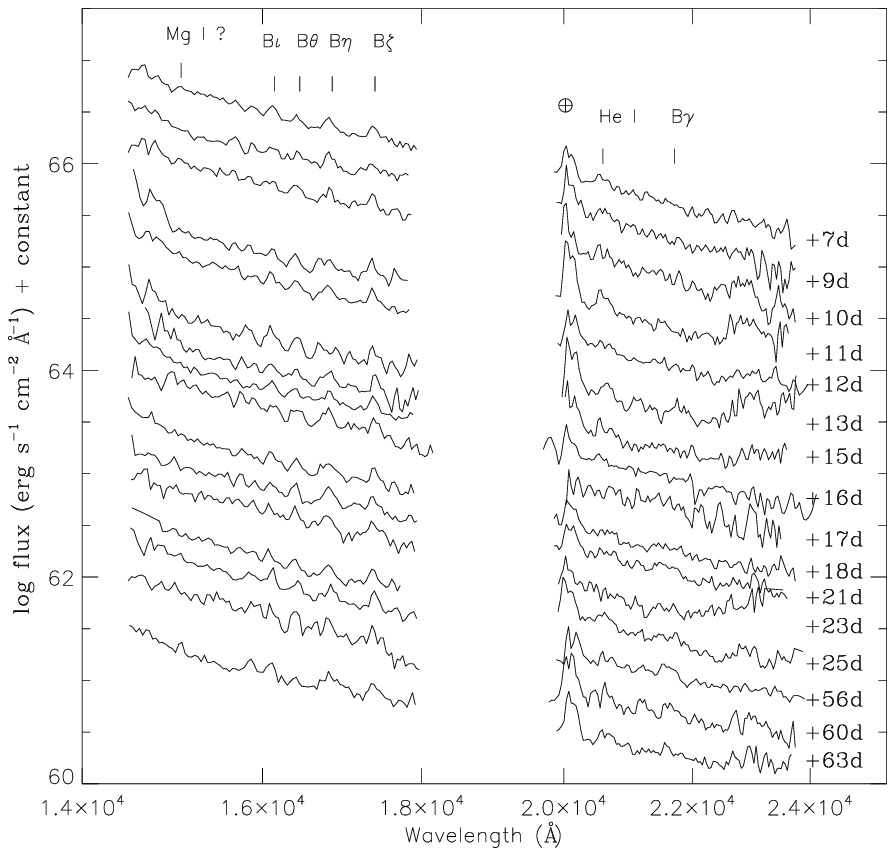}
\caption{\textit{HK} band spectroscopy during the photospheric phase from +7 days to +63 days post explosion. The positions of the most prominent spectral features are marked.}
\label{HK_bands}
\end{figure}

\begin{figure}
\includegraphics[width=8.4cm]{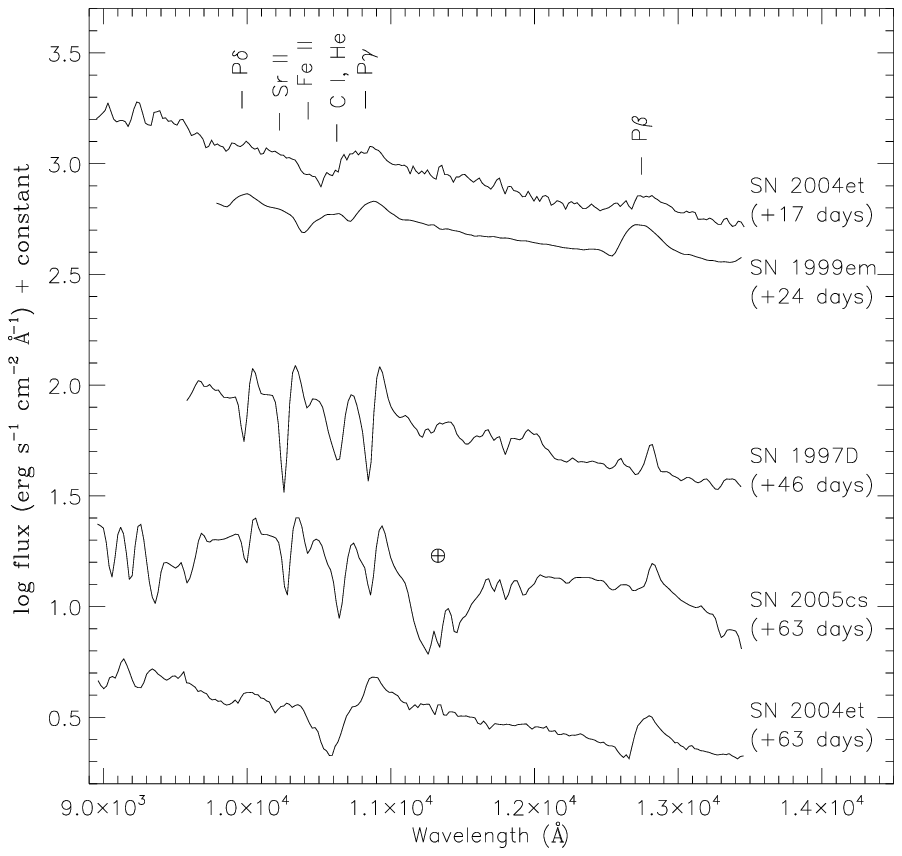}
\caption{Comparison of the \textit{IJ} band spectra of SN 1999em, SN 1997D and SN 2005cs with two spectra of SN 2004et at epochs roughly matching the comparison SNe. The positions of the most prominent spectral features are marked.}
\label{IJ_comp}
\end{figure}

\begin{figure}
\includegraphics[width=8.4cm]{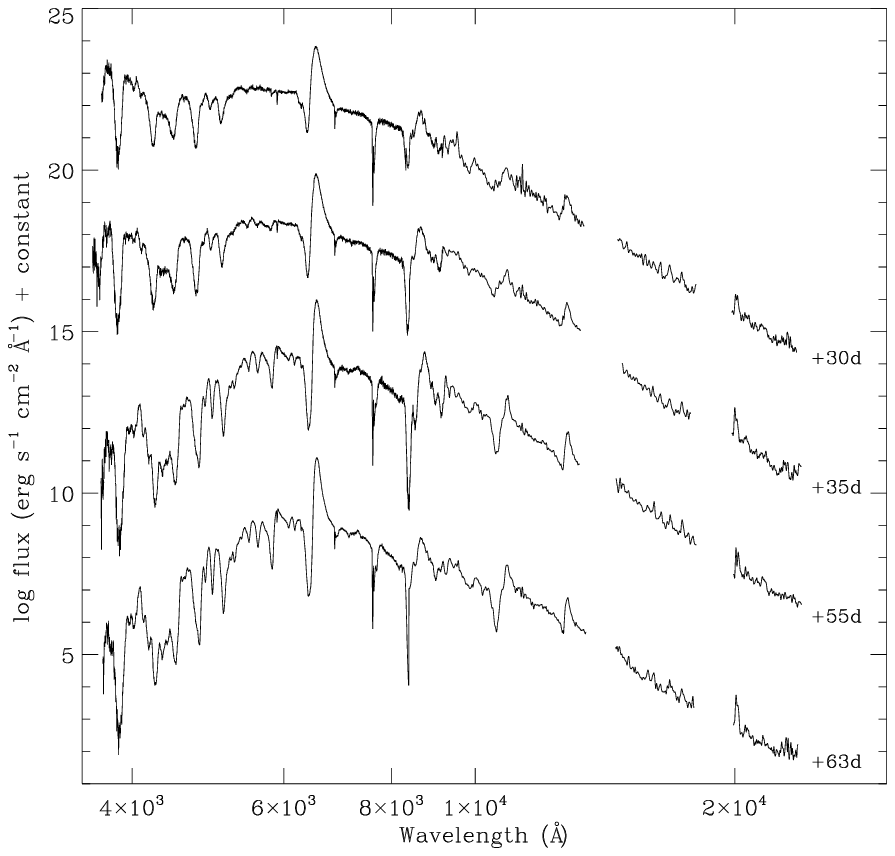}
\caption{Combined optical and NIR spectra of SN 2004et. For the epochs at +55 days and +30 days, the NIR spectra were obtained one day later than the optical observation. The NIR spectra were combined with optical spectra taken from \protect \cite{sah06} and this paper.}
\label{all_bands}
\end{figure}

 \begin{figure}
 \includegraphics[width=8.4cm]{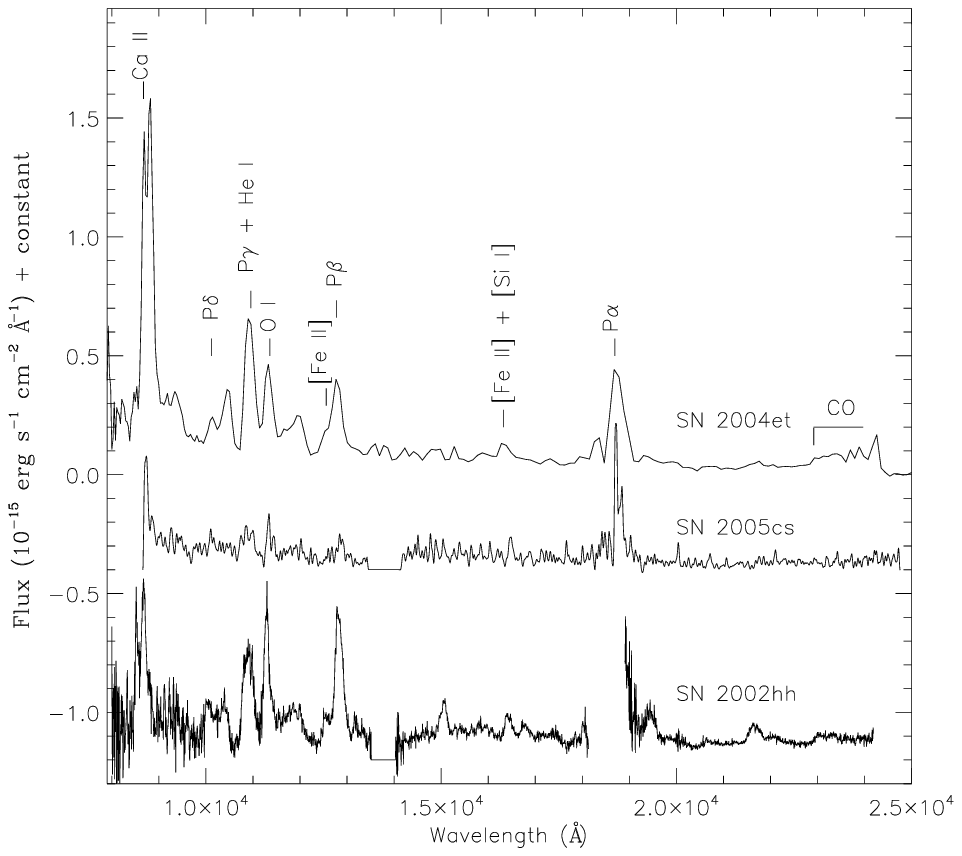}
 \caption{The late time nebular ($\sim$ 300 days) NIR spectrum of SN 2004et is compared to that of SN 2005cs at $\sim$ 281 days \protect \citep{pas09} and SN 2002hh at + 266 days \protect \citep{poz06}. }
\label{amici_comp}
\end{figure}

\begin{table*}
\begin{minipage}{126mm}
 \caption{Log of near infrared spectroscopic observations.}
 \label{tab_near}
 \begin{tabular}{@{}lcccccc}
  \hline
  \hline
  Date  &  JD (2450000+)  & Phase* (days)   &   Telescope +instrument	& Grism  &	Range ($\mu$m)    	&	Resolution (\AA)    \\
    \hline
29/09/2004& 53277.9& 7.4&    AZT-24+SWIRCAM    &  \textit{I+J}, \textit{H+K}      &   0.84--1.32, 1.45--2.38 &  38, 72\\
30/09/2004& 53278.8& 8.3&  AZT-24+SWIRCAM    &  \textit{I+J}, \textit{H+K}      &   0.84--1.32, 1.45--2.38 &  38, 72\\
01/10/2004& 53279.9& 9.4&  AZT-24+SWIRCAM    &  \textit{I+J}, \textit{H+K}      &   0.84--1.32, 1.45--2.38 &  38, 72\\
02/10/2004& 53280.8&10.3 &  AZT-24+SWIRCAM    &  \textit{I+J}, \textit{H+K}      &   0.84--1.32, 1.45--2.38 &  38, 72\\
03/10/2004& 53281.9&11.4&  AZT-24+SWIRCAM    &  \textit{I+J}, \textit{H+K}      &   0.84--1.32, 1.45--2.38 &  38, 72\\
04/10/2004& 53282.8&12.3&  AZT-24+SWIRCAM    &  \textit{I+J}, \textit{H+K}      &   0.84--1.32, 1.45--2.38 &  38, 72\\
05/10/2004& 53283.9&13.4& AZT-24+SWIRCAM    &  \textit{I+J}, \textit{H+K}      &   0.84--1.32, 1.45--2.38 &  38, 72\\
07/10/2004& 53285.8& 15.3&  AZT-24+SWIRCAM    &  \textit{I+J}, \textit{H+K}      &   0.84--1.32, 1.45--2.38 &  38, 72\\
08/10/2004& 53286.8&16.3&  AZT-24+SWIRCAM    &  \textit{I+J}, \textit{H+K}      &   0.84--1.32, 1.45--2.38 &  38, 72\\
09/10/2004& 53287.8&17.3&AZT-24+SWIRCAM    &  \textit{I+J}, \textit{H+K}      &   0.84--1.32, 1.45--2.38 &  38, 72\\
18/10/2004& 53296.8&26.3&AZT-24+SWIRCAM    &  \textit{I+J}  &   0.84--1.32 &  38\\
20/10/2004& 53298.9&28.4& AZT-24+SWIRCAM    &  \textit{I+J}, \textit{H+K}      &   0.84--1.32, 1.45--2.38 &  38, 72\\
21/10/2004& 53299.8&29.3&AZT-24+SWIRCAM    &  \textit{I+J}  &   0.84--1.32 &  38\\ 
22/10/2004& 53300.8&30.3&AZT-24+SWIRCAM    &  \textit{I+J}  &   0.84--1.32 &  38\\
23/10/2004& 53301.7&31.2&  AZT-24+SWIRCAM    &  \textit{I+J}, \textit{H+K}      &   0.84--1.32, 1.45--2.38 &  38, 72\\
25/10/2004& 53303.8&33.3& AZT-24+SWIRCAM    &  \textit{I+J}, \textit{H+K}      &   0.84--1.32, 1.45--2.38 &  38, 72\\ 
27/10/2004& 53305.8&35.3& AZT-24+SWIRCAM    &  \textit{I+J}, \textit{H+K}      &   0.84--1.32, 1.45--2.38 &  38, 72\\
17/11/2004& 53326.8&56.3& AZT-24+SWIRCAM    &  \textit{I+J}, \textit{H+K}      &   0.84--1.32, 1.45--2.38 &  38, 72\\
21/11/2004& 53330.8&60.3& AZT-24+SWIRCAM    &  \textit{I+J}, \textit{H+K}      &   0.84--1.32, 1.45--2.38 &  38, 72 \\
24/11/2004& 53333.7&63.2& AZT-24+SWIRCAM    &  \textit{I+J}, \textit{H+K}      &   0.84--1.32, 1.45--2.38 &  38, 72\\
06/12/2004& 53345.7&75.2&AZT-24+SWIRCAM    &  \textit{I+J}  &   0.84--1.32 &  19\\
25/07/2005& 53576.6&306.1&TNG+NICS                 & AMICI prism & 0.8--2.5 & (see text) \\
\hline
\end{tabular}
\begin{flushleft}
* relative to the epoch of date of explosion (JD = 2453270.5)    
\end{flushleft}
 \end{minipage}
\end{table*}

\subsection{Expansion velocity}
\label{expansion}

\begin{figure}
\includegraphics[width=8.4cm]{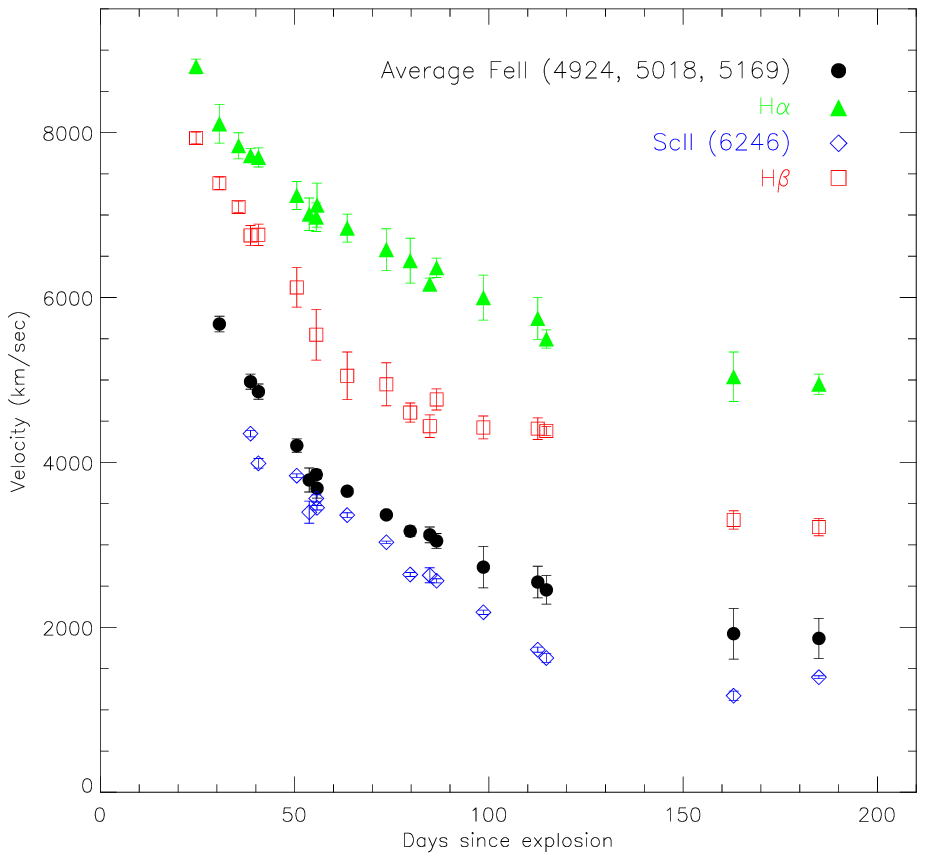}
\caption{Velocity evolution of some prominent lines of SN 2004et, measured from the minima of their absorption profiles.}
\label{vel_plot}
\end{figure}

\begin{figure}
\includegraphics[width=8.4cm]{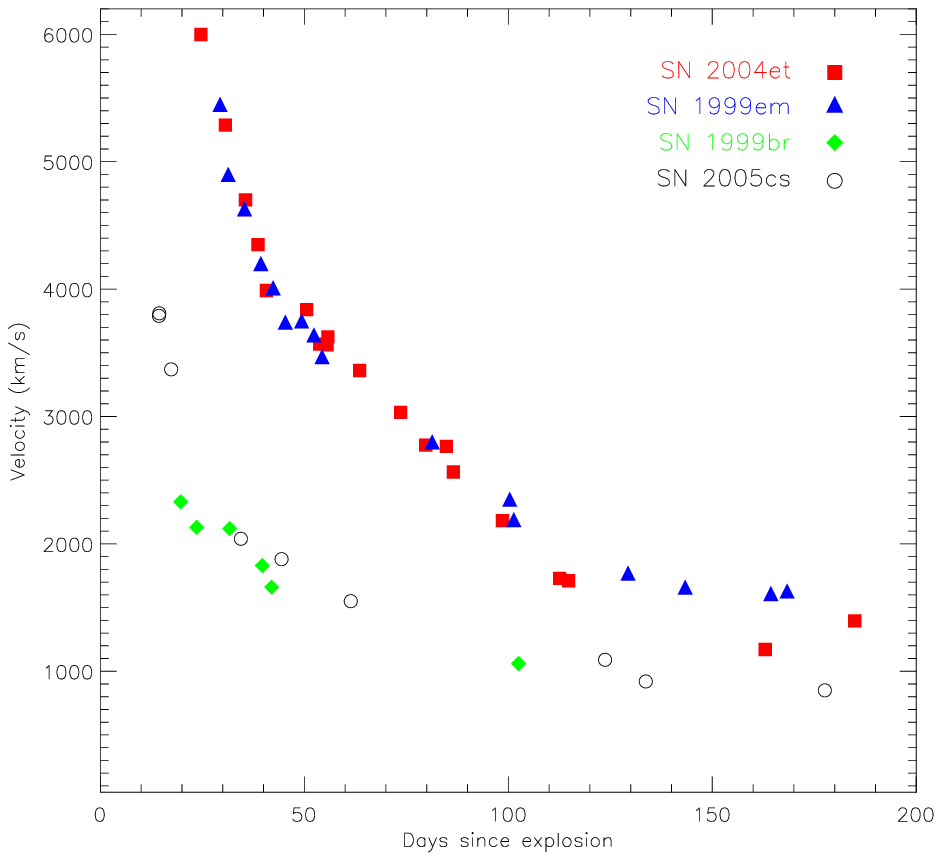}
\caption{Comparison of the expansion velocities of SN 2004et and three other type IIP SNe: SN 1999br, SN 1999em and SN 2005cs, using the \Scii\ (6246 \AA) line. The \Scii\ line was not visible in the three earliest spectra of SN 2004et and so the expansion velocity for these epochs has been estimated from 0.95 times the velocity of the average of the \Feii\ 4924, 5018, 5169 \AA\ lines. }
\label{vel_comp}
\end{figure}

The expansion velocities of \ha\ (6563 \AA), \hb\ (4861 \AA), \Scii\ (6246 \AA) and the average of the \Feii\ triplet at 4924, 5018, 5169 \AA\ are shown in Figure \ref{vel_plot} as determined by fitting a Gaussian to the absorption profile of each line and measuring the position of the minimum.  As is typical for type IIP SNe, \ha\ gives higher velocities, indicating line formation at larger radii. Lines with lower optical depths such as \Feii\ 4924, 5018, 5169 \AA\ are better indicators of the photospheric velocity. Even better in this respect is the \Scii\ 6246 \AA\ line, which has an even lower optical depth and hence is the most suitable line for determining the expansion velocity of the photosphere. 

Figure \ref{vel_comp} shows the photospheric velocity estimated from the \Scii\ (6246 \AA) line of SN 2004et compared with the velocities for three other type IIP SNe, SN 1999br \citep{pas04}, SN 1999em \citep{leo02} and SN 2005cs \citep{pas09}. The \Scii\ velocities of the comparison objects were taken from the literature. The photospheric velocity of SN 2004et is seen to have a similar evolution to that of SN 1999em. The velocities of SN 2004et and SN 1999em are, as expected, higher than the velocities of the low luminosity type IIP SNe 1999br and 2005cs.

\section{Dust formation}
\label{dust}

Dust can form in SN ejecta as they cool during the nebular phase. One of the first signs of dust formation is the formation of molecules containing carbon and silicon, which are the precursors to dust formation \citep{ger00,spy01}. The first overtone of CO emission is expected at NIR wavelengths longer than $\sim$ 2.3 $\mu$m before the onset of dust. This signature of dust formation is seen in the spectrum of SN 2004et taken at $\sim$ 300 days post explosion, which is shown in Figure \ref{amici_comp}. The SN 2002hh also shows a similar CO emission at these wavelengths, unlike SN 2005cs where no significant CO emission band is seen. No earlier NIR nebular spectra of SN 2004et are available to constrain the epoch of molecular formation and no attempt is made to estimate the quantity of CO formed due to the low resolution of the spectrum. The presence of the CO molecular band has been seen in a good fraction of type IIP SNe with observations at NIR wavelengths such as SN 1987A \citep{spy88}, SN 1995ad \citep{spy96}, SN 1998dl \citep{spy01}, SN 1999em \citep{spy01}, SN 2002hh \citep{poz06} and now SN 2004et. This suggests that type IIP SNe must cool significantly within a few hundred days of explosion to allow first molecular formation and then dust formation. 

Other indicators of dust formation include a blueshift of the spectral emission lines of elements such as \ha\ and \Oii\ 6300, 6364 \AA, an optical luminosity decrease, a NIR excess at late times or thermal emission from newly formed dust grains at MIR wavelengths. 

A blueshift of the peaks of emission lines is caused by residual opacity in the central part of the ejecta due to dust particles and results in an attenuation of the red wing of the line profile. \cite{sah06} noted that this blueshift in the peaks of the \ha\ and \Oi\ 6300, 6363 \AA\ occurred after $\sim$ 300 days. Using the spectra of \cite{sah06} and the additional spectra published in this work, the blueshift of the emission lines was quantified. The size of the shift can be determined by measuring the wavelength position of the centroid of the \ha\ emission line. For the spectra between 163 days and 300 days post explosion, the peak of the \ha\ emission line was found to be at +280 $\pm$ 50 km s$^{-1}$. The spectrum at 313 days and all subsequent spectra out to the last at 464 days show a constant blueshift to -137 km s$^{-1}$. 

 Figure \ref{blueshift} and Figure \ref{blueshifto} show the temporal evolution of the \ha\ and \Oi\ 6300, 6364 \AA\ line profiles in SN 2004et compared with those of SN 1999em. \cite{elm03} showed for SN 1999em that there was an observable blueshift of the emission lines at $\sim$ 500 days, which they suggest is caused by dust formation. We measured the rest frame position of the \ha\ peak in SN 1999em to be +182 km s$^{-1}$ at 466 days shifting to -91 km s$^{-1}$ at 642 days using spectra taken from \cite{elm03}. \cite{dan91} showed that for SN 1987A, an observable blueshift was seen in the emission peaks of the \Oi\ doublet, beyond approximately 400 days post explosion. This can again be measured from the \ha\ peak as +164 km s$^{-1}$ at +440 days, -156 km s$^{-1}$ at +501 days and then increasing to -320 km s$^{-1}$ at +628 days. The larger blueshift in the emission lines of SN 1987A compared to SN 1999em can be attributed to either a larger mass of dust forming in the ejecta or to a geometrical effect. The emission lines of SN 2004et experienced a blueshift at an earlier epoch than either SN 1987A and SN 1999em, and had intermediate values to those SN 1987A and SN 1999em. 

\begin{figure}
\includegraphics[width=8.4cm]{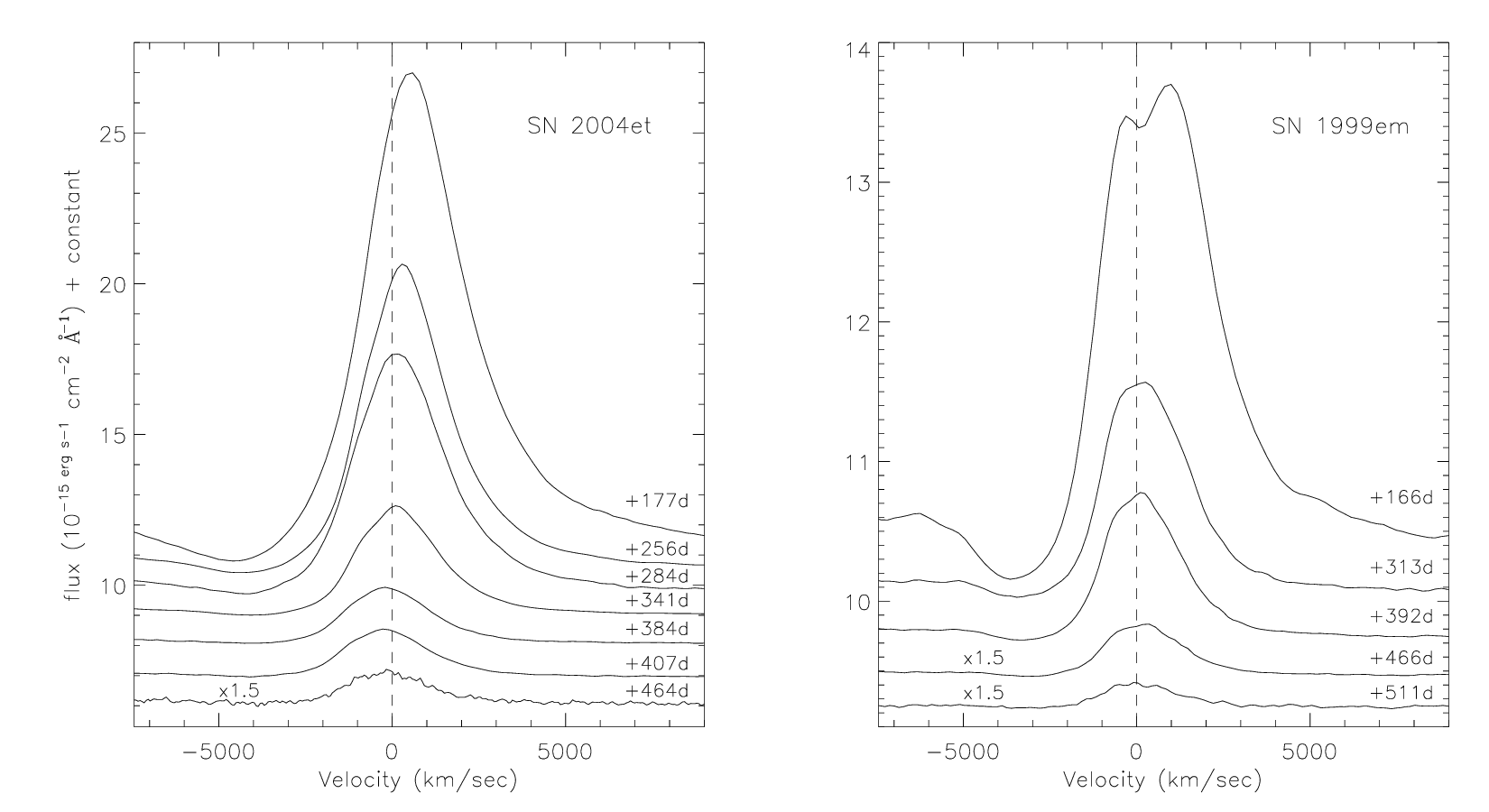}
\caption{Evolution of the line profiles of \ha\ for SN 2004et (left panel) and for SN 1999em (right panel). The vertical line in the panels corresponds to the zero velocity of \ha\  at 6563 \AA.}
\label{blueshift}
\end{figure}

\begin{figure}
\includegraphics[width=8.4cm]{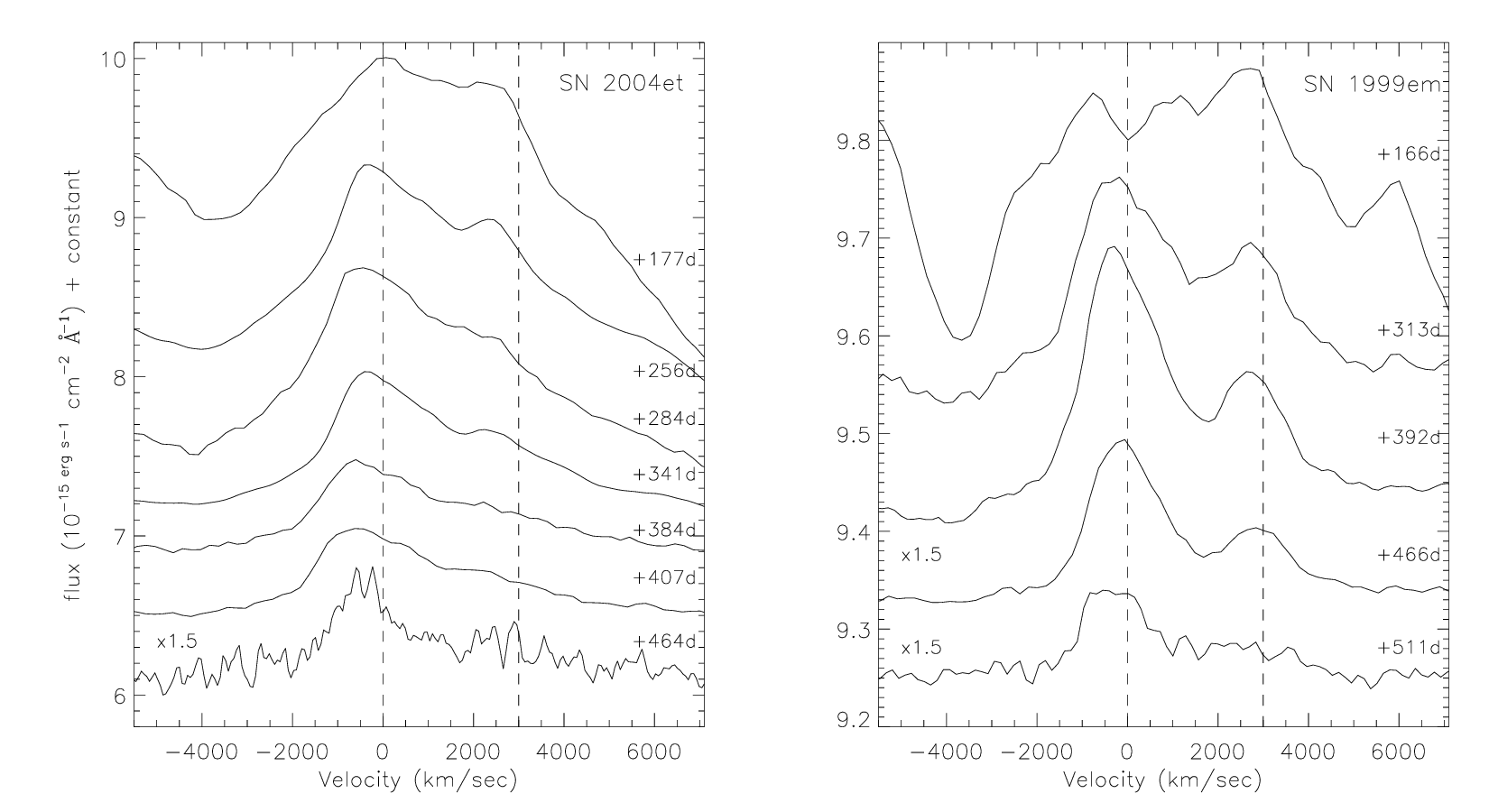}
\caption{Evolution of the line profiles of \Oi\ 6300, 6364 \AA\ for SN 2004et (left panel) and for SN 1999em (right panel). The vertical lines in the panels correspond to the zero velocities of the \Oi\ doublet at 6300, 6364 \AA\ respectively.}
\label{blueshifto}
\end{figure}

A decrease in the luminosity of the optical bands at late time is also thought to be indicative of scattering of optical photons off dust particles formed in the inner or outer envelope. Figure \ref{latetimeslope} showed a steepening in the slope of the optical \textit{BVR} bands, as detailed in Section \ref{lightcurve101} and suggests that the steepening of the slope occurs at $\sim$ 300 days. This epoch of 300 days for SN 2004et is consistent with the dust formation range estimated from the blueshift of nebular emission lines. Finally, \cite{kot09} observed thermal emission from newly formed dust grains at MIR wavelengths and presented evidence for an IR echo from the interstellar dust of the host galaxy that manifests itself as a cold component of the spectral energy distribution of SN 2004et.

\section{Estimate of O Mass}
\label{ooo}
The mass of O in type IIP SNe can be estimated from an analysis of the \Oi\ 6300, 6364 \AA\ lines \citep{uom86,chu94,li92,elm03}. The luminosity of the \Oi\ doublet in SN 2004et can be compared to that of SN 1987A during the late nebular phase but before dust formation. Dust formation did not occur until at least 400 days post explosion for 1987A \citep{dan91} but for SN 2004et dust formation appeared to occur at $\sim$ 300 days post explosion. To make a consistent comparison, spectra from $\sim$ 285 days post explosion were used. The luminosities of the \Oi\ 6300, 6364 \AA\ doublet along with the luminosities of what is thought to be the \Caii\ 7291, 7324 \AA\ doublet are given in Table \ref{oxy_table} for a selection of type IIP SNe. The feature at 7300 \AA\ could also have a contribution from the \Oii\ 7319, 7330 \AA\ doublet, particularly as SN 2004et is likely to be an O-rich SN (see below).

The spectrum of SN 1987A from 1987 December 09 was taken from \cite{ter88}. Spectra of SN 1999em and SN 2005cs were not available at the epoch of $\sim$ 285 days post explosion so spectra before and after this epoch were used and an interpolated value was derived. The spectra of SN 1999em were taken from \cite{leo02} at +168 days and +313 days post explosion. The spectra of SN 2005cs were obtained from \cite{pas09} at +281 days and +333 days post explosion. Flux calibration of the spectra used has been carried out by comparing to photometry at the same epoch.

The luminosity of the \Oi\ doublet is lower for SN 2004et than for SN 1987A. The \Oi\ doublet luminosity at late epochs is powered by $\gamma$-rays being deposited in O-rich material and a relation between O mass and line luminosity can be written as \citep{elm03}: 

\begin{equation}
\label{lumo}
\displaystyle{L_{\Oi} = \eta\ \frac{M_{O}}  {M_{exc}} L_{Co}},
\end{equation}

where M$_{exc}$ is the `excited' mass in which the $\gamma$-ray deposition takes place, $\eta$ is the efficiency of the transformation of the energy deposited in the O mass into the luminosity of the \Oi\ doublet, L$_{\Oi}$ is the O luminosity and L$_{Co}$ is the luminosity of $^{56}$Co. The ratio of the mass of O in SN 2004et to SN 1987A can be estimated, making the assumptions that $\eta$ and the `excited' mass are similar for both SNe. 

The L$_{Co}$ is directly related to the mass of $^{56}$Ni, which was estimated from the nebular phase bolometric light curve in Section \ref{bolsection} to be 0.056 $\pm$ 0.04 \msun. The O mass for SN 1987A was found to be in the range, 1.2--1.5 \msun\ \citep{li92, chu94, koz98}. In SN 2004et, the \Oi\ doublet luminosity is $\sim$ 35 per cent lower than in SN 1987A and at the same time the $^{56}$Ni mass is $\sim$ 25 per cent lower. The two differences almost match and following Equation \ref{lumo} a similar O mass is derived. The values obtained for the O mass of SN 1999em and SN 2005cs are also given in Table \ref{oxy_table}, where the $^{56}$Ni mass of SN 1999em was estimated to be 0.05 \msun\  from a comparison of its bolometric luminosity with that of SN 1987A using the Cepheid distance of 11.7 Mpc taken from \cite{leo03}. The $^{56}$Ni mass of SN 2005cs was taken as 0.003 \msun\ from \cite{pas09}.

\cite{uom86} estimated the minimum mass of O needed to produce the \Oi\ emission lines using the equation:

\begin{equation}
\label{lumo2}
\displaystyle{M_{\Oi} =  10^{8}  {F_{\Oi}} {D^{2}} {e^{2.28/T_{4}}} }
\end{equation}

where F$_{\Oi}$ is the \Oi\ doublet flux in units of ergs s$^{-1}$, D is the distance to the SN in units of Mpc and T$_{4}$ is the temperature in units of 10$^4$ K. From \cite{liu95}, the O temperature of SN 1987A at 300 days was $\sim$ 4200 K. Assuming a similar O temperature for SN 2004et at a comparable epoch, the M$_{\Oi}$ of SN 2004et was calculated for temperatures in the range 3500--4500 K. The minimum O masses determined using Equation \ref{lumo2} are given in Table \ref{oxy_table}. For SN 1999em and SN 2005cs the O temperature was also assumed to be similar to that of SN 1987A and the lower limit of the mass of O needed to produce the \Oi\ emission lines are also estimated from Equation \ref{lumo2}.

The lower limit of the O mass for SN 1987A is consistent with the O mass range set by \cite{li92} and \cite{chu94}. Equation \ref{lumo2} is most sensitive to the O temperature and this is the main source of uncertainty in the calculation. The mass of O in SN 2004et is very similar to that of SN 1987A, while the O masses of the 'normal' type IIP, SN 1999em and the low luminosity SN 2005cs are found using both methods to be lower. Despite the low  luminosity of the \Oi\ doublet seen for SN 2005cs, Equation \ref{lumo} is inversely proportional to the L$_{Co}$, which is also low for SN 2005cs and this results in similar O masses for SN 1999em and SN 2005cs. The similar O masses of SN 2004et and SN 1987A could suggest that they have similar main sequence progenitor masses, while SN 2005cs and SN 1999em have lower O masses and could be associated with lower mass progenitor stars. \cite{woo95} suggested that a main sequence star of mass of 20 \msun\ would produce an O mass of 1.5 \msun\ and so points toward relatively high mass stars as the progenitors of type IIP SNe. The discrepancies between direct imaging of SN progenitors and other estimates of the mass will be discussed in Section \ref{prog} for SN 2004et and more generally for type IIP SNe in Section \ref{exp_para}.

\begin{table*}
\begin{minipage}{\textwidth} 
 \caption{Spectral information used in Equation \ref{lumo} and Equation \ref{lumo2}. The spectra of SN 1987A and SN 2004et are both taken at $\sim$ 285 days post explosion. The luminosities for SN 1999em and SN 2005cs were interpolated as detailed in Section \ref{ooo}.}
 \label{oxy_table}
 \begin{center}
 \begin{tabular}{@{}lccccc}
 \hline
SN        &  Luminosity \Oi\ 6300, 6364 \AA\     &  Luminosity \Caii\ 7291, 7324 \AA\                                        &                                      Min. O mass from    &  O mass from      \\
           &             ( $\times$ 10$^{39} $erg s$^{-1}$)      &  ( $\times$ 10$^{39} $erg s$^{-1}$)                 &  Equation \ref{lumo2}    (\msun)         & Equation \ref{lumo}  (\msun)                \\
\hline
1987A  &  2.0 $\pm$ 0.1  &  5.6  $\pm$ 0.2           	&		0.3--1.1           &		1.2--1.5\footnote{Mass range estimate taken from \cite{li92} and \cite{chu94}.}	            \\
 1999em &0.7 $\pm$ 0.4  & 1.4 $\pm$ 0.5                &           0.1--0.4                       & 0.6--0.8                                                                                     	\\
 2004et & 1.6  $\pm$  0.4 &  3.6   $\pm$ 0.5      	&		0.2--0.9           &	1.2--1.5		\\
 2005cs&  0.05 $\pm$ 0.02     &  0.15 $\pm$ 0.02     &         0.003--0.01         & 0.8--0.9 \\
\hline
 \end{tabular}
 \end{center}
 \end{minipage}
\end{table*}

\section{Very Late time photometry of SN 2004\lowercase{et}}

\begin{figure}
\includegraphics[width=8.4cm]{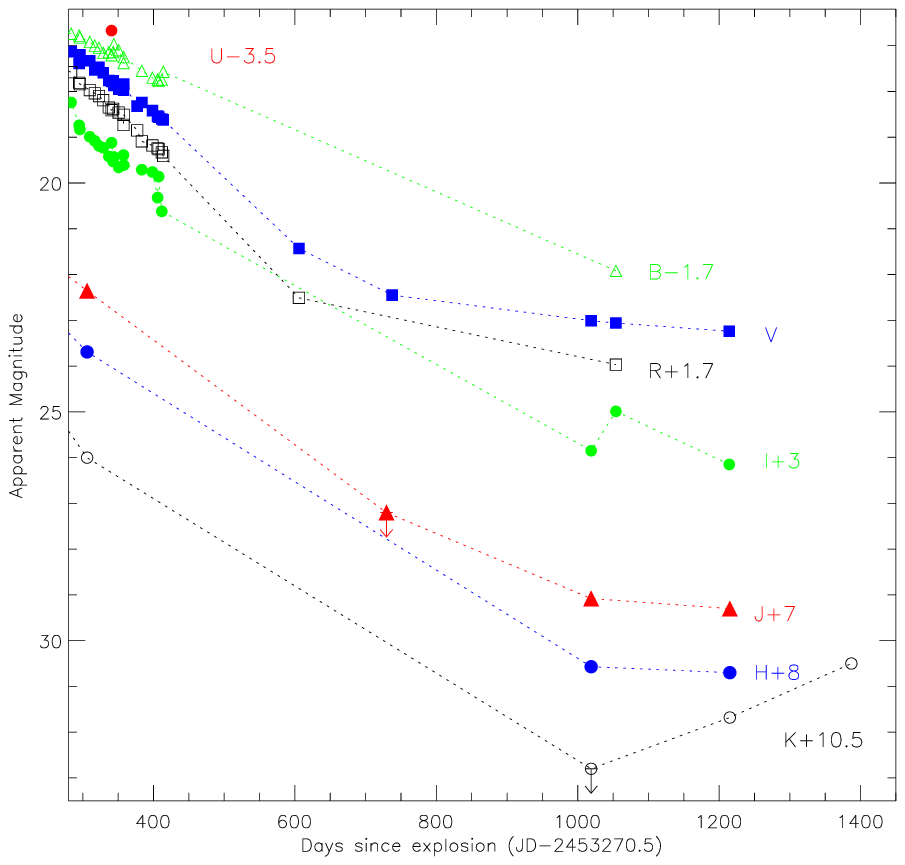}
\caption{Late time multi-band photometry of SN 2004et from +300 days post explosion to the latest Gemini North image taken on 2008 July 10 (+1387 days).}
\label{latehst}
\end{figure}

The late time optical and NIR HST data described in Section \ref{photoptnir} and detailed in Table \ref{opt_phot} and Table \ref{nir_phot} are discussed here in more detail. The two epochs of HST data were obtained at +1019 and +1214 days post explosion in five filters (two optical and three NIR). \cite{cro09} also obtained observations of SN 2004et with the WHT in four filters, \textit{BVRI} on 2007 August 12 (+1054 days), along with a ground-based adaptive optics image with the Gemini North Telescope using Altair/NIRI on 2008 July 10 (+1387 days), which gave K $\approx$ 20.0 mag. 

The late time evolution of SN 2004et at optical and NIR wavelengths is shown in Figure \ref{latehst}. The HST magnitudes have not been converted to standard \textit{VIJHK} filters because of the uncertainties in the transformation equations for SN spectra at such late phases. In this respect, the increase in the magnitude between the +1019 and +1054 days observations in the \textit{I} band should not be regarded as significant.

Ignoring this small increase, the trend between +1019 and +1214 days in all the bands is a levelling off of the magnitude decrease. The light curves at early times are powered by the radioactive decay of $^{56}$Co to $^{56}$Fe but at around +1000 days other radioactive nuclides become important. In particular \cite{fra02} showed for SN 1987A that the decay of $^{57}$Co became dominant after $\sim$ 1100 days, while at epochs greater than 2000 days, $^{44}$Ti is the dominant source of energy. This could account for some of the flattening in the light curves between +1019 and +1214 days but not for the clear increase the magnitude that is seen in the Gemini \textit{K} band magnitude at +1387 days. 

The late time optical spectra (after $\sim$ 800 days) of \cite{kot09} showed signs of an interaction between the ejecta and CSM in the wide boxy profiles of the \ha, \Caii\ and \Oi\ emission lines. This may be the source for the additional luminosity at late times. \cite{kot09} also suggested that there could be late time contributions to the luminosity from dust condensation in the cool dense shell left behind the reverse shock wave. 
The presence of interaction between the SN ejecta and the progenitor CSM is further backed up by \cite{kot09} by the discovery of a significant re-brightening in mid infrared Spitzer data. They also suggested a flux increase at optical wavelengths, which is however less convincing. They estimated the broad-band magnitudes of the \textit{BVRI} bands by fitting a black body template to late time Keck spectra. Their magnitudes estimated from a spectrum at +1146 days are $\sim$ 0.5 mag brighter in the \textit{V} band and $\sim$ 0.3 mag brighter in the \textit{I} band than the WHT images at +1054 days. In addition, their data set does not include the HST photometry from +1214 days, which does not show any increase and are actually slightly fainter than those taken with HST at +1019 days. Therefore Figure \ref{latehst} does not include the magnitudes estimated from the spectra presented in \cite{kot09}.

\section{Progenitor Mass: High mass or low mass?}
\label{prog}

The focus of this work is to present and analyse the extensive data set of SN 2004et and therefore detailed modelling of this SN is outside the purpose of this investigation. For most of the results on hydrodynamic modelling reported in the following sections, we will refer to work already published in the literature.
\cite{utr09} used a detailed 1-D hydrodynamical code to estimate the progenitor mass of SN 2004et. They modelled the bolometric light curve and the photospheric velocities of SN 2004et using hydrodynamical simulations to obtain an ejected mass of 24.5 $\pm$ 1 \msun\ and an estimated progenitor mass on the main sequence in the range 25--29 \msun.  \cite{utr09} suggested that the mass for SN 2004et obtained from the hydrodynamical modelling could be overestimated due to their use of a one-dimensional approximation. This approximation only artificially includes the effects of mixing between the He core and the H envelope and clearly cannot take into account any asphericity in the explosion. We stress that evidence for asymmetries have been found from polarimetric measurements for five other IIP SNe (e.g. \citealp{leo06}). 

\cite{you04} performed a hydrodynamical parameter study of type II SNe and SN 2004et can be compared to the models in this study. SN 2004et is found to be most similar to model B, which is a 20 \msun\ main sequence mass star with a pre-explosion radius of 431 \rsun, explosion energy of 10$^{51}$ ergs and an ejected nickel mass of 0.07 \msun. The ejected mass taking into account mass loss and remnant mass is 16 \msun. The resulting observational parameters of this model are a plateau duration of $\sim$ 100 days and an absolute \textit{V} band magnitude of $\sim$ -16.7 mag, which are both similar to the values measured observationally for SN 2004et. 

In Section \ref{exp_para} the ejected mass of SN 2004et is estimated using the hydrodynamical equations of \cite{lit85} and observational properties of SN 2004et during the plateau phase, to be 14 $\pm$ 6 \msun. The mass of the neutron star remnant of $\sim$ 1.4 $M_{\odot}$ \citep{bog07} and the mass loss due to stellar winds of $\sim$ 1 \msun, which gives a main sequence mass of  $\sim$ 16.4 $\pm$ 6 \msun. \cite{chu94} used the nucleosynthesis models of \cite{woo95} to estimate the main sequence mass of SN 1987A to be $\sim$ 20 \msun\ from their calculated O mass. The O mass range obtained for SN 2004et from the analysis of the nebular spectra is seen to be of a comparable size to the O mass of SN 1987A and hence we may guess a similar main sequence mass for SN 2004et of $\sim$ 20 \msun.

\cite{che06} showed that the radio and X-ray properties of SN 2004et were similar to those of other `normal' Type IIP SNe. SN 2004et had the highest radio luminosity of the sample studied. It was also shown using radio properties that the mass loss rate of $\sim$ 2 $\times$ 10$^{-6}$ \msun\ yr$^{-1}$ before explosion for SN 2004et is consistent with the mass-loss rate expected for a progenitor star of $\sim$ 20 $\msun$. 

These estimates of the progenitor mass of SN 2004et suggest that the main sequence mass is toward the higher end of masses for Type IIP SNe. This seems consistent with the bolometric light curve, which has one of the highest luminosities of the sample of IIP SNe shown in Figure \ref{com_opt}. The $^{56}$Ni mass ejected by SN 2004et was found to have a value of 0.056 $\pm$ 0.04 \msun, which is relatively high for a type IIP SN.  

However \citep{cro09} have studied pre-explosion images of the site of SN 2004et and find a progenitor star with a main sequence mass of 8$_{-1}^{+5}$ \msun. This value is lower than the estimates determined from studying the explosion parameters. The discrepancies between the progenitor mass estimates for type IIP SNe obtained using these two approaches has already been noted by \cite{sma09} and \cite{utr08,utr09}. Section \ref{exp_para} discusses in more detail these discrepancies for a sample of ten well-studied type IIP SNe.

\section{Explosion parameters of type IIP SNe}
\label{exp_para}

The progenitor and explosion properties of type IIP SNe can be studied in a number of ways. The observational properties of a SN such as the magnitude, expansion velocity and plateau length can be measured and models can then be used to determine the explosion parameters such as the ejected mass, explosion energy and pre-explosion radius of the star \citep[e.g.][]{zam03}. \cite{lit83, lit85} used these observational properties of SNe at mid-plateau as inputs to hydrodynamical equations to estimate their explosion parameters. \cite{ham03} determined the physical properties of 24 type IIP SNe, using the equations of \cite{lit85} but obtained much higher values than those obtained by pre-explosion imaging, particularly for the ejected mass with values in the range of 14--56 \msun. \cite{nad03} also derived the properties of a sample of type IIP SNe, using a subset of the observational data presented by \cite{ham03} and found values for the ejected mass of 10--30 \msun, which are more consistent with those obtained using direct imaging of the pre-explosion stars. 

More complex hydrodynamical models have also been developed for the analysis of the explosion parameters of type IIP SNe. SNe with good photometric and spectroscopic coverage have been studied such as SN 1987A \citep{utr93, bli00, utr04}, SN 1997D \citep{chu00,zam03}, SN 1999br \citep{zam03,pas04}, SN 1999em \citep{utr07}, SN 2003Z \citep{utr07b}, SN 2004et \citep{utr09} and SN 2005cs \citep{utr08,pas09}. They have used models that include not only the explosion energy, pre-explosion radius and ejected mass as the input parameters but also the $^{56}$Ni mass. This leads to a different set of relations for determining the physical parameters of the explosion than those determined by \cite{lit85}. \cite{zam05, zam07} presented a systematic analysis of type IIP SNe using a 1-D, Lagrangian radiation hydrodynamics code and a semi-analytic code. The SN parameters were estimated by performing a simultaneous comparison of the observed and simulated light curves, the evolution of line velocities and the continuum temperature. In some cases, these more complex models also find significantly higher masses than those obtained from direct imaging but for other SNe there is reasonably good agreement (cf Table \ref{nad_table}). SN 1987A shows excellent agreement between the mass of its progenitor and hydrodynamical modelling \citep[e.g.][]{utr93, bli00,pas05}. 

 \cite{sma09} presented estimates of the masses (and upper mass limits) of the progenitors of 20 of the nearest IIP SNe in the last 10 years. These mass estimates came from deep high resolution images of the SN field before explosion and gave an initial mass from the progenitors in the range of 8--17 \msun. If it is assumed that a typical red supergiant loses around 1 \msun\ due to stellar winds and a neutron star remnant is left behind with an average mass of 1.4 \msun\ \citep{bog07}, then the ejected masses would be in the range of 6--15 \msun, which are in general lower than those obtained from hydrodynamical modelling.

 Although the models of \cite{lit83,lit85} are based on simple hydrodynamics and use assumptions that do not completely describe the conditions of the explosion, they can be used to compare the properties of sample of type IIP SNe if they are used in a consistent manner. In this paper, we use the sample of type IIP SNe from \cite{sma09}, which have either had their progenitor star identified or have had a detection limit for the mass determined, to further investigate the apparent discrepancy between the low masses obtained from pre-explosion imaging with the high masses obtained with models. The models of \cite{lit83,lit85} are used with the same host galaxy distances and extinctions that were applied in the analysis of \cite{sma09}. We have collected photometric and spectroscopic data for as many of these objects as possible and measured the relevant parameters to calculate the physical properties of the ejecta. Table \ref{nad_table} details the distances, extinctions, observational and estimated explosion parameters from modelling along with the direct mass estimates from pre-explosion images for the sample of SNe. \cite{mis07} also carried out a similar analysis for a compilation of II-P SNe using the equations of \cite{lit85} to estimate the explosion parameters of the SNe. The advantage of our compilation is that the masses determined should be directly comparable with the progenitor masses and limits now available.

 As well as using the same distances and values of extinction of \cite{sma09}, the input parameters (\textit{V} band magnitude mid-plateau, photospheric velocity mid-plateau and the length of the plateau) for the ten SNe were reanalysed for consistency. The photospheric velocities of the 10 SNe were recalculated using the \Scii\ (6246 \AA) line, since this line probes deep into the inner regions of the SNe, and so is a good indicator of the photospheric velocity. These photospheric velocities were when possible remeasured from available spectra or the values were taken from the references listed in Table \ref{nad_table}. The plateau lengths were also measured again along with the \textit{V} band magnitudes mid-plateau.

For 2004A, the photospheric velocity could not be calculated from the \Scii\ (6246 \AA) and so was taken from the weak iron lines with a ratio of 0.95 to the \Scii\ line. This ratio is the average of those obtained when the velocities of the \Scii\ and weak iron lines are compared at epochs when both sets of lines are visible in the spectra. Both SN 2006my and SN 2006ov were only discovered at the end of the plateau and so to estimate the observational parameters, they were compared to other well constrained type IIP SNe. The light curve properties of SN 2006my were seen to be similar to those of normal type IIP SN, SN 1999em and so its parameters were estimated by a direct comparison. While SN 2006ov was found to be similar to the low-luminosity SN 2005cs.

 It can be seen in Table \ref{nad_table} that the estimated explosion energies vary by a factor of 10 between the low luminosity and `normal' SNe, with the less luminous SNe having lower explosion energies. A comparison between the explosion energies calculated using the equations of \cite{lit85}, the more complex models and the simple relation that the internal energy is assumed to be converted into kinetic energy is shown in Table \ref{energies}. The kinetic energy (1/2mv$^2$) is a rough approximation since it assumes that all the material ejected is moving at the same velocity, v. The mass, m is the ejected mass obtained from the progenitor analysis of the pre-explosion images. Although the energies from the \cite{lit85} models are smaller than those obtained from the more complex models with references given in Table \ref{energies}, they show a similar trend of energies with the low luminous SNe having lower energies and the more luminous SNe having higher explosion energies. The inferred radii of the progenitors, apart from SN 1999br, span the range $\sim$ 200--600 \rsun\ in the majority of cases, which is the range expected for extended red supergiant atmospheres.

\cite{kas09} used numerical simulations to determine how the light curves and spectra of type IIP SNe vary with their mass, metallicity and explosion energy. They investigated the standard candle relationship of type IIP SNe proposed by \cite{ham02} and showed that the correlation between the plateau luminosity and photospheric velocity holds for their set of model data. They also explored the relationship between the plateau length and the mid-plateau luminosity and found a linear relationship where the brighter SNe have shorter plateau lengths. However this correlation is not seen for our sample of 10 SNe, where the plateau length range is 100--120 days and the absolute \textit{V} band magnitude at +50 days has values between -13 and -17 and no correlation between these parameters is seen. \cite{kas09} suggested that a SN of plateau length of 110 days (the average plateau length we found) should have an absolute \textit{V} band magnitude of at least -18, which is not even reached by the brightest SN in our sample, SN 2004et at -17.15 in the \textit{V} band.

\begin{table*}
 \caption{Type IIP SNe parameters, calculated using the equations of \protect \cite{lit85}.}
 \label{nad_table}
 \begin{tabular}{@{}lccccccccccc}
 \hline
SN        &  D (Mpc)   & A$_{v}$   & $\Delta$t  &  vel$_{ph}$&  M$_{V}$& Energy &   Radius    & Mass$_{ej}$$^{1}$  & ZAMS$_{mod}$$^{2}$   &  ZAMS$_{img}$$^{3}$ &Ref. \\
             &                    &                    &                 &       (km/s)    &                &  ($\times10^{51}$ erg)  &    (R$_{\odot}$)&   (\msun)           &      (\msun)         &           (\msun)& \\
\hline
1999br &   14.1 $\pm$   2.6	& 0.06 $\pm$  0.06  &  100 $\pm$ 15  & 1541 $\pm$ 150 &  -13.16 $\pm$ 0.45  & 0.20 $\pm$ 0.09  &    31 $\pm$ 19 &   20  $\pm$ 10   & 12 $\pm$ 2 &   $<$ 15	&  1, 2\\ 
1999em &   11.7 $\pm$   1.0	& 0.31 $\pm$  0.16  &  120  $\pm$ 10 &  3046 $\pm$ 150 &  -16.68 $\pm$ 0.27  & 0.84 $\pm$ 0.29  &   437 $\pm$ 173 &   18  $\pm$ 7  &11,  21--29&  $<$ 15	&	2, 3, 4, 5  \\ 
1999gi &   10.0 $\pm$   0.8	& 0.65 $\pm$  0.16  &  115  $\pm$ 10 &  2717 $\pm$ 150 &  -15.67 $\pm$ 0.26  & 0.64 $\pm$ 0.25  &   183 $\pm$ 69 &   21  $\pm$ 9   & -- &  $< $ 14&	 6 \\ 
2003gd &   9.3  $\pm$  1.8	& 0.43 $\pm$  0.19  &  113  $\pm$ 20 &  3210 $\pm$ 200 &  -16.06 $\pm$ 0.47  & 1.04 $\pm$ 0.49  &   179 $\pm$ 122 &   24  $\pm$ 13 & 10 $\pm$ 1&    7$_{-2}^{+6}$& 2, 7   \\ 
2004A  &   20.3 $\pm$  3.4	& 0.19 $\pm$  0.09  &   107  $\pm$ 20 &  3200 $\pm$ 200 &  -16.33 $\pm$ 0.39  & 0.68 $\pm$ 0.37  &   328 $\pm$ 197 &   15	$\pm$ 9  & 12 $\pm$ 2&    7$_{-2}^{+6}$ & 2, 8, 9, 10\\ 
2004dj &   3.3  $\pm$  0.3		& 0.53 $\pm$  0.06  &  105  $\pm$ 20 &  2957 $\pm$ 150 &  -16.16 $\pm$ 0.23  & 0.65 $\pm$ 0.30  &   277 $\pm$ 107&   16 $\pm$ 9    &-- &    15  $\pm$ 3 &11, 12, 13 \\ 
2004et &   5.9 $\pm$   0.4		& 1.3  $\pm$  0.2   &  110  $\pm$ 15 &  3462 $\pm$ 150 &  -17.15 $\pm$ 0.27  & 0.88 $\pm$ 0.31    &   631 $\pm$ 251 &   14 $\pm$ 6   &25--29 &    8$_{-1}^{+5}$& 9, 14, 15  \\ 
2005cs &   7.1 $\pm$   1.2	        & 0.16 $\pm$  0.1  &  118  $\pm$ 15 &  1500 $\pm$ 150 &  -14.66 $\pm$ 0.39  & 0.17 $\pm$ 0.08     &   208 $\pm$ 123 &    13 $\pm$ 6  & 11--18&    8  $\pm$ 2	& 16,17,18,19,20 \\ 
2006my &   22.3 $\pm$ 2.6           	& 0.08              & 120 $\pm$ 20 &  2953 $\pm$ 300& -16.26 $\pm$ 0.28  & 0.86 $\pm$ 0.44     & 274 $\pm$ 135 & 22 $\pm$ 12   &--&    $ <$ 13	 & 9, 15\\ 
2006ov &   12.6 $\pm$  2.4	& 0.07  & 118 $\pm$ 30 & 1410 $\pm$ 200 & -15.12 $\pm$ 0.47  & 0.12 $\pm$ 0.09     &465 $\pm$ 363 & 9 $\pm$ 7   & --  &  $<$ 10	 & 15, 21 \\ 
 \hline
 \end{tabular}
   \medskip \\
 $^{1}$Ejected mass calculated from the equations of \cite{lit85}.\\
 $^{2}$Zero age main sequence mass from the hydrodynamical modelling, see references for further information.\\
 $^{3}$Zero age main sequence mass from direct imaging of the progenitor star.\\
       REFERENCES -- (1) \cite{pas04}; (2) \cite{zam07} (3) \cite{ham01}; (4) \cite{leo02}; (5) \cite{utr07}; (6) \cite{leo02b}; (7) \cite{hen05}; (8) \cite{hen06}; (9) this paper; (10) \cite{tsv08b};  (11) \cite{chu05}; (12) \cite{tsv08}; (13) \cite{vin06}; (14) \cite{utr09}; (15) \cite{cro09}; (16) \cite{pas09}; (17) \cite{tsv06}; (18) \cite{tak06}; (19) \cite{li06}; (20)\cite{utr08}; (21) Spiro et al. (in prep.)
\end{table*}

\begin{figure}
\includegraphics[width=8.4cm]{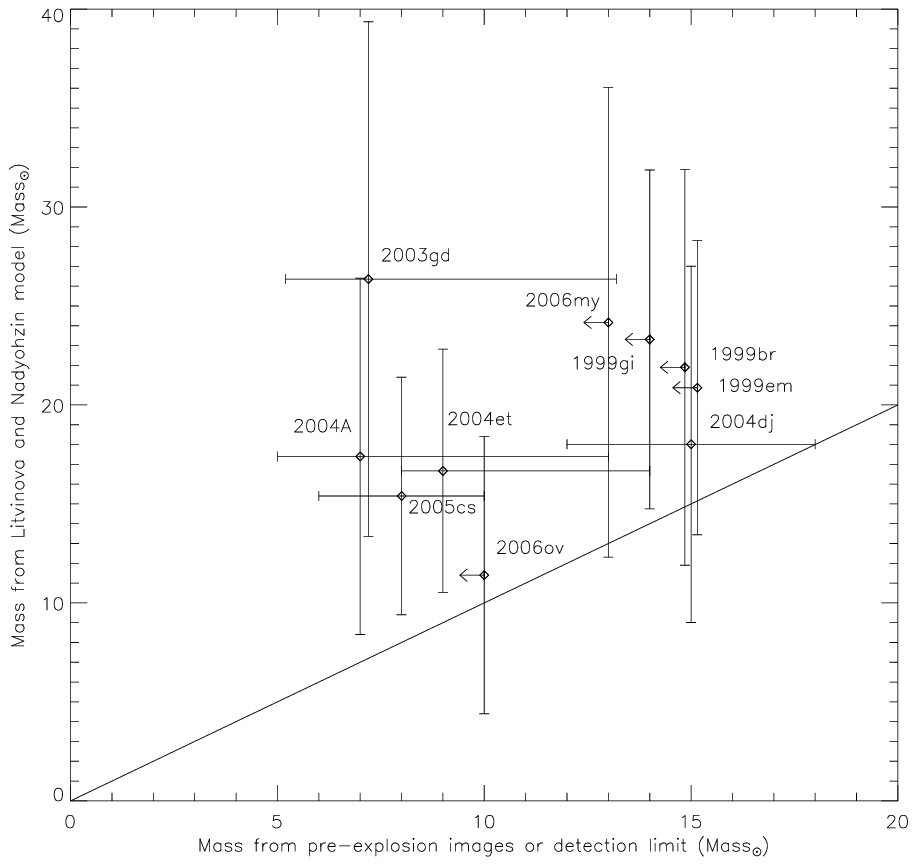}
\caption{The progenitor masses obtained from the hydrodynamical equations of \protect \cite{lit83, lit85} are plotted against the progenitor masses  obtained from the pre-explosion images. Upper mass limits are plotted where detection limits have been set \protect \citep{sma09}. The values are given in Table \ref{nad_table}, along with the associated errors. A one-to-one correspondence line is shown.}
\label{nad_mass}
\end{figure}

\begin{table}
 \caption{Comparison of explosion energies of well-studied type IIP SNe. The kinetic energy is measured using an approximation of 1/2mv$^2$, where v is the photospheric velocity and m is the ejected mass obtained from the analysis of pre-explosion images of the progenitor.}
 \label{energies}
 \begin{tabular}{@{}lcccccccccc}
 \hline
SN        &  Explosion energy$^{1}$  &   Explosion energy$^2$   &  Kinetic energy               &Ref. \\
             &  ($\times10^{51}$ erg)   &   ($\times10^{51}$ erg)    &    ($\times10^{51}$ erg)   & \\
\hline
1999br  &   0.3   & 0.1 $\pm$ 0.09   &$<$ 0.3 &1                       \\ 
2005cs &   0.4--0.8 & 0.17 $\pm$ 0.08   &0.05 -- 0.16 &1, 2 \\ 
1999em &   1.3 $\pm$ 0.3& 0.84 $\pm$ 0.29   &       $<$ 1.2  &1, 3\\ 
2003gd &   1.6 $\pm$ 0.2 & 1.04 $\pm$ 0.49   &  0.3 -- 1&1 \\ 
2004et  &   2.3 $\pm$ 0.3 & 0.88 $\pm$ 0.31   &    0.5 -- 1.3 &4\\ 
 \hline
 \end{tabular}
   \medskip \\
 $^{1}$Explosion energy from detailed hydrodynamical models (see references for more details).\\
 $^{2}$Explosion energy from the equations of \cite{lit85}.\\
       REFERENCES -- (1) \cite{zam07}; (2) \cite{utr08}; (3) \cite{utr07}; (4) \cite{utr09}
\end{table}

A plot of the progenitor masses of a selection of recent IIP SNe obtained with the equations from \cite{lit83, lit85} against the progenitor masses obtained from analysing pre-explosion images of the SN sites is shown in Figure \ref{nad_mass}. The data with upper mass limits were calculated using the detection limits of the images and the stellar evolution codes, STARS \citep{eld04} and are detailed in \cite{sma09}. To calculate the pre-explosion mass from the ejected mass, 1.4 \msun\ was added to account for the neutron star remnant along with 1 \msun\ to account for the mass lost from the star through stellar winds. It can be seen clearly in Figure \ref{nad_mass} that the hydrodynamical models give consistently higher main sequences masses that the masses/mass limits measured from pre-explosion imaging. Using the equations of \cite{lit85}, SN 2003gd shows a model mass over three times greater than the value obtained from direct imaging, which is thought to be due to the uncertainty in the plateau length because of the late phase during the plateau at which this SN was discovered. A shorter plateau for this object would give a smaller progenitor mass and would be closer to the progenitor mass detected by \cite{sma04}.

\cite{heg03} showed that a median progenitor mass value of 13 \msun\ for type IIP SNe is obtained for a Salpeter initial mass distribution in the mass range 9--25 \msun. The masses obtained from direct imaging are in reasonable agreement with this median value, spanning a range of $\sim$ 8--17 \msun\ \citep{sma09}.  It is predicted by theory that red supergiants between 17--30 \msun\ also produce type IIP SNe but no progenitor in this mass range has been detected to date. This apparent contradiction has been termed the ``red supergiant problem" by \cite{sma09} and was tentatively suggested that the higher mass stars could form black holes with no explosion. (Although \cite{sma09b} gives alternative explanations for this issue.) The mass obtained from the hydrodynamical models of \cite{lit83, lit85} are not consistent with the median progenitor mass value, while the more complex models, in some cases such as SN 2004et with a mass of 25--29 \msun, are nearly at the theoretical mass limit for red supergiants that produced type IIP SNe.

Furthermore, the discovery of several progenitor stars with low luminosities and hence fairly low initial masses (inferred from stellar evolutionary tracks; \citep{sma04,mau05,li05,mat08}) leads to a consistency question. Can a star with an initial mass of 8 \msun\ produce a progenitor with enough envelope mass to sustain a light curve plateau of 100 day duration given the relatively high observed expansion velocities (see \cite{hen05,hen06,sma09})? Also if the higher luminosity SNe with even higher expansion velocities are also produced by stars at the low end of the mass scale, what is happening during the explosion that cause the vastly varying events that are observed and how are the higher mass stars ending their lives? It appears that it is extremely difficult to reproduce a long plateau phase and high expansion velocities with low progenitor masses from any of the previously discussed hydrodynamical models. This could either point to a systematic error in the direct progenitor masses perhaps due to circumstellar dust dimming the progenitor luminosity (summarised in \citealp{sma09b}) or deficiencies in the physics of the hydrodynamical calculations as discussed recently in \cite{utr09}.

\section{Summary and conclusions}

Extensive photometric and spectroscopic data of SN 2004et at both optical and NIR wavelengths have been presented to give one of the most comprehensive data sets of a normal type IIP SN to date. Analysis of the bolometric light curve shows that SN 2004et has one of the highest luminosities of type IIP SNe and consequently a relatively large ejected mass of $^{56}$Ni of 0.056 $\pm$ 0.04 \msun. The importance of including the flux from the NIR bands when calculating the bolometric light curve is demonstrated and shown to account for up to 50 per cent of the flux during the plateau phase. Parametrised bolometric corrections including the NIR contributions were detailed and these can be used as templates for futures type IIP SNe if their spectral range coverage is incomplete. 

Excellent spectral coverage of SN 2004et at optical and NIR during the plateau and nebular phase are shown, with the optical spectra extending to +464 days and the NIR spectra to +306 days. The final epoch NIR spectrum allowed a clear detection of the first overtone band of CO at $\sim$ 2.3 $\mu$m, which is a signature of dust formation. Other signatures of dust formation were also observed such as a significant blueshift in the \ha\ and \Oi\ emission lines and an increase in the rate of decline of the optical bands. The epoch of dust formation for SN 2004et (post 300 days) was significantly earlier than for SN 1999em and SN 1987A, while the observed blueshift of the emission lines of SN 2004et had a intermediate shift between that of SN 1987A and SN 1999em. Very late time HST and WHT observations ($>$1000 days) showed a levelling off in the decline rate of the optical and NIR bands, which is thought to be mainly caused by the interaction of the ejecta with the CSM. Signatures of this late time interaction were also seen in approximately coeval optical spectra and MIR Spitzer data of \cite{kot09}. A contribution from the decay of $^{57}$Co and $^{44}$Ti could also be present at these epochs along with contributions from dust condensation in a cool dense shell around the SN.

The physical parameters of SN 2004et were compared to those of other type IIP SNe. By studying the strengths of the nebular phase \Oii\ 6300, 6364 \AA\ lines, the ejected O mass of SN 2004et was estimated to be $\sim$ 0.5--1.5 \msun, which is comparable to that of SN 1987A. The kinetic energies of a sample of well studied type IIP SNe were found to span a range of 10$^{50}$--10$^{51}$ ergs, with SN 2004et having the highest kinetic energy of the sample. 
The explosion parameters of SN 2004et were also calculated using the hydrodynamic models of \cite{lit85} and compared to other type IIP SNe, for which the progenitor star had either been identified in pre-explosion images or an upper mass limit had been set. In some cases, the masses determined from the previously discussed hydrodynamical modelling are seen to be consistently higher than those obtained from direct imaging of the progenitor. SN 2004et showed a particularly large discrepancy with a mass range determined from modelling of 16--29 \msun, while the mass obtained by \cite{cro09} from direct imaging had a value of  8$_{-1}^{+5}$ \msun. With the current models, it appears difficult in some cases to reconcile these high luminosity and high velocity events with the low progenitor masses of 7--8 \msun\ obtained from pre-explosion imaging. 

{\small{ \textit{Acknowledgements}. This work, conducted as part of the award `Understanding the lives of massive stars from birth to supernovae' (S. J. Smartt) made under the European Heads of Research Councils and European Science Foundation EURYI (European Young Investigator) Awards scheme, was supported by funds from the Participating Organisations of EURYI and the EC Sixth Framework Programme. SB and EC acknowledge some support from contract ASI/COFIS. We thank Nikolai Chugai and Melina Bersten for helpful discussions on progenitor models.

This paper is based on observations made with the following facilities: the 1.82-m Copernico Telescope of the Asiago Observatory (Asiago), the 3.58-m Italian National Telescope Galileo operated by the Fundacin Galileo Galilei of the INAF (La Palma), the 0.72-m TNT telescope of the Teramo Astronomical Observatory (Teramo), the 2.6-m Nordic Optical Telescope (La Palma), 0.7-m AZT2 telescope of the Sternberg Astronomical Institute (Moscow), 0.6-m Z600 telescope of the Crimean Laboratory of SAI (Crimea), the 0.38-m KGB telescope at the Crimean Astrophysical Observatory (Crimea) and the 1-m SAO Z1000 telescope of the Russian Academy of Sciences (Zelenchuk). Near infrared data were collected with AZT-24 telescope (Campo Imperatore, Italy), operated jointly by Pulkovo observatory (St. Petersburg, Russia) and INAF Observatorio Astronomico di Roma/Collurania. Observations were also used from NASA/ESA HST, obtained from the data archive at the Space Telescope Institute. STScI is operated by the association of Universities for Research in Astronomy, Inc. under the NASA contract NAS 5-26555.}}

\newpage

\appendix

\section{Bolometric corrections}
\label{bol_app}

Here we detail the parametrised coefficients for the BC of four type IIP SNe; SN 1987A, SN 1999em, SN 2004et and SN 2005cs. The BC as a function of days since explosion for the $V$ and \textit{R} bands during the plateau phases are shown in Figure \ref{plot_bolcorr1}. The coefficients for the \textit{V} and \textit{R} bands during the photospheric phase are given in Table \ref{bc_para1} and Table \ref{bc_para2} respectively. The nebular phase \textit{V} and \textit{R} band BC is shown in Figure \ref{plot_bolcorr3} and the parametrised coefficients are given in Table \ref{bc_para3}. The parametrised equations can be used in Equation \ref{bol_corr3} and Equation \ref{bol_corr2} to calculate the luminosity of a SN with limited photometric coverage.

\begin{table}
 \caption{Parametrised coefficients for the bolometric corrections for four well studied type IIP SNe during the plateau phase using the \textit{V} band.}
 \label{bc_para1}
 \begin{tabular}{@{}lccccc}
  \hline
  \hline
a$_i$ & SN 1987A & SN 1999em & SN 2004et & SN 2005cs \\
    \hline
a$_0$ &0.908	&0.432&0.477&0.439\\
a$_1$ &-0.014	&0.025&0.023&0.030\\
a$_2$ &2.62$\times$10$^{-4}$&-4.16$\times$10$^{-4}$&-3.44$\times$10$^{-4}$&-6.37$\times$10$^{-4}$\\
a$_3$ &-1.58$\times$10$^{-6}$&1.95$\times$10$^{-6}$&1.44$\times$10$^{-6}$&3.70$\times$10$^{-6}$\\
\hline
 \end{tabular}
 \medskip
\end{table}

\begin{table}
 \caption{Parametrised coefficients for the bolometric corrections for four well studied type IIP SNe during the plateau phase using the \textit{R} band.}
 \begin{tabular}{@{}lccccc}
  \hline
  \hline
a$_i$ & SN 1987A & SN 1999em & SN 2004et & SN 2005cs \\
    \hline
a$_0$ &1.071&0.531&0.389&0.378\\
a$_1$ &5.10$\times$10$^{-3}$&0.031&0.033&0.062\\
a$_2$ &-4.03$\times$10$^{-5}$&-4.49$\times$10$^{-4}$&-4.27$\times$10$^{-4}$&-1.62$\times$10$^{-3}$\\
a$_3$ &3.81$\times$10$^{-5}$&2.07$\times$10$^{-6}$&1.80$\times$10$^{-6}$&1.79$\times$10$^{-5}$\\
\hline
 \end{tabular}
 \label{bc_para2}
\end{table}

\begin{table}
 \caption{Parametrised coefficients for the bolometric corrections for SN 1987A and SN 1999em during the early nebular phase using the \textit{V} and $R$ bands.}
 \begin{tabular}{lccccc}
  \hline
    \hline
 a$_i$ &    \multicolumn{2}{c}{SN 1987A}&   \multicolumn{2}{c}{SN 1999em} \\
 &$V$&$R$&$V$&$R$\\
    \hline
a$_0$ &0.338&0.899&0.165&1.097\\
a$_1$ &-5.46$\times$10$^{-8}$&1.76$\times$10$^{-3}$&1.07$\times$10$^{-3}$&8.71$\times$10$^{-4}$\\
\hline
 \end{tabular}
\label{bc_para3}
\end{table}

\section{SN 2006\lowercase{my} data}
\label{06my}

We present here previously unpublished optical photometry and spectroscopy of SN 2006my. SN 2006my was a type IIP SNe, which exploded in NGC 4651 at a distance of 22.3 $\pm$ 2.6 Mpc \citep{sma09}. Table \ref{2006my} details the optical photometric observations of SN 206my, while the spectroscopic observations are detailed in Table \ref{tab_opt7}. Figure \ref{2006my_spec} shows the optical spectral evolution of the SN. The bolometric light curve of SN 2006my is included in Figure \ref{com_opt} as comparison to SN 2004et. The date of explosion is not well constrained by observations but an estimate of the explosion epoch is made by shifting the end of the plateau phase of SN 2006my to match that of SN 1999em and then assuming the same plateau length. This value of JD 24553943.0 is used in Figure \ref{com_opt} as the explosion date.

\begin{table*}
 \caption{Log of optical photometric observations of SN 2006my.}
 \label{2006my}
 \begin{tabular}{@{}lccccccc}
 \hline
  Date  &    JD(2450000+)   & Phase* (days) &            \textit{B}     &          \textit{V}       &        \textit{R}    &         \textit{I}   &         Instrument  \\
  \hline
28/06/2006  &53915	    &-28&\    & \ &19.00 (limit)$^{a}$        &\ &6  \\
08/11/2006  &54048.3 &105.3&\ & \  &15.30$\pm$0.20$^a$	 &\  &6  \\
09/11/2006  &54049.3 &106.3&\  & \ &15.30$\pm$0.20$^a$	 &\  &6  \\
22/11/2006  &54062.6 &119.6& 17.50$\pm$0.02& 15.89$\pm$0.01 & 15.41$\pm$0.01 & 15.20$\pm$0.01 & 2  \\
27/11/2006  &54066.7 &123.7& 17.51$\pm$0.02& 16.01$\pm$0.01  &15.52$\pm$ 0.02 & 15.18$\pm$0.01  & 1  \\
30/11/2006  &54070.7 &127.7& \ & 16.24$\pm$0.03  &15.59$\pm$0.02&  15.35$\pm$0.01  &2 \\
04/12/2006  &54074.7 &131.7& \ & 16.45$\pm$0.19   &15.9$\pm$0.06&  15.59$\pm$0.04  &2 \\
15/12/2006  &54084.8 &141.8& 18.51$\pm$0.04&  \	& \  & 	\   &2  \\
15/12/2006  &54085.1 &142.1& 18.54$\pm$0.07&  16.94$\pm$0.02  &16.27$\pm$0.03&  \  &4 \\
16/12/2006  &54085.8 &142.8& 18.60$\pm$0.04&  \	 &\ &   \ &  2\\  
20/12/2006  &54090.1 &147.1& 19.26$\pm$0.09&  17.60$\pm$0.03  &16.81$\pm$0.02&  \  &4 \\
21/12/2006  &54090.7 &147.7& 19.30$\pm$0.21&  17.78$\pm$0.07  &16.96$\pm$0.04&  \ & 1\\ 
21/12/2006  &54091.1 &148.1& 19.17$\pm$0.04&  17.79$\pm$0.02  &16.96$\pm$0.02&  \ &4 \\
20/01/2007  &54120.6 &177.6&\  & 18.26$\pm$0.03  &17.30$\pm$0.02&  17.17$\pm$0.02  &2 \\
21/01/2007  &54121.6 &178.6&\  & 18.31$\pm$0.18  &17.25$\pm$0.06&  17.14$\pm$0.06  &2 \\
22/01/2007  &54122.6 &179.6&\  & 18.28$\pm$0.03  &17.30$\pm$0.02&  17.15$\pm$0.01  &2 \\
08/02/2007  &54140.6 &197.6&\  & 18.43$\pm$0.03  &17.50$\pm$0.02&  17.29$\pm$0.01  &2 \\
11/02/2007  &54142.7 &199.7&\ 19.61$\pm$0.23&  18.56$\pm$0.21  &17.64$\pm$0.13&  17.27$\pm$0.08  &1\\
11/02/2007 & 54142.8 &199.8&  \ &  18.48$\pm$0.03  &17.60$\pm$0.02&    17.30$\pm$0.01	&3\\
13/02/2007  &54146.0 &203.0& 19.74$\pm$0.13&  18.62$\pm$0.05 & 17.58$\pm$0.03&  \ &4 \\
15/02/2007  &54147.9 &204.9& 19.72$\pm$0.12&  18.60$\pm$0.04  &17.61$\pm$0.03&  \  & 4 \\
07/03/2007  &54168.0 &225.0& 20.01$\pm$0.04&  18.82$\pm$0.05  &17.79$\pm$0.03&  \ &4 \\
11/03/2007  &54171.8 &228.8&  \ & 18.85$\pm$0.03  &17.84$\pm$0.01&  17.56$\pm$0.02  &5  \\
15/04/2007  &54205.6 &262.6& 20.22$\pm$0.23&  19.15$\pm$0.06  &18.22$\pm$0.07&  17.81$\pm$0.05  &1 \\
 \hline
 \end{tabular}
 \begin{flushleft}
  * since explosion (JD 2453943.0)\\
  $^{a}$ unfiltered \\	
 1 = Ekar 1.82m+AFOSC\\
 2 = Liverpool Telescope+RATCam \\ 
 3 = NOT+ALFOSC\\
 4 = Faulkes Telescope North 2m+HawkCam1\\
 5 = WHT+AUXPort\\
 6 = IAUC  Circular No. 8773 by K. Itagaki (Teppo-cho, Yamagata, Japan, 0.60-m f/5.7 reflector)\\
 \end{flushleft}
\end{table*}

\begin{figure*}
 \includegraphics[width=8.4cm]{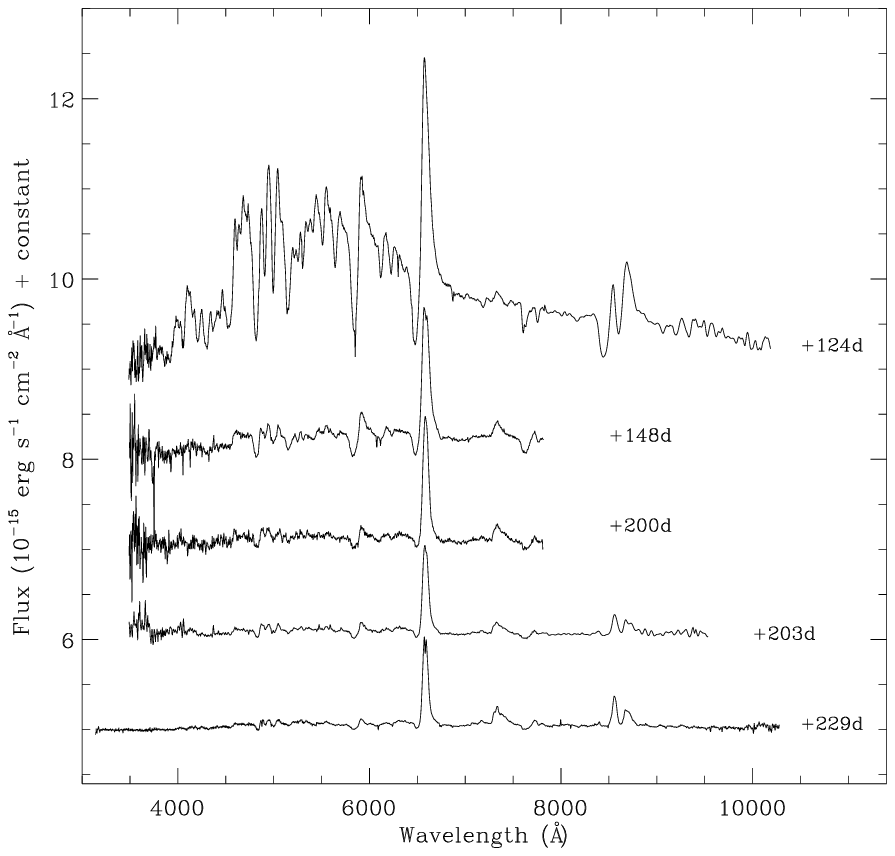}
 \caption{Optical spectral evolution of SN 2006my.}
\label{2006my_spec}
\end{figure*}

\begin{table*}
 \caption{Log of optical spectroscopic observations of SN 2006my.}
 \label{tab_opt7}
 \begin{tabular}{@{}lcccccc}
  \hline
  \hline
  Date  &  JD (2450000+)& Phase* (days)    &   Telescope +instrument	&   Grism& 	Range (\AA)    	&	Dispersion (\AA\ pixel$^{-1}$)    \\
    \hline
27/11/2006& 54066.7&123.7&  Mt. Ekar 1.82m+AFOSC  & gm2, gm4    &  3720--10200, 3480--8450&15.67, 4.99\\
21/12/2006& 54090.7&147.7&   Mt. Ekar 1.82m+AFOSC  & gm4    &  3480--8450&4.99\\
11/02/2007& 54142.7&199.7&   Mt. Ekar 1.82m+AFOSC  & gm4    &  3480--8450&4.99\\
14/02/2007& 54145.7&202.7&   Mt. Ekar 1.82m+AFOSC  & gm2, gm4    &  3720--10200, 3480--8450&15.67, 4.99\\
12/03/2007& 54171.8&228.8&  WHT+ISIS  & R158R, R300B         &   3200--5400, 5060--10200  &  1.81, 0.86 \\
\hline
 \end{tabular}
 \begin{flushleft}
   * since explosion (JD 2453943.0)\\
\end{flushleft} 
\end{table*}

\section{SN 2004A data}

\begin{table*}
 \caption{Log of previously unpublished optical photometric observations of SN 2004A.}
 \label{2006A}
 \begin{tabular}{@{}lccccccc}
 \hline
 \hline
   Date  &    JD(2450000+)   & Phase* (days)     &      \textit{B}     &          \textit{V}       &        \textit{R}    &         \textit{I}   &         Instrument  \\
  \hline
13/01/2004  &53018.2 & 7& 15.71$\pm$0.05  & 15.38$\pm$0.02 & 15.14$\pm$0.04 & 15.03$\pm$0.07 & 1  \\
01/02/2004  &53037.2 &26& 16.12$\pm$0.02  & 15.45$\pm$0.02  &15.06$\pm$ 0.12 & 14.63$\pm$0.02  & 1  \\
14/02/2004  &53050.0 &39& 16.33$\pm$0.04   & 15.44$\pm$0.03  & 15.00$\pm$0.10& 14.73$\pm$0.02    &2  \\
16/03/2004  &53080.6 & 70 & \ & \ & 15.14$\pm$0.11  & \  &1 \\
 \hline
 \end{tabular}
 \begin{flushleft}
 * since explosion (JD 2453011)\\
1 = Ekar 1.82m+AFOSC\\
2 = TNG 3.58m+LRS \\ 
 \end{flushleft}
\end{table*}

\begin{table*}
 \caption{Log of optical spectroscopic observations of SN 2004A.}
 \label{tab_opt8}
 \begin{tabular}{@{}lcccccc}
  \hline
  \hline
  Date  &  JD (2450000+)& Phase* (days)    &   Telescope +instrument	&   Grism& 	Range (\AA)    	&	Dispersion (\AA\ pixel$^{-1}$)    \\
    \hline
13/01/2004&53018.4& 7 & Mt. Ekar 1.82m+AFOSC  &  gm4    & 3480--8450&4.99\\
01/02/2004 &53037.4&  26 &Mt. Ekar 1.82m+AFOSC  & gm4    &  3480--8450&4.99\\
15/02/2004  &53051.7& 41 & TNG 3.58m+LRS  & LR-R, LR-B    & 4470--10073, 3000--8430&2.61, 2.52\\
1/03/2004&53080.6& 70 &Mt. Ekar 1.82m+AFOSC       &    gm2, gm4    &  3720--10200, 3480--8450&15.67, 4.99\\
\hline
 \end{tabular}
 \begin{flushleft}
* since explosion (JD 2453011)\\
\end{flushleft} 
\end{table*}
SN 2004A was a type IIP SN discovered in NGC 6207 about two weeks post explosion \citep{hen06}. Four new epochs of optical photometry and spectroscopy during the plateau phase are detailed in Table \ref{2006A} and Table \ref{tab_opt8} respectively. These photometry data are combined with those of \cite{hen06} and a bolometric light curve is formed. It is compared to the light curve of other IIP SNe in Figure \ref{com_opt}. Figure \ref{2004A_spec} shows the spectral evolution in the optical during the plateau phase.

\begin{figure*}
 \includegraphics[width=8.4cm]{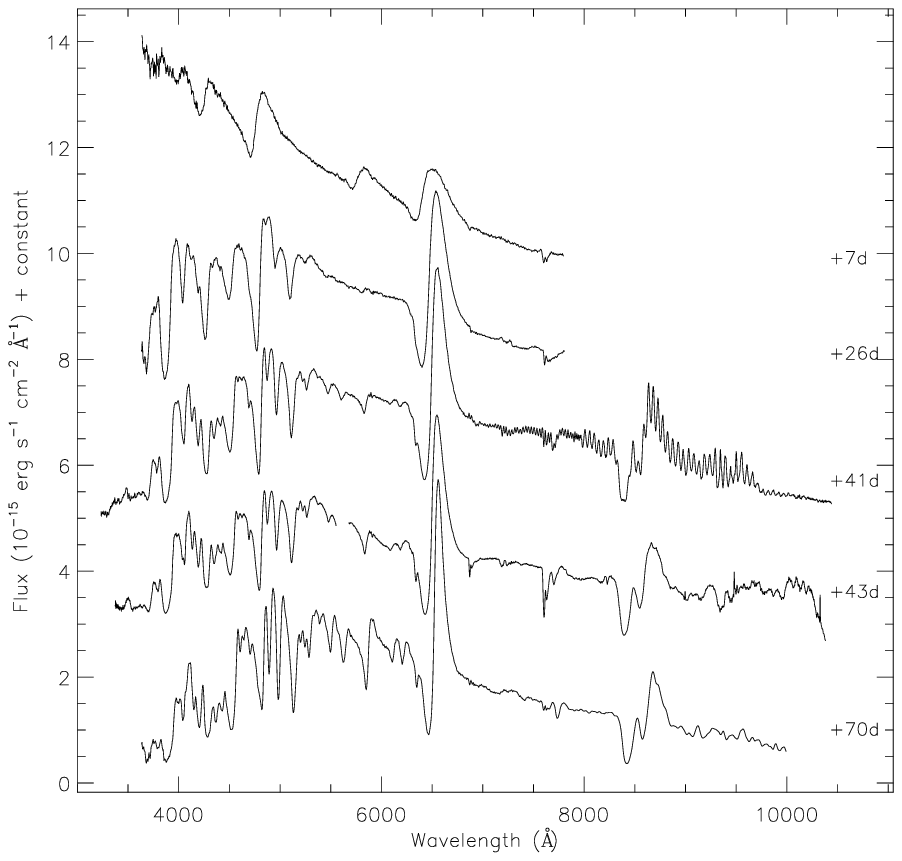}
 \caption{Optical spectral evolution of SN 2004A during the photospheric phase from +7 days to +70 days post explosion. The spectrum from +43 days is taken from \protect \cite{hen06}.}
\label{2004A_spec}
\end{figure*}


\begin{thebibliography}{99}

\bibitem[Arnett(1996)]{arn96}Arnett D., 1996, Supernovae and Nucleosynthesis: An Investigation of the History of Matter from the Big Bang to the Present, Princeton University Press, Princeton, NJ
\bibitem[Ashoka et al.(1987)]{ash87}Ashoka, B. N., Anupama J. C., Prabhu T. P., Giridhar S., Jain S. K., Pati A. K., Kameswara Rao N., 1987, JA\&A, 8, 195
\bibitem[Benetti et al.(1994)]{ben94}Benetti S., Cappellaro E., Turatto M., Della Valle M., Mazzali P. A., Gouiffes C., 1994, A\&A, 285, 147
\bibitem[Benetti et al.(2001)]{ben01}Benetti S., Turatto M., Balberg S. et al., 2001, MNRAS, 322, 361
\bibitem[Bersten \& Hamuy(2009)]{ber09}Bersten M., Hamuy M., 2009, ApJ, 701, 200
\bibitem[Beswick et al.(2004)]{bes04}Beswick R. J., Muxlow T. W. B., Argo M. K., Pedlar A., Marcaide J. M., 2004, IAU Circ., 8435 
\bibitem[Blinnikov et al.(2000)]{bli00}Blinnkov S., Lundqvist P., Bartunov O., 2000, ApJ, 532, 1132 
\bibitem[Bogdanov, Rybicki \& Grindlay(2007)]{bog07}Bogdanov S., Rybicki G. B., Grindlay J. E.,  2007, ApJ, 670, 668
\bibitem[Botticella et al.(2009)]{bot09}Botticella M. T., Pastorello A., Smartt S. J. et al., 2009, MNRAS, 398, 1041
\bibitem[Brown et al.(2007)]{bro07}Brown P., Dessart L., Holland S. T. et al., 2007, ApJ, 659, 1488
\bibitem[Chevalier(1986)]{che86}Chevalier R. A., 1986, ApJ, 308, 225
\bibitem[Chevalier, Fransson \& Nymark(2006)]{che06}Chevalier, R. A., Fransson C., Nymark T. K., 2006, ApJ, 641, 1029
\bibitem[Chugai(1994)]{chu94}Chugai N. N., 1994, ApJ, 428, L17
\bibitem[Chugai(2006)]{chu05}Chugai N. N., 2006, Astron. Lett., 32, 739
\bibitem[Chugai \& Utrobin(2000)]{chu00}Chugai N. N., Utrobin V. P., 2000, A\&A, 354, 557
\bibitem[Chugai et al.(2007)]{chu07}Chugai N. N., Chevalier R. A., Utrobin V. P., 2007, ApJ, 662, 1136
\bibitem[Crockett et al.(2009)]{cro09}Crockett R. M. et al., 2009, submitted
\bibitem[Danziger et al.(1991)]{dan91}Danziger I. J., Bouchet P., Gouiffes C., Lucy L. B., 1991, in Haynes R., Milne D., eds, Proc. IAU Symp., 148, The Large Magellanic Clouds, Sydney, Australia, p315
\bibitem[Dessart et al.(2008)]{des08}Dessart L., Blondin S., Brown P. J. et al., 2008, ApJ, 675, 644
\bibitem[Dolphin(2000)]{dol00}Dolphin A. E., 2000, PASP, 112, 1383
\bibitem[Eldridge \& Tout(2004)]{eld04}Eldridge J. J., Tout C. A., 2004, MNRAS, 353, 87
\bibitem[Elmhamdi, Chugai \& Danziger(2003)]{elm03a}Elmhamdi A., Chugai N. N., Danziger I. J., 2003a, A\&A, 404, 1077
\bibitem[Elmhamdi et al.(2003)]{elm03}Elmhamdi A., Danziger I. J., Chugai N. et al., 2003b, MNRAS, 338, 939
\bibitem[Fassia et al.(2001)]{fas01}Fassia A., Meikle W. P. S., Chugai N. et al., 2001, MNRAS, 325, 907
\bibitem[Filippenko(1997)]{fil97}Filippenko A. V., 1997, ARA\&A, 35, 309
\bibitem[Fransson \& Chevalier(1987)]{fra87}Fransson C., Chevalier, R. A., 1987, ApJ, 322, 15
\bibitem[Fransson \& Chevalier(1989)]{fra89}Fransson C., Chevalier, R. A., 1989, ApJ, 343, 323
\bibitem[Fransson et al.(1993)]{fra93}Fransson C., Houck J., Kozma C. et al., 1996, in McCray R., Wang Z., eds, Colloq. 145, Supernova and Supernova Remnants. Cambridge Univ. Press, Cambridge, p211
\bibitem[Fransson \& Kozma(2002)]{fra02}Fransson C., Kozma C., 2002, New AR, 46, 487
\bibitem[Fruchter \& Hook(2002)]{fru02}Fruchter, A. S., Hook, R. N., 2002, PASP, 114, 144 
\bibitem[Gerardy et al.(2000)]{ger00}Gerardy C. L., Fesen R. A., H\"oflich P., Wheeler J. C., 2000, AJ, 119, 2968
\bibitem[Hamuy et al.(2001)]{ham01}Hamuy M., Pinto P. A., Maza J. et al., 2001, ApJ, 558, 615
\bibitem[Hamuy \& Pinto(2002)]{ham02}Hamuy M., Pinto P. A., 2002, ApJ, 566, L63
\bibitem[Hamuy(2003)]{ham03}Hamuy M., 2003, ApJ, 582, 905
\bibitem[Heger et al.(2003)]{heg03}Heger A., Fryer C. L., Woosley S. E., Langer N., Hartmann D. H., 2003, ApJ, 591, 288
\bibitem[Hendry et al.(2005)]{hen05}Hendry M. A., Smartt S. J., Maund J. R. et al., 2005, MNRAS, 359, 906
\bibitem[Hendry et al.(2006)]{hen06}Hendry M. A., Smartt S. J., Crockett R. M. et al., 2006, MNRAS, 369, 1303
\bibitem[Hirschi et al.(2004)]{hir04}Hirschi R., Meynet G., Maeder A., 2004, A\&A, 425, 649
\bibitem[Krisciunas et al.(2009)]{kri08}Krisciunas K., Hamuy M., Suntzeff N. B. et al., 2009, AJ, 137, 34
\bibitem[Kasen \& Woosley(2009)]{kas09}Kasen D., Woosley S. E., 2009, ApJ, 703, 2205
\bibitem[Karachentsev, Sharina \& Huchtmeier(2000)]{kar00}Karachentsev I. D., Sharina M. E., Huchtmeier W. K., 2000, A\&A, 362, 544 
\bibitem[Kotak et al.(2009)]{kot09}Kotak R., Meikle W. P. S., Farrah D. et al., 2009, ApJ, 704, 306
\bibitem[Kozma \& Fransson(1998)]{koz98}Kozma C., Fransson C.,1998, ApJ, 497, 431 
\bibitem[Leonard et al.(2002a)]{leo02}Leonard D. C., Filippenko A. V., Gate E. L. et al., 2002a, PASP, 114, 35
\bibitem[Leonard et al.(2002b)]{leo02b}Leonard D. C., Filippenko A. V., Li W. et al., 2002b, AJ, 124, 2490
\bibitem[Leonard et al.(2003)]{leo03}Leonard D. C., Kanbur S. M., Ngeow C. C., Tanvir N. R., 2003, ApJ, 594, 247
\bibitem[Leonard et al.(2006)]{leo06}Leonard D. C., Filippenko A. V., Ganeshalingam M. et al., 2006, Nature, 440, 505
\bibitem[Li \& McCray(1992)]{li92}Li H., McCray R., 1992, ApJ, 387, 309L
\bibitem[Li et al.(2005)]{li05}Li W., Van Dyk S. D., Filippenko A. V., Cuillandre J. -C., 2005, PASP, 117, 121
\bibitem[Li et al.(2006)]{li06}Li W., Van Dyk S. D., Filippenko A. V., Cuillandre J. -C., Jha S., Bloom J. S., Riess A. G., Livio M., 2006, ApJ, 641, 1060
\bibitem[Limongi \& Chieffi(2003)]{lim03}Limongi M., Chieffi A., 2003, ApJ, 592, 404
\bibitem[Liu \& Dalgarno(1995)]{liu95}Liu W., Dalgarno A., 1995, ApJ, 454, 472
\bibitem[Litvinova \& Nad\"ezhin(1983)]{lit83}Litvinova I. Y., Nad\"ezhin D. K., 1983, Ap\&SS, 89, 89
\bibitem[Litvinova \& Nad\"ezhin(1985)]{lit85}Litvinova I. Y., Nad\"ezhin D. K., 1985, SvAL, 11, 145
\bibitem[Mart\'i-Vidal et al.(2007)]{mar07}Mart\'i-Vidal I. Marcaide J. M., Alberdi A. et al., 2007, A\&A, 470, 1071
\bibitem[Mattila et al.(2008)]{mat08}Mattila S., Smartt S. J., Eldridge J. J., Maund J. R., Crockett R. M., Danziger I. J., 2008, ApJ, 688, 91
\bibitem[Maund et al.(2005)]{mau05}Maund J. R., Smartt, S. J., Danziger, I. J., 2005, MNRAS, 364, 33
\bibitem[Meikle et al.(1989)]{mei89}Meikle W. P. S., Allen D. A., Spyromilio J., Varani G. -F., 1989, MNRAS, 283, 193
\bibitem[Misra et al.(2007))]{mis07}Misra K., Pooley D., Chandra P., Bhattacharya D., Ray A. K., Sagar R., Lewin W. H. G., 2007, MNRAS, 381, 280
\bibitem[Nad\"ezhin(2003)]{nad03}Nad\"ezhin D. K., 2003, 346, 97
\bibitem[Nugent et al.(2006)]{nug06}Nugent P., Sullivan M., Ellis R. et al., 2006, ApJ, 645, 841
\bibitem[Olivares(2008)]{oli08}Oilvares F., 2008, MSc thesis, University of Chile
\bibitem[Pastorello et al.(2004)]{pas04}Pastorello A., Zampieri L., Turatto M. et al., 2004, MNRAS, 347, 74
\bibitem[Pastorello et al.(2005)]{pas05}Pastorello A., Baron E., Branch D. et al., 2005, MNRAS, 360, 950
\bibitem[Pastorello et al.(2006)]{pas06}Pastorello A., Sauer D., Taubenberger S. et al., 2006, MNRAS, 370, 1752
\bibitem[Pastorello et al.(2009)]{pas09}Pastorello A., Valenti S., Zampieri L. et al., 2009, MNRAS, 394, 2266
\bibitem[Patat et al.(1994)]{pat94}Patat F., Barbon R., Cappellaro E., Turatto M., 1994, A\&A, 282, 731
\bibitem[Patat(1996)]{pat96}Patat F., 1996, PhD thesis, University of Padova
\bibitem[Poznanski et al.(2009)]{poz09}Poznanski D., Butler N., Filippenko A. V. et al., 2009, ApJ, 694, 1067
\bibitem[Pozzo et al.(2006)]{poz06}Pozzo M., Meikle W. P. S., Rayner J. T. et al., 2006, MNRAS, 368,1169
\bibitem[Pun et al.(1995)]{pun95}Pun C. S. J., 1995, ApJS, 99, 223
\bibitem[Rho et al.(2007)]{rho07}Rho J., Jarrett T. H., Chugai N. N., Chevalier R. A., 2007, ApJ, 666, 1108
\bibitem[Sahu et al.(2006)]{sah06}Sahu D. K., Anupama G. C., Srividya S., Muneer S., 2006, MNRAS, 372, 1315
\bibitem[Schmidt et al.(1993)]{sch93}Schmidt B. P., Kirshner R. P., Schild R. et al., 1993, AJ, 105, 6
\bibitem[Smartt et al.(2004)]{sma04}Smartt S. J., Maund J. R., Hendry M. A., Tout C. A., Gilmore G. F., Mattila S., Benn C. R., 2004, Sci, 303, 499
\bibitem[Smartt(2009)]{sma09b}Smartt S. J., 2009, ARA\&A, 47, 63
\bibitem[Smartt et al.(2009)]{sma09}Smartt S. J., Eldridge J. J., Crockett R. M., Maund J. R., 2009, MNRAS, 395, 1409
\bibitem[Smith et al.(2009)]{smi09}Smith N., Ganeshalingam M., Chornock R. et al., 2009, ApJ, 697, L49
\bibitem[Spyromilio et al.(1988)]{spy88}Spyromilio J., Meikle W. P. S., Learner R. C. M., Allen D. A., 1988, Nat, 334, 327
\bibitem[Spyromilio \& Leibundgut(1996)]{spy96}Spyromilio J., Leibundgut B., 1996, MNRAS, 283, L89
\bibitem[Spyromilio, Leibundgut \& Gilmozzi(2001)]{spy01}Spyromilio J., Leibundgut B., Gilmozzi R., 2001, A\&A, 376, 188
\bibitem[Suntzeff $\&$ Bouchet(1990)]{sun90}Suntzeff N. B., Bouchet P., 1990, AJ, 99, 650
\bibitem[Stockdale et al.(2004)]{sto04}Stockdale C. J., Weiler K. W., Van Dyk S. D., Sramek R. A., Panagia N., Marcaide J. M., 2004, IAU Circ., 8415
\bibitem[Tak\'ats \& Vink\'o(2006)]{tak06}Tak\'ats K., Vink\'o J., 2006, MNRAS, 372, 1735
\bibitem[Terndrup et al.(1988)]{ter88}Terndrup D. M., Elias J. H., Gregory B., Heathcote S. R., Phillips M. M., Suntzeff N. B., Williams R. E., 1988, Proc. Astron. Soc. Australia, 7, 412
\bibitem[Tsvetkov et al.(2006)]{tsv06}Tsvetkov D. Y., Volnova A. A., Shulga A. P., Korotkiy S. A., Elmhamdi A., Danziger I. J., Ereshko M. V., 2006, A\&A, 460, 769
\bibitem[Tsvetkov et al.(2007)]{tsv07}Tsvetkov D. Y., , Muminov M. M., Burkhanov O. A., Kahharov B. B., 2007, Peremennye Zvezdy (Variable Stars), 27, 4
\bibitem[Tsvetkov, Goranskiy \& Pavlyuk(2008)]{tsv08}Tsvetkov D. Y., Goranskiy V. P., Pavlyuk N. N., 2008, Peremennye Zvezdy (Variable Stars), 27, 8
\bibitem[Tsvetkov(2008)]{tsv08b}Tsvetkov D. Y., 2008, Peremennye Zvezdy (Variable Stars), 27, 9
\bibitem[Uomoto(1986)]{uom86}Uomoto A., 1986, ApJ, 310, L35
\bibitem[Utrobin(1993)]{utr93}Utrobin V. P., 2003, A\&A, 270, 249
\bibitem[Utrobin(2004)]{utr04}Utrobin V. P., 2004, Astron. Lett., 30, 293
\bibitem[Utrobin(2007)]{utr07}Utrobin V. P., 2007, A\&A, 461, 233
\bibitem[Utrobin \& Chugai(2008)]{utr08}Utrobin V. P., Chugai N. N., 2008, A\&A, 491, 507
\bibitem[Utrobin \& Chugai(2009)]{utr09}Utrobin V. P., Chugai N., N., 2009, preprint (arXiv:0908.2403)
\bibitem[Utrobin, Chugai \& Pastorello(2007)]{utr07b}Utrobin V. P., Chugai N. N., Pastorello A., 2007, A\&A, 475, 973
\bibitem[Valenti et al.(2008)]{val08}Valenti S., Benetti S., Cappellaro E. et al., 2008, MNRAS, 383,1485
\bibitem[Vinko et al.(2006)]{vin06}Vink\'o J., Tak\'ats K., S\'arnecsky K. et al., 2006, MNRAS, 369, 1780
\bibitem[Walborn et al.(1987)]{wal87}Walborn N. R., Lasker B. M., Laidler V. G., Chu Y-H, 1987, ApJ, 321, 41
\bibitem[Woosley \& Weaver(1995)]{woo95}Woosley S. E., Weaver T. A., 1995, ApJ, 101, 181
\bibitem[Yamaoka et al.(2004)]{yam04}Yamaoka H., Itagaki K., Klotz A., Pollas C., Boer M., 2004, IAU Circ., 8413, 2
\bibitem[Young(2004)]{you04}Young T. R., 2004, ApJ, 617, 1233
\bibitem[Zampieri et al.(2003)]{zam03}Zampieri L., Pastorello A., Turatto M., Cappellaro E., Benetti S., Altavilla G., Mazzali P., Hamuy M., 2003, MNRAS, 338, 711
\bibitem[Zampieri(2005)]{zam05}Zampieri L., 2005, in Turatto M., Benetti S., Zampieri L., Shea W., eds., ASP Conf. Ser., Vol 342, 1604--2004: Supernovae as Cosomlogical Lighthouses, Astron. Soc. Pac., San Francisco, p. 358 
\bibitem[Zampieri(2007)]{zam07}Zampieri L., 2007, AIPC, 924, 358
\bibitem[Zwitter, Munari $\&$ Moretti(2004)]{zwi04}Zwitter T., Munari U., Moretti S., 2004, IAU Circ., 8413

\end{thebibliography}
\end{document}